\documentclass[11pt,a4paper]{article}
\usepackage{jheppub}

\usepackage{graphicx}
\usepackage{amsmath}
\usepackage{amssymb}
\usepackage{braket}
\usepackage{hyperref}
\usepackage{frontespizio}
\usepackage{comment}
\usepackage{epsfig}
\usepackage{float}
\usepackage{color}
\usepackage[utf8]{inputenc}
\usepackage[T1]{fontenc}
\usepackage{enumitem}
\usepackage{stackengine}
\usepackage{scalerel}

\usepackage{subfigure}
\newcommand{\bea}{\begin{eqnarray}}
\newcommand{\eea}{\end{eqnarray}}
\newcommand{\be}{\begin{equation}}
\newcommand{\ee}{\end{equation}}
\newcommand{\ba}{\begin{align}}
\newcommand{\ea}{\end{align}}

\renewcommand{\d}{{\rm d}}
\newcommand{\lp }{\left (}
\newcommand{\llp }{\left [}
\newcommand{\rp }{\right )}
\newcommand{\rrp }{\right ]}
\newcommand{\PBH}{\text{\tiny PBH}}
\newcommand{\kpc}{{\rm kpc}}

\newcommand*{\TitleFont}{%
      \fontsize{15.8}{20}%
      \selectfont}

\newcommand{\sapienza}{Dipartimento di Fisica, Sapienza Universita 
	di Roma, Piazzale Aldo Moro 5, 00185, Roma, Italy}
\newcommand{\infn}{INFN, Sezione di Roma, Piazzale Aldo Moro 2, 00185, Roma, Italy}

\title{\TitleFont 
Hunt for light primordial black hole dark matter with ultra-high-frequency gravitational waves
}
\author[1,2]{Gabriele Franciolini,}
\author[3]{Anshuman Maharana,}
\author[4]{Francesco Muia}
\affiliation[1]{\sapienza}
\affiliation[2]{\infn}
\affiliation[3]{Harish-Chandra Research Institute, A CI of Homi Bhabha National Institute, Allahabad, Uttar Pradesh 211019, India}
\affiliation[4]{DAMTP, Wilberforce Road, Cambridge CB3 0WA, United Kingdom}
\emailAdd{gabriele.franciolini@uniroma1.it}
\emailAdd{anshumanmaharana@hri.res.in}
\emailAdd{fm538@cam.ac.uk}
\abstract{
Light primordial black holes may comprise a dominant fraction of the dark matter in our Universe. This paper critically assesses whether planned and future gravitational wave detectors in the ultra-high-frequency band could constrain the fraction of dark matter composed of sub-solar primordial black holes. Adopting the state-of-the-art description of primordial black hole merger rates, we compare various signals with currently operating and planned detectors. As already noted in the literature, our findings confirm that detecting individual primordial black hole mergers with currently existing and operating proposals remains difficult. Current proposals involving gravitational wave to electromagnetic wave conversion in a static magnetic field and microwave cavities feature a technology gap with respect to the loudest gravitational wave signals from primordial black holes of various
orders of magnitude. However, we point out that one recent proposal involving resonant LC circuits represents the best option in terms of individual merger detection prospects in the range $(1\div 100) \, \text{MHz}$. In the same frequency range, we note that alternative setups involving resonant cavities, whose concept is currently under development, might represent a promising technology to detect individual merger events. We also show that a detection of the stochastic gravitational wave background produced by unresolved binaries is possible only if the theoretical sensitivity of the proposed Gaussian beam detector is achieved. Such a detector, whose feasibility is subject to various caveats, may be able to rule-out some scenarios for asteroidal mass primordial black hole dark matter. We conclude that pursuing dedicated studies and developments of gravitational wave detectors in the ultra-high-frequency band remains motivated and may lead to novel probes on the existence of light primordial black holes.
}

\begin{document} 

\maketitle

\section{Introduction}

\textit{Dark Matter} (DM) constitutes around $\sim 25\%$ of the current energy density of our Universe~\cite{Bertone:2016nfn}. 
Yet, we understand very few of its properties -- it  interacts very weakly with standard matter and most of it cannot be relativistic. Its fundamental nature remains unknown.
The mass range of the currently proposed DM constituents varies from the $10^{-20} \, \text{eV}$ of ultra-light axions~\cite{Hu:2000ke, Amendola:2005ad, Hui:2016ltb} to the tens of solar masses of the heaviest \textit{Primordial Black Holes} (PBHs)~\cite{zel1967hypothesis, Hawking:1971ei, Carr:1974nx, Carr:1975qj, Chapline:1975ojl}. PBHs, in particular, represent an appealing candidate as this scenario may not require any particle beyond the Standard Model.

\begin{figure}[t!]
\centering
\includegraphics[width=.85\textwidth]{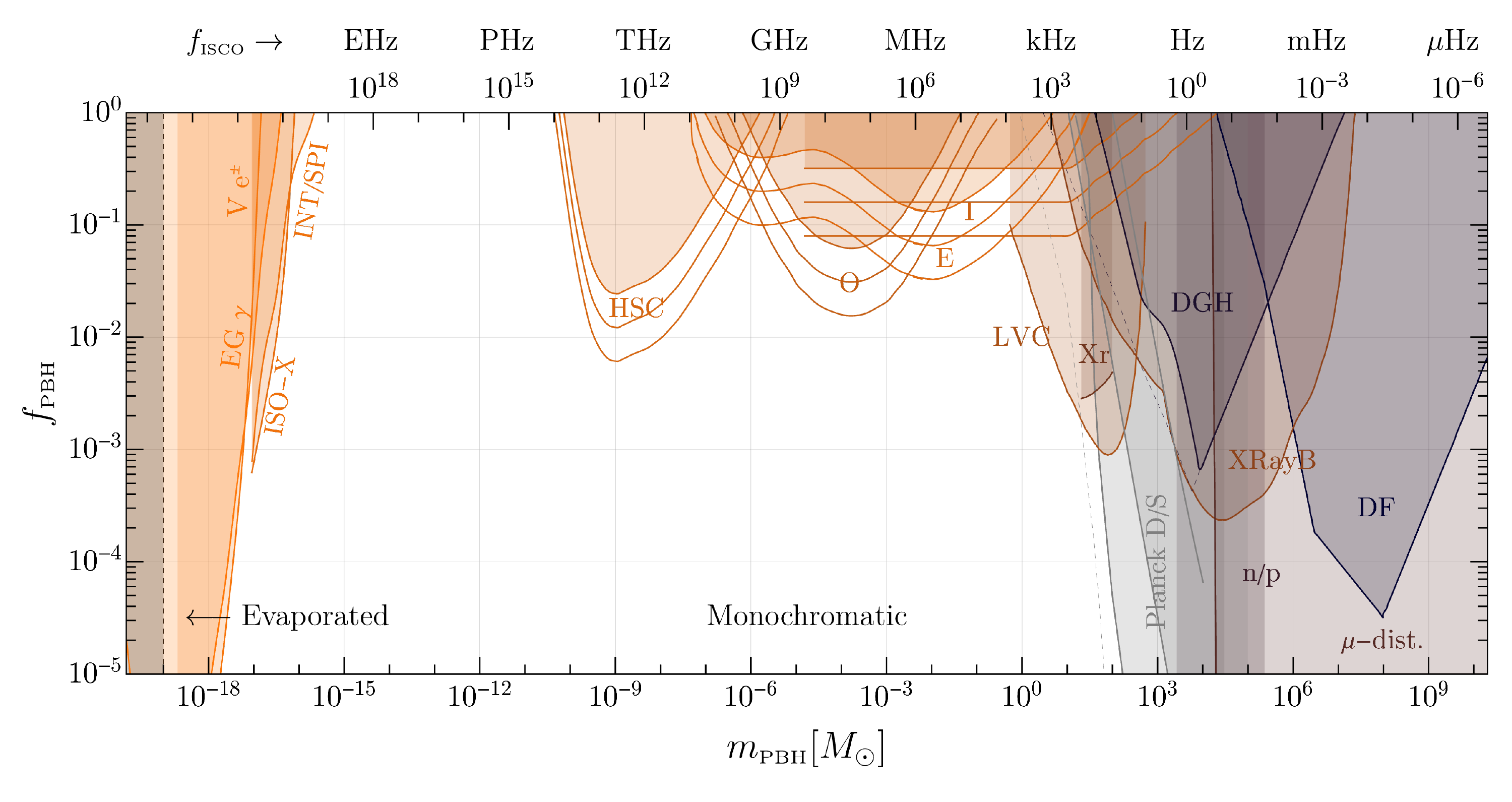}
\caption{
Most stringent costraints on the PBH abundance assuming a narrow mass distribution. 
Above the frame, we also show the corrisponding ISCO frequency for an equal mass merger with $m_1 = m_2 = m_\text{\tiny PBH}$. 
We indicate costraints from Hawking evaporation producing
extra-galactic $\gamma$-ray (EG $\gamma$)
\cite{Arbey:2019vqx}, $e^\pm$ observations by Voyager 1 (V $e^\pm$) \cite{Boudaud:2018hqb}, 
X-Ray~\cite{Iguaz:2021irx},
INTEGRAL/SPI observations~\cite{Berteaud:2022tws},
for other constraints in this mass range see also Refs.~\cite{Carr:2009jm, Ballesteros:2019exr,Laha:2019ssq, Poulter:2019ooo,Dasgupta:2019cae,Laha:2020vhg,DeRocco:2019fjq,Laha:2020ivk,Kim:2020ngi}).
We show microlensing searches by Subaru HSC \cite{Niikura:2017zjd, Smyth:2019whb}, MACHO/EROS \cite{Alcock:2000kd, Allsman:2000kg}, OGLE \cite{Niikura:2019kqi} and Icarus \cite{Oguri:2017ock}. 
We also plot various line corresponding to different assumptions on the fraction of PBHs in clusters \cite{Auclair:2022lcg}, where the most stringent line corresponds to the standard scenario where PBH clusters are seeded by Poisson initial conditions \cite{Petac:2022rio,Gorton:2022fyb}.
Observations of galactic 
X-rays (Xr)~\cite{Manshanden:2018tze} and X-Ray binaries (XRayB)~\cite{Inoue:2017csr},
and those coming from CMB distortions by spherical or disk accretion (Planck S and Planck D, respectively) \cite{Ali-Haimoud:2016mbv, Serpico:2020ehh},
Dwarf Galaxy heating (DGH)~\cite{Lu:2020bmd,Takhistov:2021aqx}, dynamical friction (DF)~\cite{Carr:2018rid}, the neutron-to-proton ratio (n/p)~\cite{Inomata:2016uip} and CMB $\mu$-distortions~\cite{Nakama:2017xvq}.
LVC stands for LIGO/Virgo Collaboration merger rate measurements
\cite{Ali-Haimoud:2017rtz,Raidal:2018bbj,Vaskonen:2019jpv,DeLuca:2020bjf, Hall:2020daa,DeLuca:2020sae,Wong:2020yig,Hutsi:2020sol,DeLuca:2021wjr}.
 We neglect the role of accretion affecting constraints on masses larger than ${\cal O}(10) M_\odot$ \cite{DeLuca:2020fpg,DeLuca:2020qqa,DeLuca:2020bjf}.%
}\label{fig: Constraints}
\end{figure}

 A necessary ingredient for PBHs to constitute DM is the existence of sizeable small-scale perturbations in the early Universe.
In the standard PBH formation model, the amplitude of curvature perturbations is enhanced by some mechanism operating during inflation (see, for example, Ref.~\cite{Sasaki:2018dmp} for a review). 
Explicit examples include: 
double inflation~\cite{Silk:1986vc, Kawasaki:1997ju, Kannike:2017bxn,Ashoorioon:2020hln}, inflection points in the inflationary potential~\cite{Garcia-Bellido:1996mdl, Alabidi:2009bk, Clesse:2015wea, Garcia-Bellido:2017mdw, Ballesteros:2017fsr, Hertzberg:2017dkh, Motohashi:2017kbs, Ballesteros:2020qam}, 
curvaton models~\cite{Kawasaki:2012wr, Carr:2017edp}, 
axion inflation models~\cite{Garcia-Bellido:2016dkw, Domcke:2017fix,Ozsoy:2018flq}. 
Other non-standard scenarios invoke, for instance, 
preheating effects~\cite{Green:2000he, Bassett:2000ha, Martin:2019nuw, Muia:2019coe, Cotner:2019ykd, Nazari:2020fmk}, 
early matter-domination~\cite{Green:1997pr, Harada:2016mhb, Krippendorf:2018tei, deJong:2021bbo, Padilla:2021zgm, DeLuca:2021pls, Das:2021wad}, collapse of cosmic strings~\cite{Polnarev:1988dh, Garriga:1993gj, Caldwell:1995fu, MacGibbon:1997pu, Helfer:2018qgv, Jenkins:2020ctp}
and late-forming PBHs \cite{Chakraborty:2022mwu}. 
As modes are stretched on super-horizon scales during inflation, curvature perturbations 
freeze-out until they re-enter the horizon during the post-inflationary epoch. At this point, if the amplitude of perturbations is larger than the threshold~\cite{Musco:2004ak, Polnarev:2006aa, Musco:2008hv, Musco:2012au, Musco:2018rwt, Kehagias:2019eil, Musco:2020jjb, Musco:2021sva}, they can collapse forming a population of PBHs with masses controlled by the energy contained in a Hubble volume. 
Various observations already constrain the fraction of DM composed of PBHs (usually denoted $f_{\text{\tiny PBH}} \equiv \Omega_\text{\tiny PBH}/\Omega_\text{\tiny DM}$). The bounds cover a wide spectrum of phenomenologically interesting range of PBH masses. We provide a summary of these in Fig.~\ref{fig: Constraints}, see  Ref.~\cite{Carr:2020gox} for a detailed recent review. 
The possibility of DM being constituted fully by PBHs is still allowed only in the so-called \textit{asteroidal} mass range $m_\text{\tiny PBH} \subset(10^{-16}\div 10^{-10}) \, M_\odot$~\cite{Katz:2018zrn, Montero-Camacho:2019jte},
where $M_{\odot} \simeq 2 \times 10^{33} \, {\rm g}$ denotes the mass of the Sun.
However, one should  keep in mind that many of the constraints in Fig.~\ref{fig: Constraints} are derived under specific assumptions, hence it is important to provide complementary and independent probes of potential light PBH populations.

The recent detection of \textit{Gravitational Waves} (GWs) by the LIGO/Virgo collaboration~\cite{TheLIGOScientific:2014jea,TheVirgo:2014hva} has turned out to be a novel powerful tool for the investigation of PBHs as DM. 
Soon after the very first GW detections, it was shown that PBHs could explain the observed GW signals while complying with the cosmological bound requiring them to be at most as abundant as the entirety of the DM~\cite{Bird:2016dcv,Clesse:2016vqa,Sasaki:2016jop}. 
By now, results from the various runs of observations set the most stringent constraints on $f_\text{\tiny PBH}$ in the solar mass range~\cite{Ali-Haimoud:2017rtz,Raidal:2018bbj,Vaskonen:2019jpv,DeLuca:2020bjf, Hall:2020daa,DeLuca:2020sae,Wong:2020yig,Hutsi:2020sol,DeLuca:2021wjr}, 
while they may still account for a fraction of the GW events~\cite{Franciolini:2021tla,Franciolini:2021xbq,Franciolini:2022iaa}.

The frequency of GWs emitted from BH mergers is tied to the \textit{Innermost Stable Circular Orbit} (ISCO) frequency
 \begin{equation}
\label{eq:fISCO}
f_\text{\tiny ISCO} \simeq  4.4 \times 10^3 \, {\rm Hz} \lp \frac{M_{\odot}}{m_1 + m_2} \rp,
\end{equation} 
where $m_1$ and $m_2$ are the masses of the two BHs.\footnote{The top frame of Fig.~\ref{fig: Constraints} reports the ISCO frequency that corresponds to a given BH mass (taking $m_1 = m_2$ for simplicity).}
Given Eq.~\eqref{eq:fISCO}, it is clear that ground-based interferometers such as LIGO, Virgo and KAGRA can only probe the final phase of mergers with masses slightly above the stellar mass.
One sees from Fig.~\ref{fig: Constraints} that high-frequency GW experiments extending beyond the kHz range are best suited to search for light PBHs
characterized by masses below $1 \, M_{\odot}$ corresponding to ISCO frequencies in the \textit{Ultra-High-Frequency} (UHF) band, i.e. to $f_\text{\tiny ISCO} > 10^3 \, \rm{Hz}$. 
In particular, the asteroidal mass window where PBHs are currently allowed to be the entirety of the DM corresponds to a maximum frequency above $f\gtrsim 10^{15} {\rm Hz}$.
Therefore, it is of great importance to extend our experimental capabilities beyond the $\rm{kHz}$.

Furthermore, there is no known astrophysical object that can produce GWs at a frequency higher than $\mathcal{O}(10) \, \rm{kHz}$. The absence of astrophysical contaminations implies that UHF-GW detectors can
potentially serve as  clean probes of new physics~\cite{Maggiore:1999vm, Aggarwal:2020olq}. Note that beyond the \textit{Stochastic GW Background} (SGWB) produced by unresolved PBH mergers, there are  a plethora of early Universe GW production mechanisms that can give rise to a stochastic signal in the UHF-GW band. All of these would  imply physics beyond the  \textit{ Standard Model} (SM) of particle physics.

The experimental status of UHF-GW searches is currently in a very preliminary stage. 
Several proposals exist, planning to cover the frequency range $(10^4 \div 10^{15}) \, \text{Hz}$ almost entirely. 
Some of the proposals have already been implemented in the form of prototypes or actual detectors, see Sec.~\ref{sec:Detectors} and Ref.~\cite{Aggarwal:2020olq} for more details. Yet, none of these are able to reach the required sensitivity to detect a cosmologically motivated GW signal. 
One of the goals of this paper is to critically assess the possibility of detecting GWs produced by a population of light PBHs potentially explaining a large fraction of the DM.

The paper is structured as follows. 
In Sec.~\ref{sec:PBHmergers} we review the theory of PBH mergers, 
including a discussion on the computation of the merger rate of binaries formed in the early Universe as well as an estimate of the maximum theoretical merger rate that may be attained in strongly clustered PBH scenarios. 
We include the enhancement due to the local (i.e. galactic) DM overdensity and recap how to estimate expected signal strain and duration. 
Then, we review the properties of the SGWB produced by unresolved PBH mergers, while also computing the memory effect and light boson superradiance.
In Sec.~\ref{sec:Detectors}, we review the current status of the experimental efforts to detect GWs at high-frequencies, including already operating detectors and various recent proposals, discussing which effort may be more promising for probing light PBHs. 
In Sec.~\ref{sec:conclusions}, we provide some future outlook and conclude. 
The data needed to reproduce the figures in this work are available upon request to the authors.

\section{Gravitational wave signatures of Primordial Black Holes}
\label{sec:PBHmergers}

A population of PBHs formed in the early Universe is expected to produce a variety of GW signals, see Refs.~\cite{Green:2020jor,Carr:2020xqk,Franciolini:2021nvv} for recent reviews. 
In this section we summarise the main predictions of the PBH model, providing a derivation for the benchmark quantities used in the upcoming section where 
an assessment of the detectability of GW signals produced by light PBHs is presented. 

\subsection{Gravitational waves from PBH formation and evaporation}

PBHs may form at very high redshift \cite{zel1967hypothesis,Hawking:1974rv,Chapline:1975ojl,Carr:1975qj,Ivanov:1994pa,GarciaBellido:1996qt,Ivanov:1997ia,Blinnikov:2016bxu} if the density perturbations overcome the threshold for collapse \cite{Musco:2020jjb,Escriva:2021aeh}. 
Their mass $m_\PBH$ is comparable to the mass contained in the cosmological horizon at the time of formation.
In particular, the scaling law relating $m_\PBH$ to the horizon mass $m_\text{\tiny H}$ for overdensities close to the critical threshold for collapse is \cite{Choptuik:1992jv, Evans:1994pj, Niemeyer:1997mt}
\be
m_\text{\tiny PBH} = \kappa \, m_\text{\tiny H} (\delta - \delta_c)^{\gamma_c},
\ee
where $\kappa = 3.3$ and $\gamma_c = 0.36$ in a radiation-dominated Universe \cite{Musco:2004ak,Musco:2008hv,Musco:2012au,Kalaja:2019uju,Escriva:2019nsa}.
Effectively, accounting for the statistical properties of curvature perturbations in the early Universe, the typical PBH mass is found to be around $m_\PBH \simeq 0.7 m_\text{\tiny H}$ \cite{Germani:2018jgr,Biagetti:2021eep}.
Thus, introducing the horizon scaling with redshift in the standard cosmological scenario, one finds a characteristic formation redshift of $z_\text{\tiny f} \approx  2 \times 10^{17} \lp {m_\PBH}/{10^{-12} M_\odot}\rp ^{-1/2}$ \cite{Sasaki:2018dmp}.
Finally, as the scale of inflation is bounded to be $H<6\times 10^{13}$ GeV by CMB observations \cite{Planck:2018jri}, the minimum PBH mass that can be formed in the early Universe is around $m_\PBH^\text{\tiny min} \simeq 10^{-33} M_\odot \approx 2 {\rm g}$.
 
It is interesting to notice that the same scalar perturbations generating PBHs are also responsible for the emission of GW at second order in perturbation theory \cite{Acquaviva:2002ud, Mollerach:2003nq, Ananda:2006af, Baumann:2007zm,Cai:2018dig,Bartolo:2018rku, Bartolo:2018evs,Unal:2018yaa,Bartolo:2019zvb,Wang:2019kaf,Cai:2019elf, DeLuca:2019ufz, Inomata:2019yww,Yuan:2019fwv,Pi:2020otn,Yuan:2020iwf,Romero-Rodriguez:2021aws,Balaji:2022rsy} (see Refs.~\cite{Yuan:2021qgz,Domenech:2021ztg} for reviews). 
As the horizon mass $m_\text{\tiny H}$ is related to the characteristic comoving frequency of perturbations by the relation \cite{Saito:2009jt}
\be
f \simeq 5\, {\rm kHz} \lp \frac{m_\text{\tiny H}}{10^{-24} M_\odot}\rp ^{-1/2},
\ee
one can immediately find that the formation of ultra-light PBHs may be associated with the emission of GW above the kHz frequency. 
Two comments, however, are in order at this point. 
Firstly, PBHs with masses below around $\lesssim 10^{-18} M_\odot$ evaporate within a timescale comparable to the age of the Universe due to the Hawking emission~\cite{Hawking:1976de, 1976PhRvD..13..198P}, and cannot account for a significant fraction of the dark matter.
This is because the evaporation timescale of a PBH with mass $m_\PBH$ is 
\be 
\tau_\PBH = \frac{10240\,\pi}{N^\text{\tiny evap}_\text{\tiny eff}}\, \frac{m_\PBH^3}{m_\text{\tiny Pl}^4}
\simeq 
10 \, {\rm Gyr } \lp \frac{N^\text{\tiny evap}_\text{\tiny eff}}{100}\rp^{-1}
\lp \frac{m_\PBH}{3 \times 10^{-19} M_\odot}\rp^{3},
\label{tau-BH}
\ee
where $m_\text{\tiny Pl} \simeq 1.22 \times 10^{19} \, \text{GeV}$ is the Planck mass and
$N^\text{\tiny evap}_\text{\tiny eff}$ is the number of particle species lighter than the BH temperature 
\begin{equation}
	T_\PBH=\frac{m_\text{\tiny Pl}^2}{8\pi m_\PBH} 
	\simeq 2 \times 10^{13}\, {\rm GHz} 
	\lp \frac{m_\PBH}{3\times 10^{-19} M_\odot}\rp^{-1}.
\end{equation}
Therefore, it is clear that UHF experiments could only probe GWs emitted by the formation mechanism of PBHs with masses so small that would have already evaporated in the early Universe. 
Such light PBHs are also expected to emit gravitons through Hawking evaporation, producing a SGWB from the early Universe~\cite{Dolgov:2011cq}. 
However, PBHs that can survive until the late-time Universe are expected to emit a negligible fraction of their mass in the form of GWs and, therefore, are not able to generate a detectable GW signature through this mechanism. 
We will not discuss further details of such a potential GW signature in this draft and we will focus on GW emission from {\it relatively} heavier (and stable) PBHs in the mass range $m_\text{\tiny PBH} \gtrsim 10^{-18} M_\odot$. 
It suffices to say that the emission of GWs from either PBH formation or evaporation would necessarily take place at a very high redshift and the maximum amplitude of such a SGWB is required to fall below the Big Bang Nucleosynthesis bound $\Omega_\text{\tiny GW} \lesssim 10^{-5}$ (see e.g. Ref.~\cite{Caprini:2018mtu}).

\subsection{Binary formation and merger rate}
\label{sec:MergerRate}

In the standard formation scenario, i.e. collapse in the radiation-dominated early Universe of Gaussian perturbations imprinted by the inflationary era, PBHs are expected to follow a Poisson spatial distribution at formation~\cite{Ali-Haimoud:2018dau,Desjacques:2018wuu,Ballesteros:2018swv,MoradinezhadDizgah:2019wjf,Inman:2019wvr}. 
We adopt this initial condition to derive the PBH merger rate, and present a discussion on the effect of initial clustering in Sec.~\ref{sec:clustering}.

PBH binaries decouple from the Hubble flow before matter-radiation equality if the distance $x$
separating two PBHs is smaller than the comoving distance~\cite{Nakamura:1997sm, Ioka:1998nz}
\begin{equation}
{x} = \lp \frac{3}{4\pi} \frac{m_1+m_2}{a_\text{\tiny eq}^3\rho_\text{\tiny eq}} \rp^{1/3},
\end{equation}
written in terms of the scale factor $a_\text{\tiny eq}$ and energy density $\rho_\text{\tiny eq}$ at matter-radiation equality.
The initial PBH spatial distribution dictates both the probability of decoupling as well as the properties of the PBH binaries. In particular, accounting for the distribution of binaries' semi-major axis and eccentricity \cite{Ali-Haimoud:2017rtz,Kavanagh:2018ggo,Liu:2018ess,Franciolini:2021xbq}, which determines the time it takes for a binary to harden and merge under the emission of GWs, one can derive a formula for the merger rate at time $t$ as~(e.g. \cite{Ali-Haimoud:2017rtz,Raidal:2018bbj}) 
\begin{align}
\frac{\d^2 R_\text{\tiny PBH} }{\d m_1 \d m_2}= 
\frac{3.8  \times 10^{-2}}{\rm kpc^3 \, yr}
f_\PBH^{\frac{53}{37}} \,
\lp \frac{t}{t_0} \rp^{-\frac{34}{37}}   
 \lp \frac{M_\text{\tiny tot}}{10^{-12}M_\odot} \rp^{-\frac{32}{37}}  
 \eta^{-\frac{34}{37}} 
S\lp M_\text{\tiny tot}, f_\PBH,\psi  \rp
\psi(m_1) \psi (m_2).
\label{PBHrate}
\end{align}
In Eq.~\eqref{PBHrate}, we introduced the current age of the Universe $t_0 = 13.8\,  {\rm Gyr}$, the
total mass of the binary $M_\text{\tiny tot}= m_1+m_2$, the symmetric mass ratio $\eta=m_1 m_2/M_\text{\tiny tot}^2$ and the PBH mass distribution $\psi (m)$ normalised such that $\int \d m \psi (m) = 1$.

We highlight the presence of the suppression factor $S\lp M_\text{\tiny tot}, f_\PBH,\psi  \rp \equiv S_1 \times S_2$
in Eq.~\eqref{PBHrate}, which corrects  the merger rate by 
introducing the effect of binary interactions with the surrounding environment in both the early- and late-time Universe. 
Specifically, the first contribution can be parameterized as~\cite{Hutsi:2020sol}
\begin{align}
	S_1 (M_\text{\tiny tot}, f_\PBH, \psi)& \thickapprox 1.42 \llp \frac{\langle m^2 \rangle/\langle m\rangle^2}{\bar N +C} + \frac{\sigma ^2_\text{\tiny M}}{f^2_\PBH}\rrp ^{-21/74} \exp \llp -  \bar N \rrp 
		\label{S1}
\end{align}
and reduces the PBH merger rate 
as a consequence of interactions close to the formation epoch between the forming binary and both the surrounding DM inhomogeneities with characteristic variance $\sigma_\text{\tiny M}^2 \simeq 3.6 \times 10^{-5}$
 as well as neighboring PBHs~\cite{Ali-Haimoud:2017rtz,Raidal:2018bbj,Liu:2018ess}, whose characteristic is
 \begin{equation}
\bar N \equiv \frac{M_\text{\tiny tot}}{\langle m \rangle } \frac{f_\PBH}{f_\PBH+ \sigma_\text{\tiny M}}.
\end{equation}
The explicit expression for the constant $C$ entering in Eq.~\eqref{S1} can be found in Ref.~\cite{Hutsi:2020sol}.
For a narrow mass function peaked at the mass scale $m_\PBH$, the expectation values simplify to become $\langle m \rangle \simeq  \langle m^2 \rangle^{1/2} \simeq  m_\PBH$, $M_\text{\tiny tot} \simeq 2 m_\PBH$, and $S_1$ is independent of the PBH mass. 
The second term $S_2$ includes the effect of successive disruption of binaries that populate PBH clusters formed from the initial Poisson inhomogeneities~\cite{Vaskonen:2019jpv,Jedamzik:2020ypm,Young:2020scc,Jedamzik:2020omx,DeLuca:2020jug,Trashorras:2020mwn,Tkachev:2020uin,Hutsi:2020sol}.
This is conservatively estimated assuming that the entire fraction of binaries included in dense environments is disrupted and reads \cite{Hutsi:2020sol}
\begin{align}\label{eq:supS2}
	S_2 (x) & \thickapprox 
	\text{min} 
	\llp 1\, ,\, 9.6 \times 10^{-3} x ^{-0.65} \exp \lp 0.03 \ln^2 x \rp  \rrp ,
\end{align}
with $x \equiv (t(z)/t_0)^{0.44} f_\PBH$. 
Recent numerical results on PBH clustering confirm the suppression factor we adopt in this work, 
provided one correctly accounts for the fraction of PBH binaries surviving outside dense PBH clusters~\cite{link}.
The suppression due to disruption in PBH sub-structures $S_2$ is at most ${\cal O}(10^{-2})$ for large $f_\PBH$, 
while it becomes negligible for small enough values of PBH abundance, i.e. $f_\PBH \lesssim 0.003$.
This estimate is compatible with the cosmological N-body simulations of Ref.~\cite{Inman:2019wvr}, where it was found that PBHs remain effectively isolated for sufficiently small $f_\PBH$.

\subsubsection{The role of late-time Universe dynamical capture for light PBHs}
We remark that other PBH binary formation mechanisms, taking place in the late-time Universe, may exist. 
For example, an alternative channel assumes PBH binary formation is induced by GW capture \cite{1989ApJ...343..725Q,Mouri:2002mc} in the present age dense DM environments. 
For initially Poisson distributed PBHs, the merger rate of binaries produced in the late-time Universe is subdominant with respect to the early Universe ones discussed above. 
This was explicitly shown for PBH binaries of masses around $m_\PBH\simeq 30 M_\odot$~\cite{Ali-Haimoud:2017rtz,Raidal:2017mfl,Korol:2019jud,Vaskonen:2019jpv,DeLuca:2020jug,link}.

Let us offer here a back-of-the-envelope computation showing why this conclusion would be even stronger for lighter PBHs. 
The cross-section for PBH capture, assuming equal mass objects, is 
$\Gamma_\text{\tiny cap} \simeq 44 \, m_\PBH^2  v_\text{\tiny rel}^{{-11/7}}$ \cite{1989ApJ...343..725Q,Mouri:2002mc}, 
while the binary formation rate in a PBH cluster takes the form $R_\text{\tiny env} \simeq r^3_\text{\tiny cl} n_\text{\tiny cl}^2 \langle  \Gamma_\text{\tiny cap} v_\text{\tiny rel}\rangle$. 
In the previous equations, we introduced the PBH characteristic relative velocity $v_\text{\tiny rel}$, the PBH cluster size $r_\text{\tiny cl}$ and local PBH number density $n_\text{\tiny cl}$ in the PBH cluster.
The merger rate of binaries produced by this channel is mostly dominated by small clusters, characterized by smaller virial velocities, that are able to survive dynamical evaporation until low redshift. 
The small scale structure, in a PBH DM scenario, is expected to be induced by the PBH initial Poisson noise (see, for example, Ref.~\cite{DeLuca:2020jug} for a more thorough discussion on PBH clustering evolution and Ref.~\cite{Inman:2019wvr} for a cosmological N-body simulation involving solar mass PBHs). Assuming the number density of such environments scales proportionally to the PBH number density, as is the case in the Press-Schechter theory \cite{1974ApJ...187..425P}, one finds that the PBH binary merger rate from PBH capture scales as
\begin{equation}
R_\text{\tiny PBH}^\text{\tiny cap} \sim m_\text{\tiny PBH}^{-11/21}.
\end{equation}
This derivation also assumes the collapsed PBH clusters are characterized by a density roughly $200$ times the mean density in the Universe at cluster formation (which does not depend on PBH masses), the size of clusters scales like $r_\text{\tiny cl} \sim  (M_\text{\tiny cl} / \rho_\text{\tiny cl})^{1/3}\sim m_\PBH^{1/3}$ and the virial velocity (i.e. approximately the characteristic PBH relative velocity) is $v_\text{\tiny rel} \sim (M_\text{\tiny cl}/r_\text{\tiny cl})^{1/2} \sim m_\text{\tiny PBH}^{1/3}$.
Therefore, taking the ratio between the merger rate for early-  and late-time Universe PBH binaries, that is 
\begin{equation}
\frac{R_\text{\tiny PBH}^\text{\tiny cap} }{R_\text{\tiny PBH}}  \sim m_\PBH^{265/777},
\end{equation}
one concludes  that capture is increasingly less relevant for light PBHs and can be safely neglected, at least in the standard PBH formation scenario discussed in this section.

\subsubsection{The role of disruptions in PBH clusters for light PBHs}
In the previous derivation of the merger rate, we included the suppression factor due to interactions with neighboring PBHs in clusters.\footnote{
The estimate of $S_2$ does not account for potential interactions with  astrophysical objects in the late-time Universe  (such as stars and astrophysically formed BHs). 
However, this phenomenon is not expected to significantly affect the PBH merger rate as the astrophysical environments are characterized by smaller densities and a larger velocity dispersion when compared to PBH small-scale clusters, reducing the probability of PBH binary disruption.} 
Let us show here that scaling the result in Eq.~\eqref{eq:supS2} obtained in the solar mass range to smaller masses leads to a consistent estimate for the merger rate at the present epoch. 
Indeed, one can show that the characteristic interaction time-scale $t_\text{\tiny p}$ in PBH clusters is only weakly dependent on the PBH mass.
In particular, we can define~\cite{Vaskonen:2019jpv}
\begin{equation}
	1/t_\text{\tiny p}  = n_\text{\tiny cl} \langle \sigma_\text{\tiny p} v_\text{\tiny rel}\rangle
\end{equation}
where $\sigma_\text{\tiny p}$ is the cross-section for scattering events that are able to increase the angular momentum of the binary by an amount comparable to its initial value. 
As the merger timescale is given by $\tau \simeq 3 a^4 j^7 / 170 m_\PBH^3$ \cite{Peters:1963ux,Peters:1964zz}, where the angular momentum is defined as $j\simeq \sqrt{1-e^2}$ in terms of the binary eccentricity $e$
and $a$ is the semi-major axis, such interactions may enhance merger time-delays, reducing the fraction of binaries that are able to merge within the age of the Universe.  
The disruptive interaction cross-section is $\sigma_\text{\tiny p} \approx 28 m^{7/4} \tau^{1/4}/v_\text{\tiny rel}^2 j^{29/12}$, and therefore one obtains
\begin{equation}
	t_\text{\tiny p} \sim  m_\PBH^{-10/111},
\end{equation}
which is only weakly dependent on PBH masses.
Additionally, the time-scale for dynamical relaxation potentially bringing PBH binaries in the cluster centers and enhancing PBH interactions is independent of the PBH mass \cite{binn,Vaskonen:2019jpv,DeLuca:2020jug}.
Based upon these considerations, we conclude that one can safely extrapolate the suppression factor computed for solar-mass PBH binaries to lower masses.

In order to bracket uncertainties on potentially modified PBH initial conditions, 
in the next section we are going to present an estimate for the maximum merger rate potentially attained in initially clustered PBH scenarios, where the binary formation rate is boosted. 
This will serve as an upper bound for the PBH merger rate used in the following sections.

\subsubsection{The effect of accretion on the PBH merger rate}
PBH binaries may experience efficient phases of accretion impacting individual masses, spins and the binary's orbital geometry \cite{DeLuca:2020bjf,DeLuca:2020fpg,DeLuca:2020qqa}. 
In this subsection, we provide an estimate showing why this potential effect is not expected to modify the merger rate of light PBHs. 

Due to the long characteristic time-scale for the accretion process compared to the characteristic binary period, one can assume that PBH masses vary adiabatically. 
Therefore, one can compute the modification of semi-major axis $a$ and eccentricity $e$ assuming 
the adiabatic invariants for the Keplerian two-body problem $I_\phi$ and $I_r $ are kept fixed, i.e.~\cite{landau1976mechanics}
\begin{align}
	I_\phi &= \frac{1}{2\pi} \int_0^{2\pi} p_\phi \d \phi = L_z \simeq \text{const.}  \\
	I_r &=  \frac{1}{2\pi} \int_{r_{\rm min}}^{r_{\rm max}} p_r \d r  = - L_z + \sqrt{M_\text{\tiny tot} \mu^2 a} \simeq \text{const.},
\end{align}
where we introduced the notation for the reduced mass $\mu = \eta M_\text{\tiny tot}$.
One obtains that the eccentricity $e$ is an adiabatic invariant while the semi-major axis evolves following
\be
\frac{\dot a}{a} + 3\frac{\dot m_\PBH}{m_\PBH} =0\,,
\ee
assuming equal mass binaries with $m_1 \simeq m_2 \simeq m_\PBH$. 
This shows that mass accretion shrinks binary orbits. 
Including this effect in the computation of the merger rate, Eq.~\eqref{PBHrate}, leads to an enhancement factor scaling as  \cite{DeLuca:2020qqa}
\begin{equation}\label{rateaccretionPBHbin}
R_\text{\tiny PBH} \propto
\lp 1 + \int \d t \frac{\dot m_\PBH}{m_\PBH} \rp^{9/37} 
\exp\llp \frac{36}{37} \int \d t \frac{\dot m_\PBH}{m_\PBH}
\rrp.
\end{equation}
At this point, the crucial ingredient is the efficiency of accretion for light PBHs. 
We are going to assume that the impact of secondary DM halos accumulating around PBHs (e.g. \cite{Ricotti:2007au,Ricotti:2007jk,Adamek:2019gns,DeLuca:2020qqa}) is small, 
which is inevitably the case when PBHs are a dominant component of the DM.\footnote{We note that much larger accretion rates may be obtained when the secondary DM halo becomes relevant. 
This has important consequences for PBH binary properties, and in particular the spin distribution when $m_\PBH \gtrsim M_\odot$ \cite{DeLuca:2020bjf,Franciolini:2021xbq,Franciolini:2022iaa} and comparison with constraints on the PBH abundance \cite{DeLuca:2020fpg,DeLuca:2020sae}.
We do not expect, however, this effect to be able to qualitatively change the results of this section as far as light (i.e. sub-solar) PBHs are concerned.
}
One can compute the accretion rate in a cosmological setting starting from high redshift (soon after binary formation epoch) using the Bondi-Hoyle formula.
As shown in Ref.~\cite{DeLuca:2020qqa}, the relevant effective velocity is modulated by the orbital evolution while the baryonic gas density $\rho$ is enhanced by the binary attracting matter on its center of mass. One finds that 
\begin{align}
\dot m_{\PBH,1} = \dot M_\text{\tiny bin}  \frac{1}{\sqrt{2 (1+q)}}, 
	\qquad
\dot m_{\PBH,2} = \dot M_\text{\tiny bin}  \frac{\sqrt{q} }{\sqrt{2 (1+q)}}, 
\label{M1M2dotFIN}
\end{align}
which simplifies to $\dot m_{\PBH,1,2} 
\sim \dot M_\text{\tiny bin}/2 $ 
for equal mass binaries, where the binary gas accretion rate is
\be \label{R1bin}
\dot M_\text{\tiny bin} = 4 \pi \lambda \rho_\text{\tiny gas} v^{-3}_\text{\tiny eff} M^2_\text{\tiny tot}\,,
\ee
in terms of the effective velocity $v_\text{\tiny eff}$ and the accretion eigenvalue $\lambda$ (see Ref.~\cite{Ricotti:2007au} for more details).
Without the effect of a DM halo, this rate was estimated in Ref.~\cite{Ali-Haimoud:2016mbv} to be
\begin{equation}
\frac{\dot M_\text{\tiny bin}}{\dot M_\text{\tiny Edd}} \sim 10^{-5} \lp \frac{M_\text{\tiny tot}}{M_{\odot}}\rp,
\end{equation}
when normalised to the Eddington rate $\dot{M}_\text{\tiny Edd} \equiv L_\text{\tiny Edd} \approx 2 m_\PBH / \text{Gyr}$ and for redshift below $z\simeq 10^2$, where the mass accretion integrals in Eq.~\eqref{rateaccretionPBHbin} are dominated.
Therefore, even conservatively assuming a prolonged accretion phase lasting until well within the reionization and structure formation epochs, one finds that
\begin{equation}
\int dt \frac{\dot{m}_\PBH}{{m}_\PBH} \sim 3 \times 10^{-4}  \lp \frac{{m}_\PBH}{M_{\odot}} \rp,	
\end{equation}
showing accretion on light PBH binaries is irrelevant and cannot affect the merger rate through the corrections shown in Eq.~\eqref{rateaccretionPBHbin}.

\subsubsection{Maximum theoretical merger rate}\label{sec:clustering}

The description of the merger rate in the previous section is based 
on the standard scenario of PBH formation where binaries are assembled from an initially Poisson distributed population of PBHs. 
In this section we explore whether changing this initial condition may give rise to larger PBH merger rates.

It is known that local non-Gaussianities of primordial curvature perturbations can modify the initial distribution of PBHs and make them clustered directly at the formation time \cite{Tada:2015noa,Young:2015kda,Young:2019gfc,Atal:2020igj,DeLuca:2021hcf}.\footnote{The clustering of PBHs with masses $\gtrsim M_\odot$ could be significantly constrained in the future through CMB distortion observations, probing the scales corresponding to the average PBH distance at formation \cite{DeLuca:2021hcf}. 
However, the clustering scales corresponding to lighter PBHs considered in this work are much smaller and not observable with this technique.} 
In this scenario, since the PBHs are closer to each other on average, the binary formation rate is enhanced due to a larger probability of decoupling from the Hubble flow. 
However, it is fair to say that little is known about the cosmological evolution of binaries in PBH halos produced in such a clustered scenario. On the one hand, the presence of dense PBH halos would likely enhance the rate of binary-PBH gravitational interaction, leading to binary disruptions and an effective suppression of the merger rate at late time. On the other hand, dense PBH clusters tend to evaporate due to the gravitational relaxation process \cite{binn,DeLuca:2020jug}, limiting the potential extent of this suppression. These dynamical processes are still poorly modeled for clustered scenarios and, therefore, we will purposely neglect potential suppression factors and adopt this as the maximum theoretical PBH merger rate. 
We will also neglect the contribution to the merger rate from disrupted binaries which are still merging within the age of the Universe and may potentially become relevant for large $f_\text{\tiny PBH}$ \cite{Vaskonen:2019jpv}. 

Clustered PBH scenarios may also boost the merger rate of binaries dynamically formed in the late-time Universe (e.g. through capture \cite{1989ApJ...343..725Q,Mouri:2002mc} or three-body interactions \cite{Ivanova:2005mi,2010ApJ...717..948I,Rodriguez:2021qhl,Kritos:2020wcl}).\footnote{We note that hyperbolic encounters \cite{Garcia-Bellido:2017knh,Garcia-Bellido:2021jlq,Morras:2021atg} give rise to very fast burst of GW radiation, which may be hard to detect with UHF GW experiments. Additionally, they give a small contribution to the SGWB compared to binary PBHs \cite{Garcia-Bellido:2021jlq}. We will therefore neglect this GW source in our considerations. }
This holds, however, provided PBH clusters are able to survive dynamical relaxation until late times \cite{binn}.
Since a proper assessment of the merger rate in such scenarios is still lacking in the literature, 
we neglect such effect and only discuss the maximum theoretical merger rate attainable from early Universe binary formation from clustered PBHs~\cite{Raidal:2017mfl}.

\begin{figure}[t!]
\centering
\includegraphics[width=0.6\textwidth]{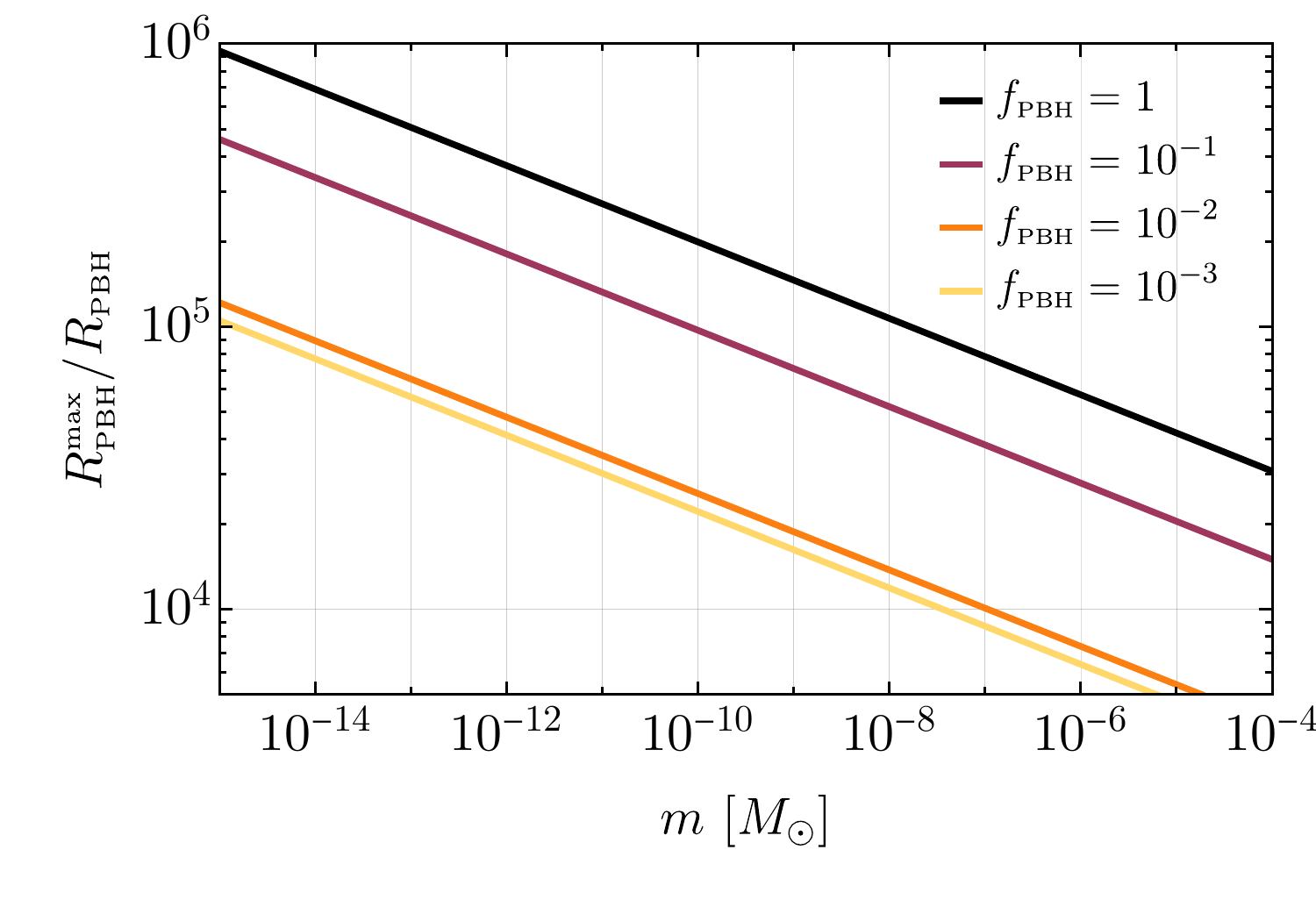}
\caption{
Ratio between the maximum attainable merger rate in clustered scenarios and the merger rate in the standard PBH formation scenario. The ratio scales non-linearly with $f_\PBH$ and grows again as $\propto 1/f_\text{\tiny PBH}$ for small enough values of the abundance. 
}\label{fig: max rate}
\end{figure}

Following the notation introduced in Ref.~\cite{Raidal:2017mfl},
we define the local PBH overdensity in the early Universe  $\delta_\text{\tiny dc}$ in terms of the PBH correlation function $\xi_\text{\tiny  PBH}$ at formation, assuming it is constant up to the binary scale $\tilde{x}$ at the decoupling epoch~\cite{Tada:2015noa, Raidal:2017mfl, Suyama:2019cst, Atal:2020yic, DeLuca:2021hcf}, that is $\delta_\text{\tiny dc} \approx 1 + \xi_\text{\tiny  PBH}  (x)$ when $x < \tilde{x}$.
In the limit in which the local fraction of PBHs is large,
 i.e. $\delta_\text{\tiny dc} f_\PBH \gg 1$, the merger rate in the clustered scenario can be written as \cite{DeLuca:2021hde}
\begin{align}
\label{eq:PBHrateclu}
R^\text{\tiny max}_\PBH
	&\simeq 
	\frac{57}{{\rm kpc}^3 {\rm yr}}
	\,
	 f_\PBH 	\lp \frac{t}{t_0}\rp^{-1}  \lp \frac{m_\PBH}{10^{-12} M_\odot}\rp^{-1}  \lp 1 + 1.7 \times 10^{-4} \Delta_\text{\tiny dc} \rp \exp\llp- 9.5 \times 10^{-5} \Delta_\text{\tiny dc}\rrp,
\end{align}
where we defined 
\be
\label{eq:critdc}
\Delta_\text{\tiny dc} =  
6 \times 10^{-5}\,
\delta_\text{\tiny dc} f_\PBH 	\lp \frac{t}{t_0}\rp^{3/16}  \lp \frac{m_\PBH}{10^{-12} M_\odot}\rp^{5/16}.
\ee
In Fig.~\ref{fig: max rate} we plot the ratio between the maximum attainable merger rate in clustered scenario and the merger rate in the standard PBH formation scenario. 
One finds that the rate can never be enhanced by more than six orders of magnitude in the mass range of our interest.
As we will see in the following, this extreme scenario would correspond to a reduction of the average distance of a
PBH merger from us in one year of observations of {\it at most} two orders of magnitude.

In the extreme scenarios where the PBH merger rate is drastically enhanced by some clustering mechanism, 
the rate of conversion of PBH binaries to freely propagating GWs would be high and could possibly be constrained from dark matter to dark radiation conversion bounds (see e.g. \cite{DES:2020mpv}).
We leave a proper study of this possibility to future work. 
On the other hand, even though $\Omega_\text{\tiny GW}$ could  become larger than around $\approx 10^{-5}$, the BBN bound would not apply, as most of the GW energy density 
would only be produced at late times.

\subsection{Impact of local DM enhancement}
\label{sec:DMenhancement}

For sources that are closer to the Earth than ${\cal O}(100)$ kpc, a correction due to the local DM overdensity needs to be taken into account. 
Following Ref.~\cite{Pujolas:2021yaw}, we model the Milky Way DM halo 
as a Navarro-Frenk-White density profile~\cite{Navarro:1995iw,Navarro:1996gj},
\be
	\rho_\text{\tiny{DM}}(r) = \frac{\rho_0}{\frac{r}{r_0} \left(1+\frac{r}{r_0}\right)^{\!2}} \,.
\ee
We fix the reference energy density $\rho_0$ 
such that the local DM density is 
$\rho_\text{\tiny DM}(r=r_\odot) = 7.9\times 10^{-3}M_\odot/{\rm pc}^3$~\cite{Cautun:2019eaf}, 
while the reference scale is $r_0 = 15.6\, \kpc$ and the solar system location is $r_\odot \simeq  8.0\,\kpc$. 
As we will compute the number density of sources within a volume centered at the solar system location, the average overdensity within a volume of radius $r$ around $\hat r \simeq  r_\odot$ can be estimated by
effectively cutting the density profile at $r_\odot$
\begin{align}
	\rho(r-\hat r)=
\begin{cases}
		\rho_\text{\tiny DM}(r_\odot)   &,   \quad   r-\hat r < r_\odot,	
\\
		\rho_\text{\tiny DM}( r-\hat r)         &,   \quad   r-\hat r \gtrsim  r_\odot,
\end{cases}
\end{align}
see Ref.~\cite{Pujolas:2021yaw} for more details.
\textcolor{black}{As the volume average of the dark matter overdensity around an observer located on the earth when $r \ll r_\odot$ is dominated by the value 
of the local density $\rho_\text{\tiny DM}(r_\odot)$, it follows that our estimates are not much sensitive to deviations from the NFW profile at small scales towards the galactic center (e.g. cored DM profiles). }
As we expect the distribution of binaries to follow the large/galactic scales, 
the local DM overdensity enhances the merger rate in Eq.~\eqref{PBHrate} by an overall factor of 
\begin{equation}
\label{eq:Rlocal}
	R_\text{\tiny PBH} ^\text{\tiny local} (r) = \delta (r) R_\text{\tiny PBH} \,,
\end{equation}
where we defined the overdensity factor $\delta(r) \equiv \rho_\text{\tiny DM}(r)/\bar \rho_\text{\tiny DM}$. 
Therefore, one finds that this correction falls within the range
$
	\delta(r) \subset(1 \div 2 \times 10^5) \,	
$.

\begin{figure}[t!]
\centering
\includegraphics[width=0.49\textwidth]{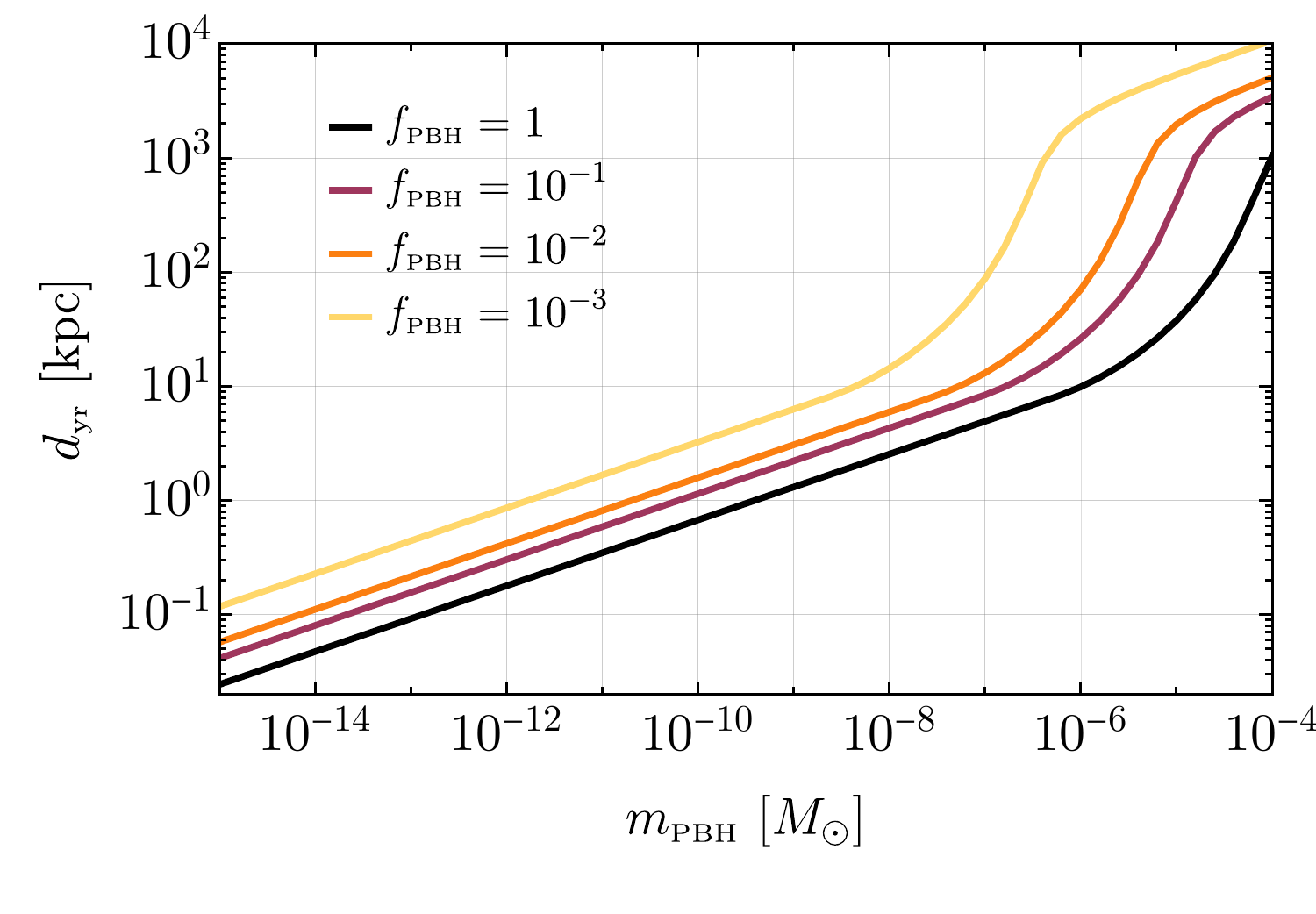}
\includegraphics[width=0.49\textwidth]{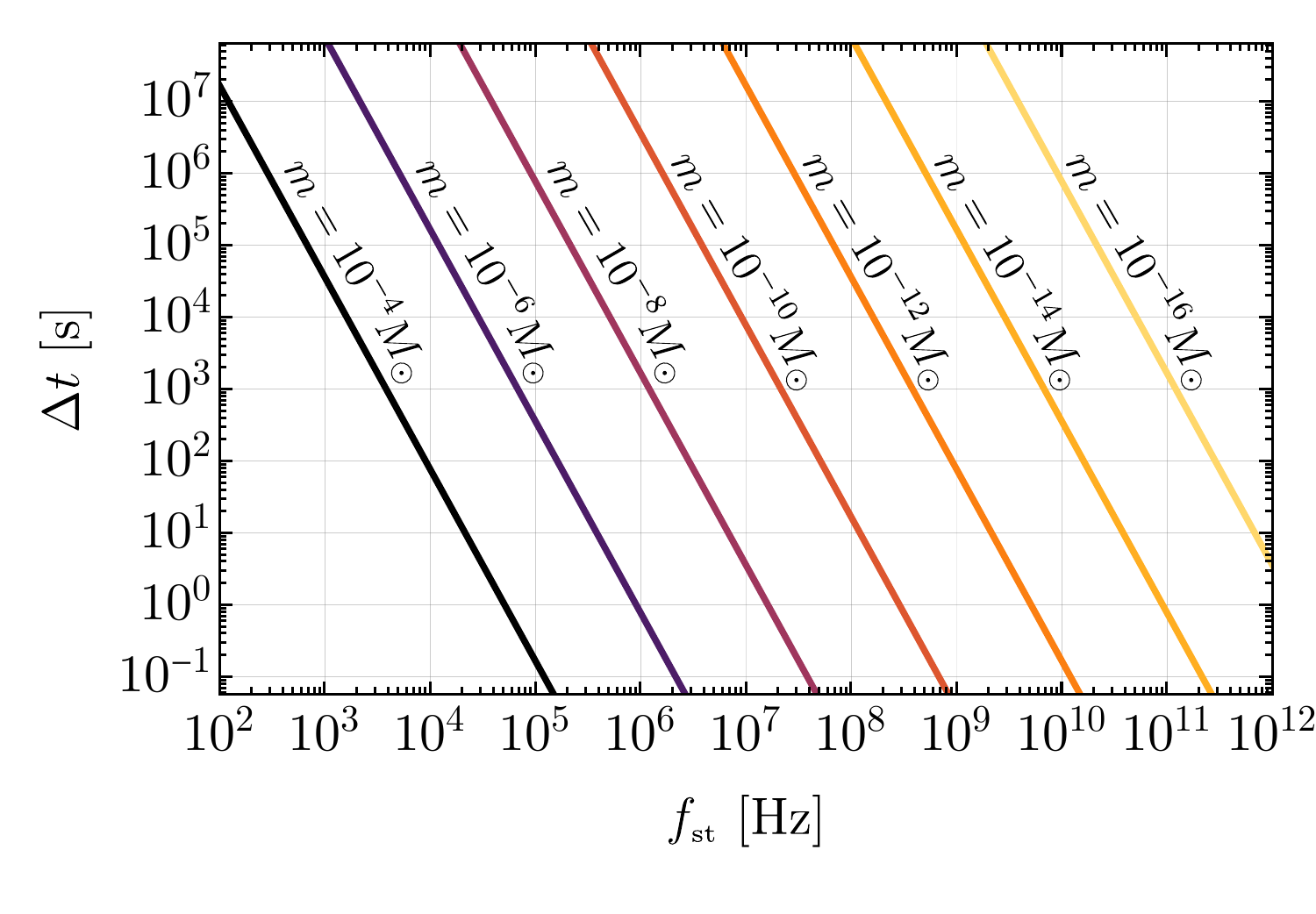}
\caption{%
{\bf Left: }
Charactetistic size of a region containing at least a merger event per year. 
The change in slope happening around $10^{-5}M_\odot$ corresponds to where the local DM 
enhancement start decreasing, i.e. for $r\gtrsim r_\odot$. 
{\bf Right:} Time it takes for a BH binary of masses $m_1 = m_2 = m$ to span a range of frequencies at least as large as half a decade above $f_\text{\tiny st}$.
}\label{fig: dist}
\end{figure}

Accounting for this local enhancement factor, we compute the volume $V_\text{\tiny yr}$, 
or equivalently the distance $d_\text{\tiny yr}  \equiv  \lp 3 V_\text{\tiny yr}/ 4 \pi  \rp ^{1/3}$, 
enclosing the region where one expects at least one merger per year, on average. 
We will neglect the effect of cosmological redshift as it is irrelevant for the small distances with which we are concerned.
We define the number of events per year $N_\text{\tiny yr}$ within the volume $V_\text{\tiny yr}$ as 
\begin{equation}
N_\text{\tiny yr} \equiv 
\Delta t \int^{d_\text{\tiny yr}}_0 \d r 4 \pi r^2 R^\text{\tiny local}_\text{\tiny PBH}(r)
\,,
\end{equation}
where we set $\Delta t = 1\,{\rm yr}$.
In Fig.~\ref{fig: dist}, we show the distance $d_\text{\tiny yr}$ as a function of PBH masses and abundance for $N_\text{\tiny yr} = 1$ and assuming a narrow PBH mass distribution. It is interesting to notice that when the characteristic merger distance becomes larger than ${\cal O}(10)$ kpc, the galactic overdensity decreases significantly and $d_\text{\tiny yr}$ changes slope, leading to steeper dependence on PBH mass. Once $\delta(r) \sim {\cal O}(1) $, the slope goes back to the one expected from the volume factor and a constant merger rate per unit volume. 

\subsection{Gravitational wave strain and signal duration}
\label{sec:SignalDuration}

As we will see in the following, two crucial properties of PBH mergers affect the binary detectability.
These are the characteristic GW strain and the GW signal duration. 
The leading-order GW signal from a BH inspiral for the two polarizations $h_{\scriptscriptstyle{+,\times}}$ in the stationary phase approximation (assuming that the frequency varies slowly) can be written as \cite{Maggiore:1900zz}
\begin{equation}
h_{\scriptscriptstyle{+,\times}}(t) = h_0 \, F_{\scriptscriptstyle{+,\times}}(\theta) \, G_{\scriptscriptstyle{+,\times}}(t) \,,
\end{equation}
where $F_{\scriptscriptstyle{+,\times}}(\theta)$ is a function that depends on the binary orientation angle $\theta$, $G_{\scriptscriptstyle{+,\times}}(t)$ corresponds to the binary oscillation phase, and
\begin{align}
\label{eq:GWamplitude}
h_0 &= 
\frac{4}{d_L} 
\left(G m_c \right)^{5/3} 
\left(\pi f \right)^{2/3} 
\nonumber \\
&\simeq 9.77 \times 10^{-34} \left(\frac{f}{1 \, \text{GHz}}\right)^{2/3} \left(\frac{m_\text{\tiny PBH}}{10^{-12} \, M_\odot}\right)^{5/3} \left(\frac{d_L}{1 \, \text{kpc}}\right)^{-1} \,,
\end{align}
where $m_c = {(m_1 m_2)^{3/5}}/{(m_1 + m_2)^{1/5}}$ is the chirp mass for two BHs with masses $m_1$ and $m_2$ (in the last step we have used $m_1 = m_2 = m_\text{\tiny PBH}$), 
and $d_\text{\tiny L}$ is the distance from the observer.
Adopting the stationary phase approximation, the GW signal in Fourier space is~\cite{Sathyaprakash:2009xs}
\be\label{eq:spawaveform}
\tilde{h}_{\scriptscriptstyle{+,\times}}(f) = {\mathcal A}_{\scriptscriptstyle{+,\times}} e^{i \Psi_{\scriptscriptstyle{+,\times}}(f)}, 
\ee
where the explicit expressions for $\Psi_{\scriptscriptstyle{+,\times}}(f)$ and $\mathcal{A}_{\scriptscriptstyle{+,\times}}$ are given e.g. in~\cite{Maggiore:1900zz}. 
Assuming equal mass binaries, the characteristic strain $h_c (f)\equiv 2 f | \tilde h(f) | $ is 
\begin{equation}\label{eq:spawaveform2}
|h_c (f) | \simeq 4.54 \times 10^{-28} \lp \frac{m_\text{\tiny PBH}}{10^{-12} M_\odot }\rp^{5/6} \lp \frac{d_\text{\tiny L}}{\rm kpc} \rp^{-1} \lp \frac{f}{\rm GHz}\rp ^{-1/6} \,,
\end{equation}
where we have used that ignoring the angular dependence one has $|\tilde{h}(f)| \approx |\tilde{h}_{\scriptscriptstyle{+}}(f)| \approx |\tilde{h}_{\scriptscriptstyle{\times}}(f)|$.
This modeling of the GW signal only includes the inspiral phase of the binary up to the ISCO frequency in Eq.~\eqref{eq:fISCO}, before the objects plunge, merge and the ringdown signal is emitted by the remnant BH reaching its stationary configuration. 
This is, however, sufficient for our purposes as only the GW signal produced during the inspiral phase can last for a sufficiently long time to allow for potential detection (see more details in Sec.~\ref{sec:Detectors}). 
We also observe from Fig.~\ref{fig: dist} that for binaries at the edge of the galactic DM enhancement (e.g. $m_\PBH \gtrsim 10^{-6} M_\odot$ and high $f_\PBH $), the characteristic distance grows roughly as $d_\text{\tiny yr}\propto m_\PBH$. On the other hand, the characteristic strain scales as $h_c \propto m_\PBH^{5/6}$. 
This means that one expects a similar strain from characteristic inspiraling sources within such a mass range, unless $m_\PBH \gtrsim 10^{-3} M_\odot$.

An important property of inspiraling sources is the GW signal duration. If we consider an equal mass PBH binary with $m_1 = m_2 = m_\PBH$, the coalescence time can be written as~\cite{Maggiore:1900zz}
\begin{equation}
\label{eq:CoalescenceTime}
\tau(f) \approx 83 \, {\rm{sec}} \, \left(\frac{m_\PBH}{10^{-12} M_\odot}\right)^{-5/3} \left(\frac{f}{\rm{GHz}}\right)^{-8/3} \,.
\end{equation}
Using Eq.~\eqref{eq:CoalescenceTime}, one can find the time spent by the inspiral phase to span a given frequency interval.
This quantity will be crucial when computing the detector sensitivities in Sec.~\ref{sec:Detectors}.
In Fig.~\ref{fig: dist}, we show the time it takes for an equal-mass binary to span at least half a decade of frequencies. 
We warn the reader, however, that the time spent spanning a very narrow resonant frequency band could be much smaller than what is estimated in Eq.~\eqref{eq:CoalescenceTime}. We will discuss this in detail in the next section.

\subsubsection{GW amplitude vs characteristic strain}

The variation of the GW frequency plays a crucial role in the definition of the characteristic strain for coherent GW signals. Let us consider for instance  two BHs in the inspiral phase: as they emit GWs, they get closer and closer to each other and eventually merge. As they are approaching, the GW frequency, which is twice the orbital frequency, grows. The number of cycles that the binary spends at a given frequency $f$ is determined by~\cite{Moore:2014lga},
\begin{equation}
\label{eq:Ncycles}
N_\text{\tiny cycles} = \frac{f^2}{\dot{f}} \simeq 2.16 \times 10^6 \, \left(\frac{f}{10^9 \, \text{Hz}}\right)^{-5/3} \left(\frac{m_\text{\tiny PBH}}{10^{-9} \, M_\odot}\right)^{-5/3} \,.
\end{equation}
$N_\text{\tiny cycles}$ is an important quantity because it determines whether the signal can be considered to be approximately monochromatic, if $N_\text{\tiny cycles} \gg 1$. 
In the stationary phase approximation, a GW signal with an approximately constant amplitude $h_0$ as defined in Eq.~\eqref{eq:GWamplitude} produces a characteristic strain
\begin{equation}
\label{eq:StrainInspirals}
h_c(f) = \sqrt{\frac{2 f^2}{\dot{f}}} h_0 \,,
\end{equation}
where $\dot{f}$ can be explicitly written as~\cite{Maggiore:1900zz}
\begin{equation}
\label{eq:dotf}
\dot{f} = \frac{96}{5} \pi^{8/3} m_c^{5/3} f^{11/3} \simeq 4.62 \times 10^{11} \, \text{Hz}^2 \left(\frac{m_\text{\tiny PBH}}{10^{-9} M_\odot}\right)^{5/3} \left(\frac{f}{\text{GHz}}\right)^{11/3}\,,
\end{equation}
and  we considered two equal mass PBHs $m_1 = m_2 = m_\text{\tiny PBH}$.
Note that only close to the ISCO frequency, 
namely at the final phase of the merger, 
the prefactor $f^2/\dot{f} \sim \mathcal{O}(1)$, and then $h_c (f)$ is of the same order of magnitude as the GW amplitude $h_0$.\\

When comparing a GW signal with a detector sensitivity curve, one has to compare the observation time $t_\text{\tiny obs}$ with the characteristic time of variation of the frequency $t_\text{\tiny f} = f/\dot{f}$. If $t_\text{\tiny obs} \ll t_\text{\tiny f}$, the observation time sets an upper bound on $N_\text{\tiny cycles}^{\text{\tiny obs}} < N_\text{\tiny cycles}$ and the characteristic strain is mainly determined by $h_0$
\begin{equation}\label{eq:StrainMonochromatic}
h_c(f) \simeq \sqrt{N_\text{\tiny cycles}^{\text{\tiny obs}}} \, h_0 \,, \quad \text{for} \quad t_\text{\tiny obs} \ll t_\text{\tiny f} \,.
\end{equation}
In the opposite limit, when $t_\text{\tiny obs} \gg t_\text{\tiny f}$, then one can observe the signal for its entire duration and the characteristic strain is enhanced by a factor $\sqrt{N_\text{\tiny cycles}}$ with respect to the GW amplitude
\begin{equation}
h_c(f) \simeq \sqrt{N_\text{\tiny cycles}} \, h_0 \,, \quad \text{for} \quad t_\text{\tiny obs} \gg t_\text{\tiny f} \,.
\end{equation}
Note that Eq.~\eqref{eq:StrainMonochromatic} is also  valid for strictly monochromatic sources, for which the prefactor $f^2/\dot{f}$ in Eq.~\eqref{eq:StrainInspirals} is not well-defined and the condition $t_\text{\tiny obs} \ll t_\text{\tiny f}$ is always satisfied.\\

In other words, for a coherent GW signal, $h_c(f)$ represents the maximum signal that can be observed at a given frequency, as it takes into account the maximum enhancement due to the intrinsic number of cycles spent by the binary at that frequency. If the observation time is smaller than the characteristic time of variation of the GW frequency, then the GW signal is suppressed by a factor $(N_\text{\tiny cycles}^\text{\tiny obs}/N_\text{\tiny cycles})^{1/2}$ with respect to $h_c$.

\begin{figure}[t!]
\centering
\includegraphics[width=0.8\textwidth]{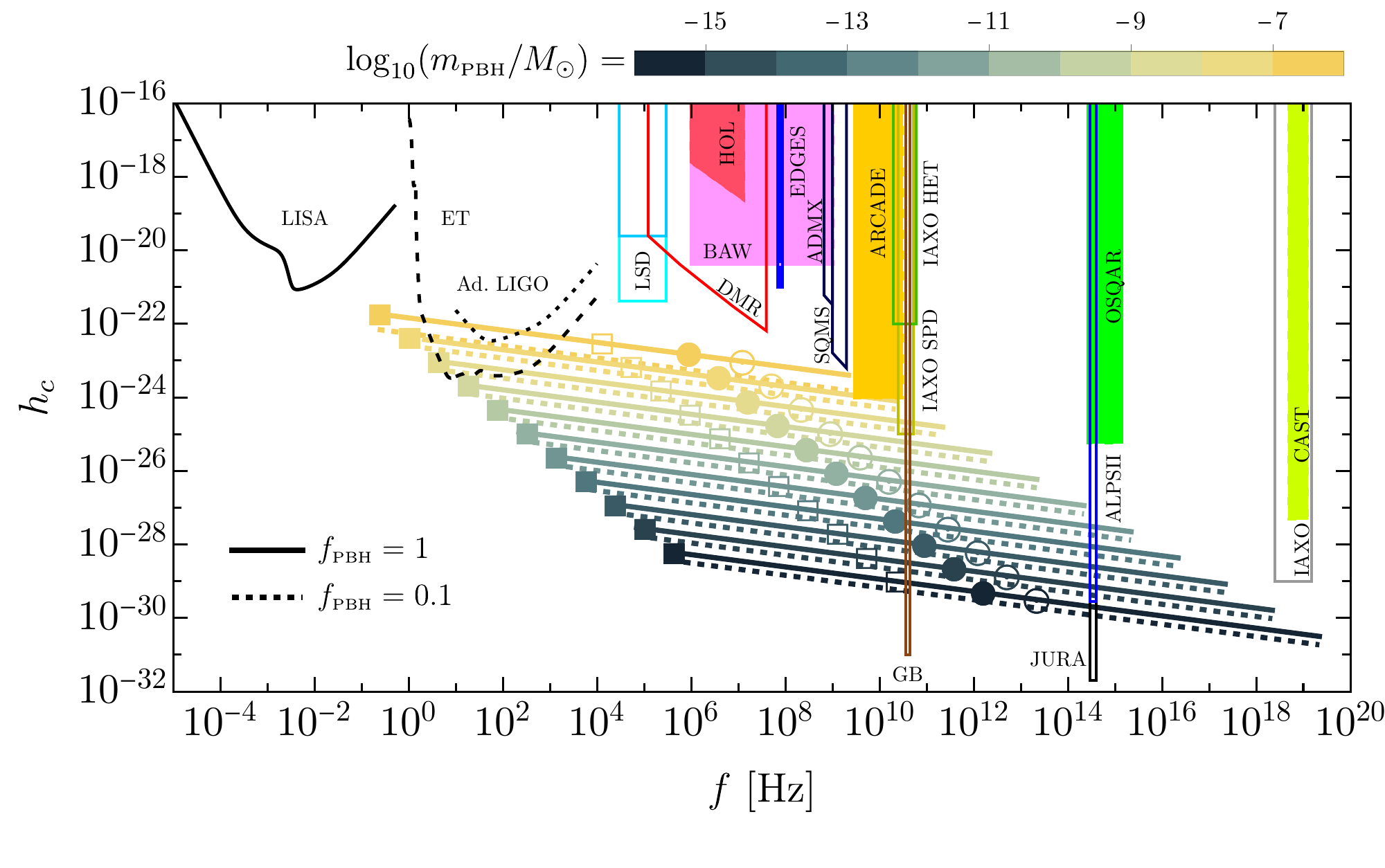}
\caption{
The solid (dashed) lines indicate the characteristic strain spanned by an inspiraling merger with component masses $m_\text{\tiny PBH}\subset(10^{-16}\div 10^{-6})M_\odot$ 
located at a distance $d_\text{\tiny yr}$ for $f_\text{\tiny PBH} = 1$ ($f_\text{\tiny PBH} = 0.1$). 
The signal is cut at the binary ISCO frequency while the empty square, filled and empty circles show the frequency emitted when 
the time to merger is $\Delta t = 1\, \text{day}$,  $\Delta t = 1\, \text{sec}$ and  $\Delta t = 10^{-3}\, \text{sec}$, respectively. 
The signal is drawn starting from the frequency spanned by a merger at time $t\simeq 10$ Gyr from the merger (filled square). 
For comparison, we show the planned and future UHF-GW experiments from Ref.~\cite{Aggarwal:2020olq} (see also Sec.~\ref{sec:Detectors}).
Additionally, we include the sub-kHz interferometric detector 
Ad. LIGO \cite{Harry_2010}, Einstein Telescope (ET-D) \cite{Hild:2010id} and LISA \cite{Robson:2018ifk}.
We do not show the result assuming $R_\PBH^\text{\tiny max}$ to avoid clutter. 
We warn the reader that the strain crossing the sensitivity band of the detector is not necessarily proof of detectability, due for instance to the short duration of the signal. We will come back to this with more details in Fig.~\ref{fig:PlotSingleEvents}. 
}\label{fig: PBH hc single}
\end{figure}

In Fig.~\ref{fig: PBH hc single} we plot the detector sensitivity curves against the characteristic strain $h_c$, which is an upper bound on the observable signal,
for a binary located at a distance $d_\text{\tiny yr}$ under various assumptions on $m_\PBH$ and $f_\PBH$.\footnote{Note that in Figs.~\ref{fig: PBH hc single},~\ref{fig: PBH hc SGWB} and \ref{fig: PBH hc supperradiance} we report the detectors as displayed in~\cite{Aggarwal:2020olq}, with the addition of some recent proposals~\cite{Berlin:2021txa, Domcke:2022rgu, Berlin:2022hfx}. In Sec.~\ref{sec:Detectors}, we will discuss in details the various detectors as well as the detection prospects.} 
In Sec.~\ref{sec:Detectors} we will discuss explicitly which quantity should be compared with the sensitivity curves of each detector.
For comparison, we also summarise the sensitivities of GW experiments, see Sec.~\ref{sec:Detectors} for more details. 
It is important to stress at this stage that even though various experiments may be able to reach strain sensitivities comparable to the one expected from light PBH binaries, due to the short duration of the signal individual binary emission may still deceive detection.
We will discuss this point in detail in Sec.~\ref{sec:Detectors}.
One promising attempt to evade this problem is to focus on the SGWB produced by unresolved mergers building up across the evolution of the Universe. 
This signal is stationary and not limited in time duration. 
As we will see in the next section, however, there is a trade-off to be paid,  as such a signal it is typically associated with a smaller characteristic strain, being dominated by sources at much further distances from the earth.

\subsection{Stochastic gravitational wave background}
\label{sec:SGWB}

Unresolved PBH mergers also contribute to a SGWB, whose spectrum at frequency $\nu$ can be computed as
\begin{equation}\label{stocOmega}
\Omega_\text{\tiny GW} (\nu)= 
 \frac{\nu}{\rho_0} 
 \iint \d m_1 \d m_2 
  \int_0^{{\nu_\text{\tiny cut}}/{\nu}-1}
  \frac{ \d z }{(1+z)H(z)} 
  \frac{\d R_\text{\tiny PBH} }{\d m_1 \d m_2} 
   \frac{\d E_\text{\tiny GW} (\nu_s)}{\d \nu_s},
\end{equation}
in terms of the redshifted source frequency $\nu_s =\nu (1+z)$, the present energy density $\rho_0 = 3 H_0^2/8\pi$ in terms of the Hubble constant $H_0$, and the energy spectrum of GWs. 
In this expression, the redshift upper integration limit corresponds to the
maximum $z$ up to which the energy spectrum can contribute to the given frequency of $\Omega_\text{\tiny GW}(\nu)$ 
while $\nu_\text{\tiny cut}$ is the maximum frequency of the GW emitted by  the binary. 
To compute the integral over the distribution of masses, we assume a log-normal PBH mass distribution 
\begin{equation}\label{lognormal mass function}
    \psi(m | M_c, \sigma ) = 
    \frac{1}{\sqrt{2\pi}\sigma  m}
    \exp 
    \llp 
    -\frac{\log ^2 (m/M_c)}{2 \sigma^2} 
    \rrp,
\end{equation}
characterised by a central mass scale $M_c$ (not to be confused with the chirp mass $m_c$ above) and a given width $\sigma$.
This model-independent parametrization of the mass function can describe
a population arising from a symmetric peak in the power spectrum of curvature perturbations in a wide variety of formation models (see e.g. Refs.~\cite{Dolgov:1992pu,Carr:2017jsz}) and is often used in the literature to
set constraints on the PBH abundance from GW measurements 
\cite{Garcia-Bellido:2017fdg,Raidal:2018bbj,DeLuca:2020sae,Wong:2020yig,Gow:2019pok,Hall:2020daa,Hutsi:2020sol,DeLuca:2021wjr,Franciolini:2021tla}.

We describe the GW energy spectrum emitted by coalescing binary BHs using the phenomenological model presented in Ref.~\cite{Ajith:2009bn}. 
The GW emission can be divided into three distinct parts, corresponding to the inspiral, merger and ringdown, respectively. 
Each stage is related to a different frequency range $\nu$, which depends on the binary BH component masses $m_1$ and $m_2$, and non-precessing spin magnitudes $\chi_1$ and $\chi_2$. 
Assuming circular orbits, one can write \cite{Zhu:2011bd}
\begin{equation}
    \frac{\mathrm{d}E}{\mathrm{d}\nu} = \frac{(G \pi)^{2/3} \mathcal{M}^{5/3}}{3} 
    \begin{cases}
     \nu^{-1/3} f_1^2  & \nu < \nu_\text{\tiny merger},
     \\
     \omega_1 \nu^{2/3} f_2^2 & \nu_\text{\tiny merger} \leq \nu < \nu_\text{\tiny ringdown},
      \\
     \omega_2 f_3^2 & \nu_\text{\tiny ringdown} \leq \nu < \nu_\text{\tiny cut}. 
    \end{cases}
    \label{eq:dEdnu}
\end{equation}
We report in Appendix~\ref{app_GW_energy} the exact expressions for the factors entering in Eq.~\eqref{eq:dEdnu}.

Translating the SGWB abundance computed in Eq.~\eqref{stocOmega} in terms of a characteristic strain, we find\footnote{
Our definition of characteristic strain $h_c$ follows Eq.~(4b) of Ref.~\cite{Aggarwal:2020olq}.
Notice a difference of a factor $\sqrt{2}$ in the definition of $h_c$ compared to \cite{Ringwald:2020ist} (see their Eq.~(2.28)). }
\begin{equation}
\label{eq:StochasticStrain}
	h_c(f) \approx \left[ \frac{3}{4 \pi^2 } \left ( \frac{H_0^2}{f^2}\right) \Omega_\text{\tiny GW} (f) \right]^{1/2},
\end{equation}
where the current Hubble rate is $H_0 = 2.18 \times 10^{-18}\,$Hz. Therefore, the characteristic strain can be written as 
\begin{equation}\label{strain SGWB}
h_c \approx 2 \times 10^{-31} \left ( \frac{f}{{\rm GHz}} \right)^{-1} \left ( \frac{\Omega_\text{\tiny GW} }{10^{-7}} \right)^{1/2} .
\end{equation}
The tail at low-frequency of a SGWB produced by inspiraling binaries is characterized by a scaling $\Omega_\text{\tiny GW}(f) \sim f^{2/3}$ \cite{Moore:2014lga}. This corresponds to a characteristic strain scaling as $h_c (f) \sim f^{-2/3}$. Note that in the case of a SGWB we do not have the same issues that we addressed in Sec.~\ref{sec:SignalDuration} for individual inspiral sources: the characteristic strain is uniquely determined by the energy density in GWs. Furthermore, a SGWB signal is stationary, implying that there is no issue related to the duration of the signal.

\begin{figure}[t!]
\centering
\includegraphics[width=0.8\textwidth]{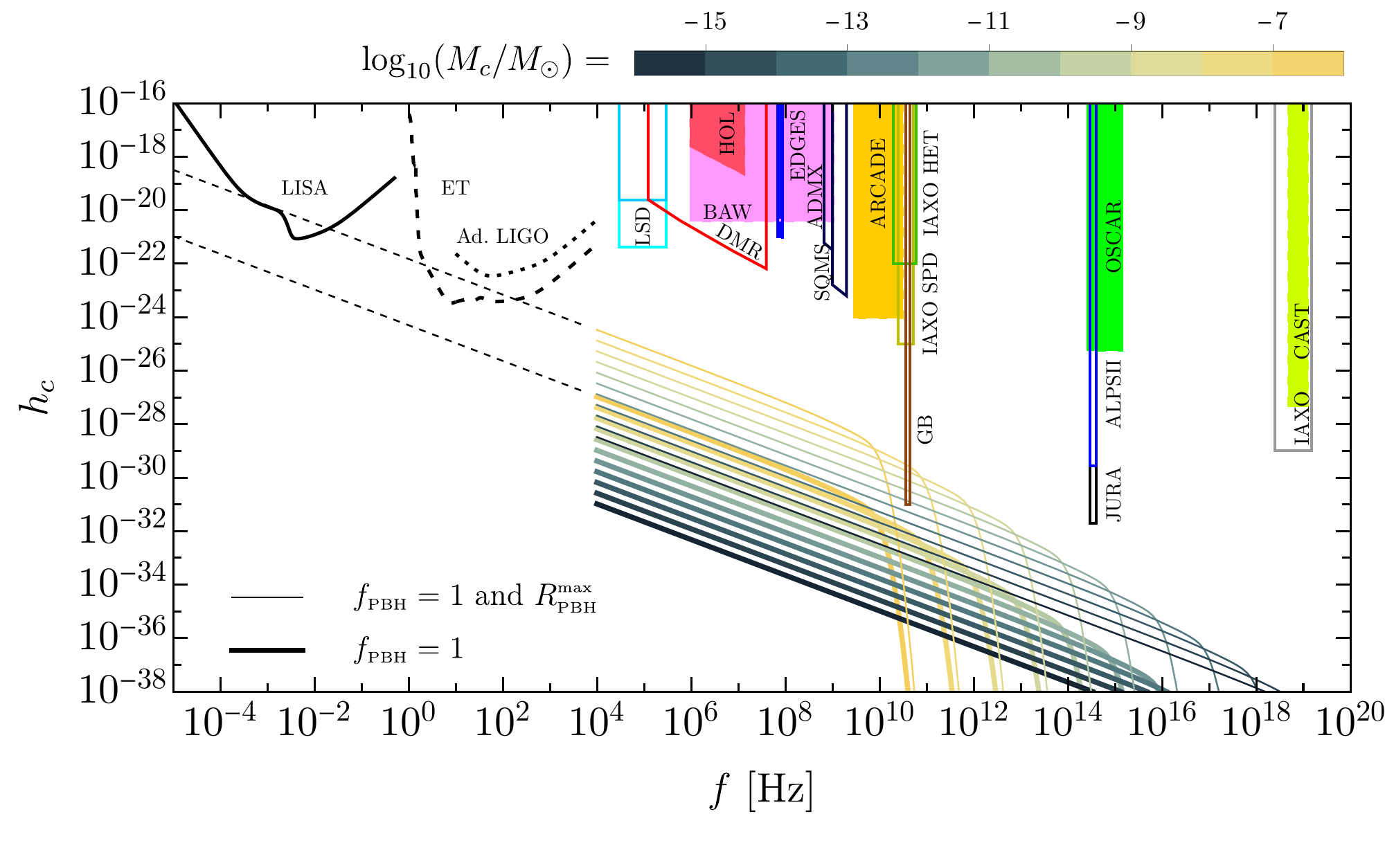}
\caption{
Same as Fig.~\ref{fig: PBH hc single} but showing the characteristic strain induced by a 
SGWB from a population of PBHs.
The PBH mass distribution is taken to be log-normal with $\sigma = 0.3 $ and the various central masses $M_c\subset(10^{-16}\div 10^{-6})M_\odot$ indicated in the inset.
The thick line accounts for both the suppression of binary formation in the early Universe ($S_1$) and the binary disruption in the late-time Universe clusters ($S_2$) and assumes $f_\text{\tiny PBH }=1$.
The thin line assumes  $f_\text{\tiny PBH }=1$ and the theoretical maximum merger rate. 
We warn the reader that some of the sensitivities reported in this plot were not derived assuming a SGWB signal, and may degrade w.r.t. what is shown in the plot.
Also, the actual performance of sub-kHz detectors at observing a SGWB is typically described in terms of "power-law integrated" sensitivity curves, see for example App. C of Ref.~\cite{Bavera:2021wmw}.
We do not discuss sub-kHz detections with such details but we refer the interested reader to Refs.~\cite{DeLuca:2021hde,Pujolas:2021yaw}. 
}\label{fig: PBH hc SGWB}
\end{figure}

In Fig.~\ref{fig: PBH hc SGWB} we show the spectrum of characteristic SGWB strain produced by a narrow PBH population (whose mass distribution is described by Eq.~\eqref{lognormal mass function}) for various values of $f_\PBH$. 
In order to set an upper bound to such a contribution, we also show the result assuming the maximum merger rate {\it potentially} obtained in clustered scenarios $R_\text{\tiny PBH}^\text{\tiny max}$.
It is interesting to notice that the SGWB tail at low frequencies is expected to possess a natural cut-off due to the minimum (red-shifted) frequency emitted by the binaries at the formation epoch. 
We do not model such a drop of the signal as it would fall, in any case, below the sensitivity reach of sub-kHz interferometric GW detectors such as ET and LISA.

As we indicate with thin black dashed lines in Fig.~\ref{fig: PBH hc SGWB}, unless the maximum rate is achieved and one considers masses above $M_c\gtrsim 10^{-8} M_\odot$, 
the low-frequency tail of the signal would be, in any case, too faint to be visible by sub-kHz GW interferometers such as ET and LISA at lower frequencies. This result was already pointed out in 
Ref.~\cite{DeLuca:2021hde},  see their Fig.~4 and related discussion.
This confirms the necessity of UHF experiments to search and constrain the SGWB of light PBH mergers.

\subsection{Gravitational wave memory}
\label{sec:Memory}

The GW memory is a permanent displacement between freely falling test masses that is induced by the passage of a GW \cite{Christodoulou:1991cr,PhysRevD.44.R2945,Blanchet:1992br,Favata:2008ti,Favata:2009ii,Pollney:2010hs,Lasky:2016knh,Hubner:2019sly,Ebersold:2020zah,Zhao:2021hmx}. 
For GW experiments based on interferometry, it was shown 
the non-linear memory could be directly detectable if a sufficient portion of the memory is induced on a timescale $\tau \approx 1/f_\text{opt}$ where $f_\text{opt}$ is the frequency of the detector's peak sensitivity \cite{PhysRevD.45.520} (see Ref.~\cite{Johnson:2018xly} for a summary of the future detection prospects at sub-kHz experiments). 
Following Ref.~\cite{Johnson:2018xly} and Refs. therein, 
we model the $h_c$ signal as a step function $\Theta $ with an UV cut-off at the ISCO frequency, i.e.
\begin{equation}
h_c (f)	\simeq h_c^\text{\tiny mem} \Theta (f_\text{\tiny ISCO} -f ).
\end{equation}
The value of the strain amplitude, averaged over source orientations and sky positions, can be estimated to be a fraction of the GW signal at peak frequency~\cite{PhysRevLett.118.181103} as
\begin{equation}
\kappa = \sqrt{\frac{\langle h_\text{\tiny mem}^2\rangle}{\langle h_\text{\tiny osc}^2 \rangle}} \simeq 1/20 \,. 	
\end{equation}

As the GW memory signal extends to much smaller frequencies with respect to $f_\text{\tiny ISCO}$, it was recently shown in Ref.~\cite{PhysRevLett.118.181103} that one could significantly outperform some UHF-GW experiments in the search for signals in the MHz frequency range by looking for the corresponding GW memory at ground-based detectors (see also Refs.~\cite{Domenech:2021odz,Lasky:2021naa}). 
Following the same logic, we check whether a population of PBH mergers shown in Figs.~\ref{fig: PBH hc single} and \ref{fig: PBH hc SGWB}
 would leave a detectable GW memory signal at GW interferometric searches at lower frequencies. 
The GW strain signals turns out to be 
\begin{equation}
h_c^\text{\tiny mem} \sim 4.8 \times 10^{-31} 
\lp \frac{m_\PBH}{10^{-12} M_\odot } \rp
\lp \frac{d_\text{\tiny L}}{\rm kpc} \rp ^{-1} \,.
\end{equation}
As one can see, the strain induced by the memory effects of PBH of mergers with masses $m \lesssim 10^{-4} M_\odot$ at a distance $d_\text{\tiny yr}$ (indicated in Fig.~\ref{fig: dist})  would fall much below the forecasted sensitivity curves of both LISA and 3G detectors. Also, we notice that the early inspiral phase shown in Fig.~\ref{fig: PBH hc single} would be associated with larger strain signals (scaling as $h_c \sim f^{-1/7}$) in the sub-kHz range. 
Therefore, we conclude that the memory signature persisting at lower frequency even for light PBH mergers may not allow for a detection at sub-kHz interferometric detectors.
We conclude this section by mentioning that the memory strain would still cross some of the sensitivity bands of UHF-GW detectors (such as GB and JURA, see Sec.~\ref{sec:Detectors} for more details on these detectors). The nature of the signal, however, is very different from the one assumed to derive sensitivity curves for these detectors (that is a plane monochromatic wave). 
Therefore, a dedicated study on the feasibility of measuring such a signature is still required.

\subsection{Black hole superradiance}
\label{sec:Superradiance}

In this section we discuss the possible interplay between a light PBH population and a light scalar field. The possible coexistence of the two would give rise to GW signatures that may be used to constrain both sectors. 
It is fair to admit, however, that such a scenario assumes the existence of two distinct extensions of the current standard model of particle physics and cosmology. Nevertheless, it is worth understanding how such a scenario could be constrained by searching for UHF-GWs. 
The existence of light pseudo-scalars that would trigger the superradiance mechanism, such as \textit{axions}, is very well motivated from the UV perspective. The QCD axion for instance is one of the most promising proposals to solve the strong CP problem~\cite{Peccei:1977hh, Wilczek:1977pj, Weinberg:1977ma}. Furthermore, other types of axions (sometimes called \textit{axion-like particles}) are very common in string theory, arising from the existence of extra-dimensions as the Kaluza-Klein zero-modes of form fields~\cite{Arvanitaki:2009fg}.

GW emission can arise from clouds of light bosons in rotating BH backgrounds as a result of gravitational superradiance~\cite{Ternov:1978gq,Zouros:1979iw,Arvanitaki:2009fg,Arvanitaki:2010sy, Arvanitaki:2012cn, arXiv:2010.13157, Detweiler:1980uk,Yoshino:2013ofa,Arvanitaki:2014wva,Brito:2014wla,Brito:2015oca}. 
We note PBHs are expected to be produced with a very small spin in the standard scenario, i.e. from the collapse of large radiation overdensities \cite{DeLuca:2019buf,Mirbabayi:2019uph}. 
However, in alternative scenarios, such as the formation from an assembly of matter-like objects (particles, Q-balls, oscillons, etc.), domain walls and heavy quarks of a confining gauge theory, larger PBH spins at formation are predicted~\cite{Harada:2017fjm,Flores:2021tmc,Dvali:2021byy,Eroshenko:2021sez,DeLuca:2021pls,Chongchitnan:2021ehn,deFreitasPacheco:2020wdg}.
In case PBHs possessed a non-vanishing spin, one would expect superradiant instabilities to take place already in the early Universe and remove most of the angular momentum, leaving a population of slowly rotating PBHs. 
However, each PBH merger generates a spinning remnant with $\chi\simeq 0.68$~\cite{Barausse:2009uz} (assuming spin-less progenitors) and a mass around $m_f \simeq 2 \, m_\PBH$. 
This process may trigger superradiant instabilities of light scalar fields in the present epoch, potentially leading to the emission of observable UHF-GW signatures~\cite{Aggarwal:2020umq}.
     
When the  Compton wavelength
of a boson (${\bf b}$) is  of the size of the BH, i.e
\begin{equation}
\label{eq:maMBH} 
m_{\bf b} \sim 10^{-4}\, \text{eV} \, \left( \frac{m_\text{\tiny PBH} }{ 10^{-6}M_\odot} \right)^{-1}   
\,,
\end{equation}
the boson  accumulates outside the BH event horizon efficiently. The characteristic  timescale for the growth of the boson cloud for the dominant $\ell = m =1$ mode is \cite{Brito:2015oca}
\begin{equation}
\tau \sim 3 \times 10^{-5}\, {\rm sec} \lp\frac{m_\PBH}{10^{-6} M_\odot} \rp .
\end{equation}
The primary process  for the production of UHF-GWs is  annihilation of bosons, e.g. (pseudo-) scalars $\bf{b}$ into gravitons $h$. The associated GW frequency is
twice the Compton frequency of the boson, i.e.  
\begin{equation}
\label{eq:fsprannihilation}
f \sim  2 \times  10^{6} \, \text{Hz} \left( \frac{m_b}{10^{-9} \, \text{eV}} \right) 
\sim
2 \times 10^2\,  {\rm GHz}\, \lp \frac{m_\text{\tiny PBH}}{10^{-6} M_\odot } \rp^{-1}
\, ,
\end{equation}
which corresponds to frequencies above 200 kHz for $m_\PBH \lesssim M_\odot$. 
Note that this GW signal is  monochromatic and coherent~\cite{Arvanitaki:2014wva}, making it distinct from  other astrophysical or cosmological sources.  The expected characteristic GW amplitude for this process is\cite{Arvanitaki:2012cn} 
\begin{align}
    h_0 &
\simeq 5 \times 10^{-30}\,  
\frac{1}{\ell}
    \lp \frac{\alpha}{0.1}\rp 
        \lp \frac{\epsilon}{10^{-3}}\rp 
\lp \frac{d_\text{\tiny L}}{{\rm kpc}}\rp ^{-1} 
\lp \frac{m_\text{\tiny PBH}}{10^{-6} M_\odot}\rp,
\label{eq:strain_axion_annihilation}
\end{align}
where $\alpha = G m_\text{\tiny PBH} \, m_b$, $\ell$ is the orbital angular momentum number of the decaying  bosons and $\epsilon < 10^{-3}$ denotes the fraction the PBH mass accumulated in the  cloud. {The superradiance condition constrains $\alpha/\ell < 0.5$~\cite{Arvanitaki:2010sy}.} See Refs. \cite{Brito:2014wla, arXiv:2010.13157} for more recent calculations of the strain. 
\begin{figure}[t!]
\centering
\includegraphics[width=0.8\textwidth]{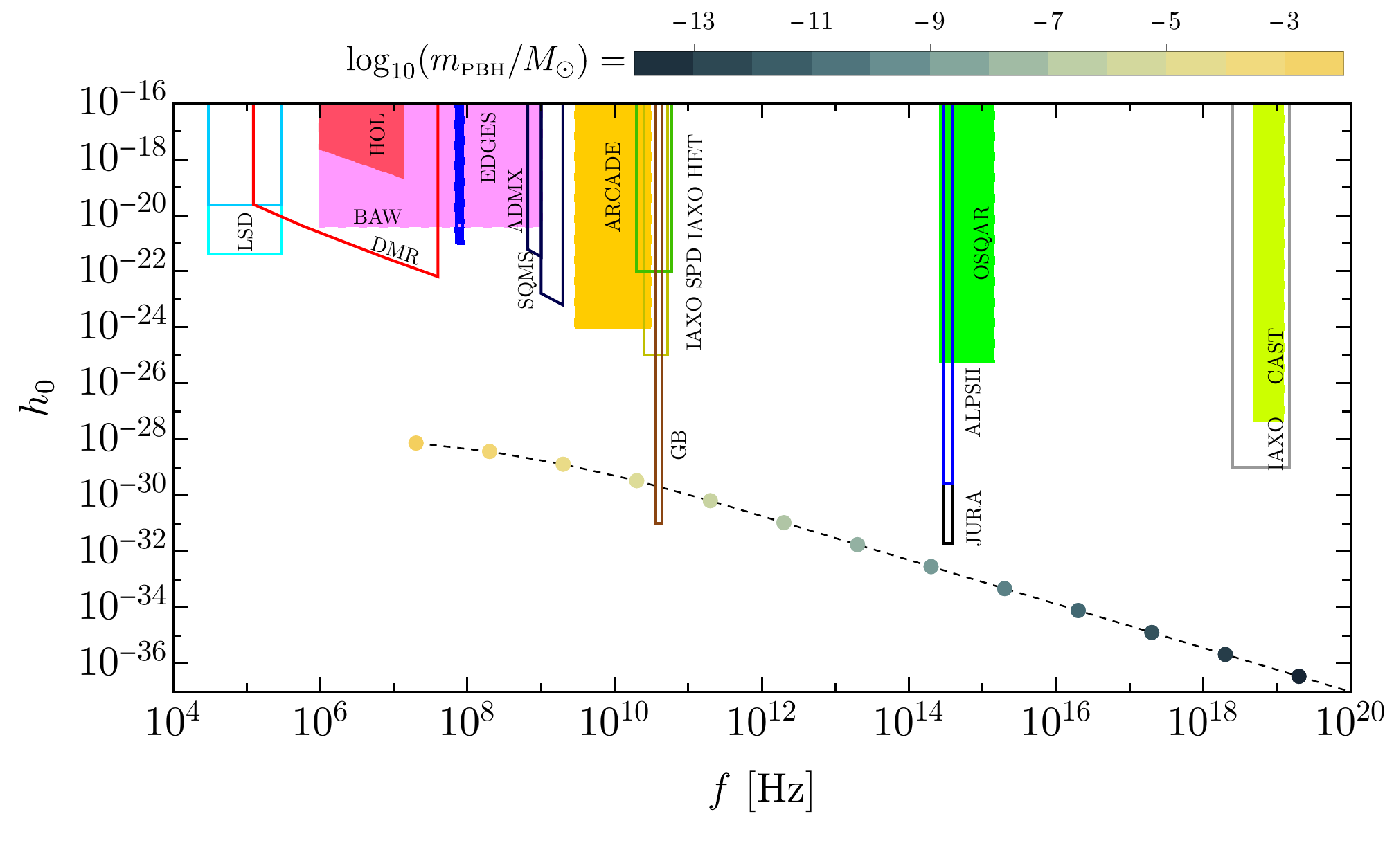}
\caption{
Same as Fig.~\ref{fig: PBH hc single} but showing the GW strain generated by scalar boson field supperadiant instabilities.
As in Fig.~\ref{fig: PBH hc single}, the binary leading to a spinning PBH remnant is assumed to be at a distance $d_\text{\tiny yr}$.
Note the change of scale for the color coding according to the PBH mass $m_\text{\tiny PBH}$ compared with previous figures.  
}
\label{fig: PBH hc supperradiance}
\end{figure}
The duration of the signal is (see \cite{Brito:2015oca} and the references therein)
\begin{equation}
\label{eq:SuperradianceDuration}
	 \tau \approx 0.13 \,  {\rm yr} \left( {m_\text{\tiny PBH} \over 10^{-6}M_\odot} \right) 
	 \left( \alpha \over  0.1 \right)^{-15} \left( { {\chi_i - \chi_f} \over  0.5 } \right)^{-1} \,,
\end{equation}
where $\chi^{i}$ and $\chi^{f}$ are the dimensionless
BH spin at the beginning and end of the superradiant growth.
We compare the expected GW signal amplitude from a source located at a distance $d_\text{\tiny yr}$ in Fig.~\ref{fig: PBH hc supperradiance} along with UHF-GW detector proposals.

Note that, despite restricting ourselves to the case of a (pseudo-) scalar, a similar phenomenon can occur in the presence of vector and tensor fields. In such cases though, the duration of the signal is much shorter than what is reported in Eq.~\eqref{eq:SuperradianceDuration},
making extremely challenging to detect PBH masses $m_{\text{\tiny PBH}} \lesssim 10^{-5} \, M_\odot$ (see Ref.~\cite{Brito:2015oca} for more details).\footnote{As reported in Ref.~\cite{Brito:2015oca}, the signal duration for vector and tensor superradiant instabilities as a function of the mass of the BH scales as
$\tau \sim 5 \times 10^{-10} \, \text{yr} \, \left({m_{\text{\tiny{PBH}}}}/{10^{-6} \, M_\odot}\right)$.}

\section{Reach of planned and future ultra-high frequency GW experiments}
\label{sec:Detectors}

In this section we discuss the detectability of the GW signals that are produced in the various scenarios discussed in Sec.~\ref{sec:PBHmergers} with current and planned technologies.

First, let us point out that
the detection of GWs becomes more difficult as their frequency increases. This is the case both for single events and for stochastic backgrounds of GWs. For instance, the amplitude of GWs at the ISCO frequency (that coincides with the characteristic strain, see Sec.~\ref{sec:SignalDuration}) for a BH merger is (see also Eq.~\eqref{eq:spawaveform2})
\begin{equation}
\label{eq:StrainMergers}
h_0^{\text{\tiny ISCO}} \sim h_c^{\text{\tiny ISCO}} \simeq 
2
 \times 10^{-23} \left(\frac{f}{{\rm GHz}}\right)^{-1} \left(\frac{d_\text{\tiny L}}{{\rm kpc}}\right)^{-1} \,.
\end{equation}
On the other hand, for a SGWB characterised by a given GW energy density $\Omega_\text{\tiny GW}$, the GW strain scales as (see Eq.~\eqref{strain SGWB})
\begin{equation}\label{strain SGWB rep}
h_c^{\text{\tiny SGWB}} \simeq 2 \times 10^{-31} \left ( \frac{f}{{\rm GHz}} \right)^{-1} \left ( \frac{\Omega_\text{\tiny GW} }{10^{-7}} \right)^{1/2} .
\end{equation}
In both Eq.~\eqref{eq:StrainMergers} and Eq.~\eqref{strain SGWB} the strain decreases as $\propto 1/f$, showing that it is increasingly difficult to detect both types of signal (e.g. coming from the chirp phase of a binary PBH merger or from a cosmic accumulation of events that produce a copious amount of energy density in GWs).

Currently, there are various proposals for detectors operating at frequencies higher than the range currently explored at LIGO/Virgo/KAGRA, that is $(10 \div 10^3) \, \text{Hz}$. These can be divided into four broad categories, as we summmarise in the following.
\begin{itemize}[leftmargin=*]
\item {\bf Mechanical resonators}
\begin{itemize}[leftmargin=*]
\item[1)] Resonant spheres~\cite{Aguiar:2010kn}, that operate in the range $(4 \div 10) \, \text{kHz}$~\cite{Harry:1996gh};
\item[2)] Levitated Sensor Detectors (LSD)~\cite{Arvanitaki:2012cn, Aggarwal:2020umq}, that operate in the range $(30 \div 300) \, {\rm kHz}$;
\item[3)] Bulk Acoustic Wave (BAW) devices~\cite{Goryachev:2014yra, Page:2020zbr, Goryachev:2021zzn}, that work in the range $(1 \div 10^3) \, \text{MHz}$.
\end{itemize}
\item {\bf Detectors based on GW-electromagnetic wave conversion}
\begin{itemize}[leftmargin=*]
\item[1)] Superconducting Radio Frequency Cavity (SRFC) based detectors, see e.g.~\cite{Berlin:2021txa, Berlin:2022hfx}, operating in the range $(1 \div 10^3) \, \text{MHz}$;
\item[2)] Resonant electromagnetic antennas, see~\cite{Herman:2020wao, Herman:2022fau}, operating in the frequency range $(1 \div 10^3) \, \text{MHz}$, see Sec.~\ref{sec:Discussion} for more details;\footnote{Note that the concept of resonant antennas is very similar to SRFCs. However, the authors of~\cite{Herman:2020wao, Herman:2022fau} suggest a way to extract the detector sensitivity that is sensitive to the GW amplitude at linear order, contrary to the case of SRFC, where the effect is at second order in $h_0$. As the proposal of~\cite{Herman:2020wao, Herman:2022fau} does not account for noise sources yet, we do not include it in the plots. See also Sec.~\ref{sec:Discussion} for more comments on this concept.}
\item[3)] Conversion of GWs into eletromagnetic waves in a static magnetic field, as for ‘light-shining-through-a-wall’ axion experiments and axion helioscopes, see Sec.~\ref{sec:GertEffect} for more details.
\item[4)] 
Conversion of GWs into electromagnetic waves in a static magnetic field equipped with an additional Gaussian Beam (GB), see Sec.~\ref{sec:GB} for more details.
\item[5)] Resonant LC circuits (DMR)~\cite{Domcke:2022rgu}, operating in the frequency range $(10^{-1} \div 10^2) \, \text{MHz}$, see Sec.~\ref{sec:Haloscopes} for more details.
\end{itemize}
\item {\bf Interferometers}
\begin{itemize}[leftmargin=*]
\item[1)] The holometer (HOL) experiment~\cite{Holometer:2016qoh}, that operates in the range $(1 \div 13) \, {\rm MHz}$;
\item[2)] A $0.75 \, \rm m$ interferometer~\cite{Akutsu:2008qv}, that works at $f \sim 100 \, {\rm MHz}$.
\end{itemize}
\item {\bf Others}
\begin{itemize}[leftmargin=*]
\item[1)] Magnon-based detectors~\cite{Ito:2019wcb}, operating at $f \sim 10 \, {\rm GHz}$;
\item[2)] It is also possible to convert radio telescope data (such as the ones from EDGES and ARCADE~\cite{Fixsen:2009xn, Bowman:2018yin}) into constraints on the presence of a stochastic background of GWs around the epoch of re-ionization~\cite{Domcke:2020yzq}. These operate at $f \sim 100 \, \text{MHz}$ and $f \sim 10 \, \text{GHz}$ respectively.
\end{itemize}
\end{itemize}
Note that not all the detectors mentioned above may be used to probe the various signals discussed in this work. 
Consider, for example, the SGWB produced by PBH mergers. 
As most of its contribution is emitted in the late-time Universe, it cannot be detected using data from radio telescopes, that can only probe the re-ionization epoch. 
In the next sections, we will provide other examples of detectors that may not be suitable to probe some of the GW signals produced by PBHs.

\textcolor{black}{
\subsection{Levitated Sensor Detectors}
\label{sec:lsd}
Levitated Sensor Detectors (LSDs) are mechanical resonators that operate as GW detectors in the frequency range $(10 \div 300) \, \text{kHz}$, with a bandwidth of $\Delta f \sim f/10$. These were initially proposed in~\cite{Arvanitaki:2012cn} and more recently developed further in~\cite{Aggarwal:2020umq}. In the simplest version~\cite{Arvanitaki:2012cn}, an LSD consists of a laser standing wave that propagates between two mirrors in a cavity. A dielectric nanoparticle placed close to an anti-node of the standing wave will experience a force that pulls the nanoparticle back to the anti-node, which is then an equililbrium position of the \textit{optical potential}: the dielectric nanoparticle is \textit{optically trapped}. The position of the anti-nodes and the intensity of the restoring force (i.e. the optical potential) depend on the intensity of the laser and on the dielectric constant of the nanoparticle, and can be tuned by varying these external parameters. A second laser, whose intensity is much lower compared to the one used to build the optical trap, can be used to read out the position of the nanoparticle. If a GW goes through the detector\footnote{Only the component of the GW that propagates orthogonally to the axis of the cavity is relevant here.}, it modifies the proper distance between the mirrors of the cavity as well as the distance between the mirrors and the nanoparticle. This causes a nanoparticle initially at rest in an equilibrium position to move, subject to the optical potential. If the GW frequency matches the optical frequency, a resonance in the oscillation of the nanoparticle is triggered: this can be observed using the second, weaker, laser. The sensitivity of LSDs depends crucially on various factors: it improves using longer cavities, more massive nanoparticles and cooler environments. At the moment, a one-meter prototype is under construction at Northwestern University, while different possible improvements have been suggested. The best design sensitivity proposed so far could be achieved with a $100$-meter cryogenic setup, which would reach a power spectral density sensitivity of $\sqrt{S_n} \simeq 10^{-23} \, \text{Hz}^{-1/2}$ at $f \gtrsim 10^5 \, \text{Hz}$~\cite{Aggarwal:2020umq}. Such an instrument would employ various technical upgrades with respect to the simplest version~\cite{Aggarwal:2020umq}, including for instance a Michelson interferometer configuration to remove common noise, and the use of a multi-layered stack of dielectric discs, to make the suspended object more massive. We will use this theoretically designed detector as a benchmark for our estimates of the sensitivity, assuming a power spectral density sensitivity of $\sqrt{S_n} = 10^{-23} \, \text{Hz}^{-1/2}$, a frequency band of $(1 \div 1.1) \times 10^5 \, \text{Hz}$, where we are using that $\Delta f = f/10 \sim 10^4 \, \text{Hz}$. We can then compute the sensitivity to the amplitude of GWs produced by a BH binary with mass $m_{\rm PBH}$. In order to convert between power spectral density $\sqrt{S_n}$ and characteristic strain $h_c$, we recall that
\begin{equation}
h_c = \sqrt{\frac{f^2}{\Delta f} S_n} \,.
\end{equation}
Once we have the characteristic strain $h_c$, we can use Eq.~\eqref{eq:StrainMonochromatic} to compute the GW amplitude $h_0$, employing Eq.~\eqref{eq:Ncycles} and Eq.~\eqref{eq:dotf} to evaluate the number of cycles spent by the signal in the detector bandwidth. The final expression for a GW signal coming from a BH binary with mass $m_{\rm PBH}$ is
\begin{equation}
h_0 = 2.6 \times 10^{-19} \, \left(\frac{m_{\rm PBH}}{M_\odot}\right)^{5/6} \,.
\end{equation}
}

\subsection{Conversion in an external static magnetic field}
\label{sec:GertEffect}

One of the most promising classes of detectors is based on the conversion of GWs into electromagnetic radiation, known as inverse Gertsenshtein effect~\cite{gertsenshtein1962wave, Boccaletti1970ConversionOP}. In its simplest implementation, a detector that exploits the inverse Gertsenshtein effect consists of a conversion region of length $L$ and area $A$ (e.g. let us assume that it is a cylinder) that hosts a static magnetic field $\mathbf{B}$, orthogonal to the axis of the cylinder (say $\hat{z}$). A plane wave GW traveling along the axis of the conversion region generates an electromagnetic wave traveling in the same direction,\footnote{Note that there is also an electromagnetic component parallel to the static magnetic field $\mathbf{B}$~\cite{Berlin:2021txa}.} whose amplitude is proportional to the square of the GW amplitude $h_c$ (or $h_0$, for coherent GW signals), to the square of the length of the conversion region $L$ and to the square of the strength of the static magnetic field $|\mathbf{B}| \equiv B$. Therefore, the electromagnetic power that is produced in the presence of a SGWB whose amplitude is given by Eq.~\eqref{eq:StochasticStrain} can be written as
\begin{equation}
\label{eq:EMpower}
P_\text{\tiny EM}(f) = \frac{\pi^2}{4 \mu_0 c} A B^2 L^2 f^2 |h_c(f)|^2 \,,
\end{equation}
where $\mu_0$ is the magnetic constant. In the case of an inspiral or a monochromatic source, such as superradiance, $h_c$ should be replaced by $h_0$ from Eq.~\eqref{eq:EMpower}, see also Eq.~\eqref{eq:StrainInspirals}. This electromagnetic power can then be revealed at the end of the conversion region through appropriate detectors whose properties depend on the frequency of interest.

In order for this effect to be efficient, it is important that the phase coherence between the GW and the produced electromagnetic wave is mantained~\cite{Ringwald:2020ist}, namely that
\begin{equation}
\label{eq:PhaseCoherence}
f \gg \frac{0.45}{\pi} \frac{L}{A} \simeq 4.3 \times 10^7 \, {\rm Hz} \, \left(\frac{L}{{1 \, \text{m}}}\right) \left(\frac{{1 \, \text{m}}}{\sqrt{A}}\right)^2 \,.
\end{equation}
For instance a hypothetical cylindrical detector with $L = 1 \, \text{m}$ and area $A = 0.785 \, \text{m}^2$ (corresponding to a radius of $0.5$ m) can operate at $f \gg 4.3 \times 10^7 \, \text{Hz}$. It is interesting to notice that the idea behind these detectors is similar to the one underpinning various axion experiments, including telescopes such as CAST~\cite{Zioutas:1998cc, GraciaGarza:2015sos} (decommissioned) and IAXO~\cite{Ruz:2018omp} (planned) and ‘light-shining-through-a-wall’ experiments such as OSQAR~\cite{OSQAR:2015qdv, OSQAR:2013jqp} (decommissioned), ALPS~\cite{ALPS:2009des, Ehret:2010mh} (decommissioned), ALPS II~\cite{Bahre:2013ywa, Albrecht:2020ntd} (under construction) and JURA~\cite{Beacham:2019nyx} (proposal).
Using data already collected in axion experiments such as OSQAR and CAST, it is possible to place bounds on the presence of a stochastic background of GWs at the frequency at which these detectors naturally operate~\cite{Ejlli:2019bqj}, which is extremely high: $f \sim 10^{15} \, {\rm Hz}$ and $f \sim 10^{18} \, {\rm Hz}$. It is, on the other hand, possible to adapt the photon receivers of such detectors in order to be sensitive to GWs at lower frequencies, see e.g.~\cite{Ringwald:2020ist}, where two options for receivers that operate around the $\mathcal{O}(10) \, {\rm GHz}$ were considered. In the following, we will only use the most promising proposal which entails the use of Single Photon Detectors
 (SPDs) to reveal the electromagnetic wave induced by the GW. For this proposal the sensitivity reported in Ref.~\cite{Ringwald:2020ist} is\footnote{Note a $\sqrt{2}$ difference with respect to~\cite{Ringwald:2020ist} due to the different definition of the characteristic strain in Eq.~\eqref{eq:StochasticStrain}.}
\begin{align}
h_c^{\text{\tiny SPD}} \simeq \,\,
&2.64 \times 10^{-24} \left(\frac{S/N}{2}\right)^{\frac{1}{2}}\left(\frac{\Delta t}{{1 \,\rm yr}}\right)^{-\frac{1}{4}} \left(\frac{\Delta f}{10^{11} \, {\rm Hz}}\right)^{-\frac{1}{2}} \nonumber \\
& \times \epsilon^{-\frac{1}{2}} \left(\frac{\Gamma_D}{10^{-3} \, {\rm Hz}}\right)^{\frac{1}{4}} \left(\frac{B}{1 \, \text{T}}\right)^{-1} \left(\frac{L}{1 \, \text{m}}\right)^{-1} \left(\frac{A}{1 \, \text{m}^2}\right)^{-\frac{1}{2}} \,,
\label{eq:SPDsensitivity}
\end{align}
where $S/N$ is the signal-to-noise ratio that one wants to achieve, $\Delta t$ is the measurement time, $f$ is the frequency of the GW, $\Delta \omega = 2 \pi \Delta f$, $\epsilon$ is the single photon detection efficiency, $\Delta f$ is the bandwidth of the receiver, and $\Gamma_D$ is the dark count rate. We invite the reader to check Ref.~\cite{Ringwald:2020ist} for more comments on the experimental feasibility of the benchmark values used in Eq.~\eqref{eq:SPDsensitivity}. Note that Eq.~\eqref{eq:SPDsensitivity} is actually the expression for $h_0^\text{\tiny SPD}$ instead of $h_c^\text{\tiny SPD}$ when it refers to coherent sources, such as inspirals and superradiance.

We stress that, while in principle it is possible to tune the central frequency of the detector, it is always necessary to be in the regime where Eq.~\eqref{eq:PhaseCoherence} is satisfied. Given the specifics of the experiments ALPS II ($B = 5.3 \,\text{T}$, $L = 210 \, \text{m}$, $A \simeq 0.02 \, \text{m}^2$), IAXO ($B = 2.5 \,\text{T}$, $L = 20 \, \text{m}$, IAXO will use $8$ tubes, each of which has area $A \simeq 0.4 \, \text{m}^2$) and MADMAX ($B = 4.83 \,\text{T}$, $L = 6 \, \text{m}$, $A \simeq 1.23 \, \text{m}^2$) using Eq.~\eqref{eq:PhaseCoherence} we can compute the minimum frequency that can be probed by these experiments:
\begin{equation}
\label{eq:CutoffFrequency}
f_\text{\tiny min}^{\text{\tiny ALPS}} 
\gg 4.6 \times 10^{12} \, {\rm Hz} \,, 
\qquad 
f_\text{\tiny min}^{\text{\tiny  IAXO}} 
\gg 2.2 \times 10^{9} \, {\rm Hz} \,, \qquad f_\text{\tiny min}^{\text{\tiny  MADMAX}} 
\gg 2.1 \times 10^{8} \, {\rm Hz} \,.
\end{equation}
Given that the amplitude of the signal drops at higher frequencies, IAXO and MADMAX appear to be the most suitable experiment to probe the signals that we are interested in.

From Eq.~\eqref{eq:SPDsensitivity}, it is clear that the sensitivity depends crucially on the combination $B L A^{1/2}$ and on the measurement time $\Delta t$. 
In particular, the dependence $\left({\Delta t}/{\rm yr }\right)^{-1/4}$ proves that the sensitivity gets better when the signal remains for a sufficiently long time 
in the observable frequency band of the detector. 
This excludes immediately the possibility of detecting the chirp phase of light PBH mergers with these types of detectors.
Indeed, combining Eqs.~\eqref{eq:fISCO} and \eqref{eq:CoalescenceTime}, one finds that the final phase of a PBH merger would only last for 
\begin{equation}
	\Delta t \sim {\cal O}(1) \times \frac{1}{f_\text{\tiny ISCO}} 
	\lesssim {\cal O}(10^{-8}) \, {\rm sec},
\end{equation} 
where the frequencies must be $f \gtrsim 10^{8} \, {\rm Hz}$ to satisfy Eq.~\eqref{eq:PhaseCoherence}. 
An ideal candidate signal would be the one coming from superradiant bosonic fields, that gives monochromatic GWs
with a long coherence time.

However, it is also possible to detect the early inspiral phase of single light PBH mergers: 
as the PBH binary gets closer to merging, the frequency of the produced GW grows, spanning the detector sensitivity range from low to high frequencies. 
If the change in frequency is slow enough, the signal remains in the detector sensitivity band for a sufficient time interval, potentially allowing for detection. 
Of course, this type of detector is well suited for the probe of SGWB signals, given that the integration time is not an issue in that case. 

Given that the graviton-to-photon conversion is sensitive to the GW amplitude propagating along the direction of the magnetic field $\mathbf{B}$, one should account 
for the effective contribution of the GW amplitude along the 
detector principal axis. 
As the photon emission depends quadratically on the GW strain, in the case of SGWBs we may consider the average over all the orientations, which amounts to a factor $\langle \cos^2(\theta)  \rangle \simeq 1/2$. 
In the case of single event mergers we should keep in mind that the sensitivities reported here are maximum values: the signal would be suppressed by a factor ${\cal O}(1)$ depending on the orientation and specific details of the experimental apparatus.

In the analyses of Sec.~\ref{sec:Comparison} we will always assume that the detectors have $S/N = 2$, $\Gamma_D = 10^{-3}$ and $\epsilon = 1$. In particular, we will consider three possibilities: \textit{i)} a hypothetical SPD detector (HSPD) that spans a frequency range of $\Delta f = 5 \times 10^8 \, \text{Hz}$ with minimum frequency $f_\text{\tiny min} = 5 \times 10^8 \, \text{Hz}$ and parameters equal to the benchmark values of Eq.~\eqref{eq:SPDsensitivity} ($B = 1 \, \text{T}$, $L = 1 \, \text{m}$, $A = 0.785 \, \text{m}^2$ (corresponding to a radius of $0.5 \, \text{m}$); \textit{ii)} MADMAX, with the specifics above and spanning the frequency range $f \subset(2 \div 4) \times 10^9 \, \text{Hz}$; \textit{iii)} IAXO, with the specifics above and spanning the frequency range $f \subset(2.8 \div 5.1) \times 10^{10} \, \text{Hz}$.
\footnote{Note that, in this section, we are assuming that the scaling of sensitivity as a function of observation time given in Eq.~\eqref{eq:SPDsensitivity} remains valid.
If the number of photons expected in the photon detector is smaller than one, the sensitivity might degrade further and would need a dedicated analysis.}

\subsection{Experiments based on a Gaussian beam}
\label{sec:GB}

A different design~\cite{Li:2000du, Li:2003tv, Li:2004df, Li:2006sx, Li:2008qr, Tong:2008rz, Stephenson:2009zz, Li:2009zzy, Li:2011zzl, Li:2013fna, Li:2014bma, Li:2015nti} that still makes use of the graviton-to-photon conversions entails the use of an additional ingredient: a \textit{Gaussian beam} (GB) with frequency $f_0$ placed in the same direction of the conversion volume axis. A GW traveling along the same $z$-axis can give rise to an induced electromagnetic wave in the direction orthogonal both to the GB and to the static magnetic field. The induced electromagnetic wave is generated at first order in the amplitude of the GW, leading to a potentially large gain in sensitivity with respect to the detectors described above. 
In turn, however, the noise due to the GB photons can be quite large. 
To mitigate this issue, it was proposed to use reflectors in order to focus the induced electromagnetic wave in the direction of the receivers~\cite{Li:2008qr, PhysRevLett.89.223901, doi:10.1063/1.1553993, Hou:05, Woods:2012upj, Ringwald:2020ist}. 
The sensitivity reported in Ref.~\cite{Ringwald:2020ist} is given by
\begin{align}
\label{eq:GBsensitivity}
&
h_c^{\text{\tiny  GB} }
\simeq 
2.8
\times 10^{-29} \, \eta^{-1} \left(\frac{S/N}{2}\right) \left(\frac{\Delta t}{10^4 \, {\rm sec}}\right)^{-1/2} \left(\frac{\Delta f_0/f_0}{10^{-6}}\right)^{-1} \nonumber \\
&\times \epsilon^{-1} \left(\frac{\Gamma_D}{10^{-3} \, {}\rm Hz}\right)^{1/2} \left(\frac{E_0}{5 \times 10^5 \, {\rm V}/{\rm m}}\right)^{-1} \left(\frac{B}{10 \, {\rm T}}\right)^{-1} \left(\frac{L}{5 \, {\rm m}}\right)^{-1} \left(\frac{\Delta A}{0.01 \, {\rm m}^2}\right)^{-1} \left(\frac{\mathcal{F}}{10^{-5}}\right)^{-1} \,,
\end{align}
where $0 < \eta < 1$ is the reflectivity of the reflectors, $\Delta t$ is the measurement time, $\Delta f_0$ is the bandwidth of the detector, $E_0$ is the amplitude of the electric field of the GB, $\Delta A$ is the surface of the electromagnetic wave detector and $\mathcal{F}$ is a function that depends on the geometry of the experimental setup, see Ref.~\cite{Ringwald:2020ist} for more details.
The benchmark values for the various experimental quantities represent the state-of-the-art values that can in principle be achieved in the laboratory.

Note that this detector works at resonance: the orthogonal electromagnetic wave is produced efficiently only if the frequency of the GW matches exactly the frequency of the GB, $f_0$. For this reason, such a detector is not suitable for the detection of coherent signals from PBH inspirals, see the discussion in Sec.~\ref{sec:GertEffect}. However, this concept can in principle be used to detect monochromatic signals such as superradiance, as long as the GW frequency matches the GB frequency to a very good accuracy ($\Delta f/f \sim 10^{-6}$), or SGWBs. 
In the former case, assuming that it is possible to tune the frequency of the detector to match the GW frequency of the source, the GB detector would be able to observe/exclude superradiance from PBHs in the mass range $(10^{-6} \div 10^{-5}) \, M_\odot$, which would correspond to a (pseudo-) scalar mass of $\sim 10^{-4} \, \text{eV}$. In that case, the sensitivity in Eq.~\eqref{eq:GBsensitivity} would refer to the GW amplitude, $h_0$.

It is necessary to emphasize that the feasibility of the experimental apparatus including the reflectors, maintaining the exceptional sensitivity that is reported in the original theoretical papers, has been questioned multiple times and seems extremely difficult to be achieved. Beyond the issues related to diffraction effects caused by the reflectors~\cite{Woods:2012upj}, the original papers do not take into account the noise due to the fact that the laser cannot be exactly linearly polarized in one direction, which would likely dominate the noise budget.\footnote{We thank Sebastian Ellis for pointing out this issue.} More studies to explore the feasibility of such a concept are a necessary step to be taken in the future, in order to assess all the possible noise sources that might deteriorate the sensitivity.

\subsection{Resonant LC circuits}
\label{sec:Haloscopes}

It has recently been shown that axion haloscopes results can be reinterpreted as limits on GWs in the UHF band~\cite{Domcke:2022rgu}. This is the case because, similarly to what happens if axions are present, a passing GW produces an effective current in Maxwell's equations, that causes the existence of oscillating electric and magnetic fields.\footnote{This class of experiments (as well as that described in Sec.~\ref{sec:MicrowaveCavities}) also makes use of the inverse Gertsenshtein effect. The main differences with respect to the type of detectors described in Sec.~\ref{sec:GertEffect} are the geometry of the setup and the way in which the electromagnetic wave generated by the passing GW is measured: while in the case of Sec.~\ref{sec:GertEffect} the detector counts the photons in the generated electromagnetic wave, resonant LC circuit detectors measure the generated magnetic flux.} In particular, data from experiments like ABRACADABRA~\cite{Kahn:2016aff, Ouellet:2018beu, Ouellet:2019tlz, Salemi:2021gck} and SHAFT~\cite{Gramolin:2020ict} already put bounds on UHF-GWs, despite their sensitivity is not yet competitive with other bounds in the same frequency range. In Ref.~\cite{Domcke:2022rgu}, the authors show that in a setup geometry similar to the ABRACADABRA one, with a toroidal static magnetic field, a passing GW produces a magnetic flux at the center of the toroid. This flux can be detected using a pickup loop for which it is demonstrated that the most efficient geometry is a figure $8$ shape.

Interestingly, this type of detector can probe very short signals. In fact, in the case of axions, a signal coherent for a time of order $\mathcal{O}(\mu\text{sec})$ can be probed, corresponding to an axion mass of order $\mathcal{O}(1) \, \text{neV}$. Such a coherence time, which can be written in terms of the quality factor of the signal $Q$: $\tau = Q/f$, corresponds exactly to that of a PBH merger in the chirp phase at $f \sim 1 \, \text{MHz}$, namely $\tau \sim 1/f_\text{\tiny ISCO} \sim \text{MHz}^{-1} \sim 1 \, \mu\text{sec}$. For such a signal one has $Q \sim \mathcal{O}(1)$. Therefore, from Eq.~\eqref{eq:fISCO}, with resonant LC circuit detectors it is possible to probe the chirp phase of PBHs with masses around $m_\text{\tiny PBH} \simeq 10^{-3} \, M_\odot$. Better sensitivities can be achieved if the signal is coherent on longer timescales, i.e. when $Q \gg 1$, as it is the case for superradiance for instance: the sensitivity scales as $Q^{-1/4}$. Ref.~\cite{Domcke:2022rgu} shows that major progress can be achieved in the future, as the sensitivity scales with the volume of the region containing a static magnetic field as $V^{7/6}$. Therefore, experiments like DMRadio~\cite{Chaudhuri:2014dla, Silva-Feaver:2016qhh} and improvements thereof will be able to probe an interesting region of the parameter space. 
For DMR detectors, the curves that we plot in Fig.~\ref{fig:PlotSingleEvents} refer to the best case future scenario: DMradio with a magnetic field volume of $100 \, \text{m}^3$ and figure 8 pick-up loop to detect the magnetic field flux. 

\subsection{Microwave cavities}
\label{sec:MicrowaveCavities}

Another class of experiments that have been developed in the context of axion searches but turn out to be equally useful for the detection of UHF-GWs is given by microwave cavity experiments, see e.g.~\cite{Berlin:2021txa, Berlin:2022hfx}. As for the concept described in Sec.~\ref{sec:Haloscopes}, a GW passing through a cavity containing a static magnetic field produces an effective current in Maxwell's equations. This, in turn, gives rise to an electromagnetic field that oscillates at the same frequency of the GW. Such induced electromagnetic field might be detected using resonant detectors such as microwave cavities. In particular, data from axion experiments can be used to place bounds on UHF-GWs, as shown in Ref.~\cite{Berlin:2021txa} where the authors computed the projected sensitivity of ADMX~\cite{ADMX:2021nhd, ADMX:2019uok, ADMX:2018ogs}, HAYSTAC~\cite{HAYSTAC:2018rwy}, CAPP~\cite{Lee:2020cfj} and ORGAN~\cite{McAllister:2017lkb}.

\begin{figure}[t!]
\centering
\includegraphics[width=0.8 \textwidth]{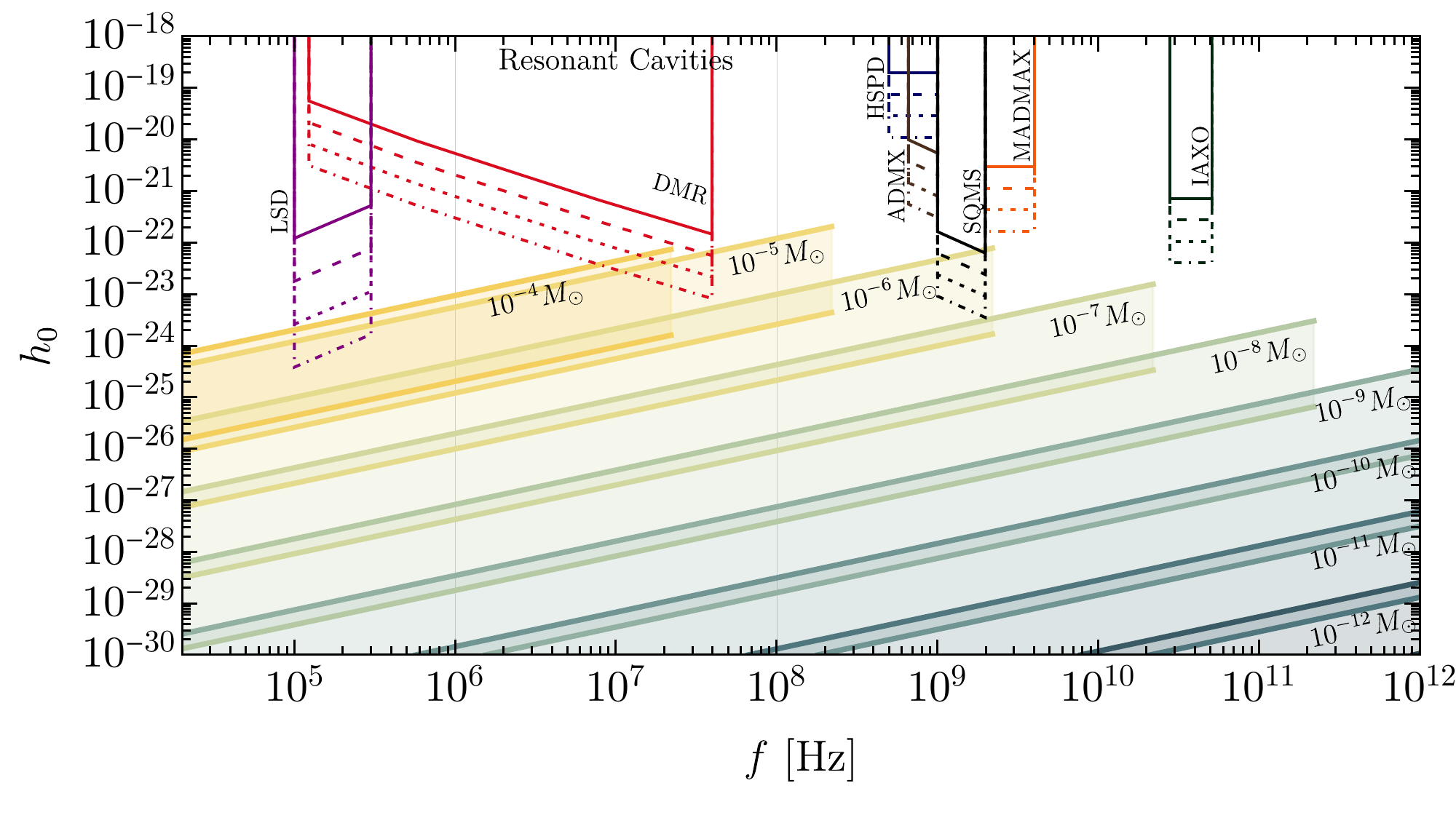}
\caption{
We plot characteristic GW amplitude $h_0$ emitted by a PBH binary merger at a distance $d_\text{\tiny L}=d_\text{\tiny yr}$.
Each color reports a different value of $m_\text{\tiny PBH} \subset (10^{-4}\div 10^{-11}) M_\odot$ as indicated in the insets. The highest part of the filled band corresponds to the strain obtained assuming the maximum theoretical merger rate $R_\PBH^\text{\tiny max}$ (see Sec.~\ref{sec:clustering}), 
while the lowest curve corresponds to the strain values obtained for $f_\text{\tiny{PBH}} = 1$ in the standard scenario using Eq.~\eqref{eq:Rlocal}. 
For each experimental apparatus, we report four different lines, corresponding to the four integration times allowed by the signal with masses spanning four decades below the heaviest observable merger. 
For example, considering for example the DMR detectors, each line from top to bottom corresponds to different integration times set by the maximum time spent by mergers of masses $m_\PBH=(10^{-5}, 10^{-6}, 10^{-7},10^{-8}) M_\odot$ around the frequency of $f\simeq 400$ MHz. See the main text for a complete description of the detector specifications.
}\label{fig:PlotSingleEvents}
\end{figure}

Following Ref.~\cite{Berlin:2021txa}, we consider a cylindrical cavity (whose volume is $V_\text{\tiny cav}$) that contains a static magnetic field $\mathbf{B}$ whose direction is parallel to the axis of the cavity. Assuming a cavity-GW coupling coefficient $\eta_n$ and system temperature $T_\text{\tiny sys}$, the sensitivity of such a cavity can be estimated as
\begin{align}
h_0 = \,\, &
3 \times 10^{-22}
\left(\frac{0.1}{\eta_n}\right)
\left(\frac{8 \, \text{T}}{|\bf B|}\right) 
\left(\frac{0.1 \, \text{m}^3}{V_\text{\tiny cav}}\right)^{5/6}
\left(\frac{10^5}{Q}\right)^{1/2} 
\nonumber \\
& \times    
\left(\frac{T_\text{\tiny sys}}{1 \, \text{K}}\right)^{1/2} 
\left(\frac{1 \, \text{GHz}}{f}\right)^{3/2}
\left(\frac{\Delta f}{10 \, \text{kHz}}\right)^{1/4} 
\left(\frac{1 \, \text{min}}{\Delta t}\right)^{1/4}
\,,
\end{align}
where the benchmark values are taken from the ADMX experiment.  Ref.~\cite{Berlin:2021txa} shows that a cavity-GW coupling coefficient of order $\eta_n \sim \mathcal{O}(1)$ is a reasonable assumption for the modes ($T_{010}$ and $T_{020}$) already used in axion experiments such as ADMX and ORGAN. 
Note that the quality factor $Q$ and the cavity bandwidth $\Delta f$ are linked through the relation $\Delta f \simeq f/Q$. In Fig.~\ref{fig:PlotSingleEvents} we plot the sensitivity curves for the two most promising detectors: ADMX and SQMS. For these curves we took the parameters reported in~\cite{Berlin:2021txa} $f \subset(0.65 \div 1.02) \, \text{GHz}$, $Q \sim 8 \times 10^4$, $|\mathbf{B}| = 7.5 \, \text{T}$, $V_\text{\tiny cav} = 136 \, \text{L}$ and $T_\text{\tiny sys} = 0.6 \, \text{K}$, while for SQMS we used $f \subset(1 \div 2) \, \text{GHz}$, $Q \sim 10^6$, $|\mathbf{B}| = 5 \, \text{T}$, $V_\text{\tiny cav} = 100 \, \text{L}$ and $T_\text{\tiny sys} = 1 \, \text{K}$. We also take $\eta_n = 0.1$ in both cases, while the integration time used is given by the intrinsic timescale dictated by the frequency evolution of the inspiraling PBH binary, see Eq.~\eqref{eq:CoalescenceTime}. The duration of the signal has to be longer than the \textit{ring-up time} of the cavity, determined by the inverse of the bandwidth. Using Eq.~\eqref{eq:CoalescenceTime}, this requirement sets an upper limit on the mass of the PBHs whose mergers can be probed by microwave cavities. 
Therefore, as we will show in Fig.~\ref{fig:PlotSingleEvents}, the ADMX and SQMS sensitivity curves apply to PBH inspirals with masses  
$m_\text{\tiny PBH} \lesssim 10^{-9} \, M_\odot$ and 
$m_\text{\tiny PBH} \lesssim 10^{-10}\, M_\odot$ , respectively \cite{Berlin:2021txa}.

\subsection{Realistic comparison with GW signals from PBH mergers}
\label{sec:Comparison}

In this section we explore how the assumptions that lie behind the most promising detector curves reported above (e.g. in Fig.~\ref{fig: PBH hc single} and Fig.~\ref{fig: PBH hc SGWB}) affect the detectability of the GW signals from PBH mergers.

\subsubsection{Detection prospects for PBH mergers (Fig.~\ref{fig:PlotSingleEvents})}

When comparing the signals from coherent sources with the various detectors, we plot the GW amplitude $h_0$ instead of the characteristic strain as the theoretical papers describing detector proposals assume a passing GW plane wave, whose amplitude is $h_0$.
Also, the sensitivity of each detector is computed accounting for the intrinsic GW signal duration, depending on the PBH masses and frequency and using Eq.~\eqref{eq:CoalescenceTime}.

In Fig.~\ref{fig:PlotSingleEvents} we plot the sensitivity curves corresponding to \textcolor{black}{the LSD detectors described in Sec.~\ref{sec:lsd}}, the SPD detectors described in Sec.~\ref{sec:GertEffect} (HSPD, MADMAX and IAXO), resonant LC circuits described in Sec.~\ref{sec:Haloscopes} (DMR) and microwave cavities, described in Sec.~\ref{sec:MicrowaveCavities} (ADMX and SQMS). Concerning the GW signals, we plot the curves corresponding to PBH inspirals with mass $m_\text{\tiny PBH} \subset (10^{-4} \div 10^{-12}) \, M_\odot$. For each band, the upper curve saturates the maximum theoretical merger rate, see Sec.~\ref{sec:clustering}, while the lower curve corresponds to $f_\text{\tiny PBH} = 1$ in the standard scenario, see Eq.~\eqref{eq:Rlocal}.

\textcolor{black}{For LSD detectors, we used the benchmark values reported in Sec.~\ref{sec:lsd}. The sensitivity curves from top to bottom refer to PBH inspirals with masses
\begin{equation}
m_\PBH\subset (10^{-4},10^{-5}, 10^{-6}, 10^{-7}) \, M_\odot \qquad \text{LSD}.
\end{equation}}

For the case of the SPD detectors described in Sec.~\ref{sec:GertEffect}, the sensitivity curves from top to bottom refer to PBH inspirals with masses
\begin{equation}
m_\text{\tiny PBH}\subset  
\begin{cases}
(10^{-6}, 10^{-7}, 10^{-8}, 10^{-9}) M_\odot \qquad \,\,\,\,\, \text{HSPD}, \\
(10^{-7}, 10^{-8}, 10^{-9}, 10^{-10})M_\odot \qquad \,\,\text{MADMAX}, \\
(10^{-8}, 10^{-9}, 10^{-10}, 10^{-11})M_\odot \qquad \text{IAXO}, 
\end{cases}
\end{equation}
respectively. Note that the short duration of the various signals in these frequency bands makes the sensitivity degrade significantly. 
We do not show smaller masses as the signal amplitude becomes increasingly distant from the detectors' reach.

For DMR detectors, the curves that we plot in Fig.~\ref{fig:PlotSingleEvents} are evaluated using the quality factors of the signals corresponding from top to bottom to 
\begin{equation}
m_\PBH\subset (10^{-5}, 10^{-6}, 10^{-7}, 10^{-8}) \, M_\odot \qquad \text{DMR}.
\end{equation}
Note, therefore, that the top curve is applicable to the signal corresponding to
$M_\text{\tiny PBH} = 10^{-5} \, M_\odot$ PBH inspirals, which are close to the chirp phase ($Q \simeq 1$) at the right end of the DMR frequency band. For this case, which represents the best case scenario in terms of detectability, the gap between the DMR sensitivity curve and the loudest signal from PBH mergers (obtained under the assumption that the signal saturates the maximum merger rate, see Sec.~\ref{sec:clustering}) is only one order of magnitude.

Concerning microwave cavities, the requirement that the time spent by the signal in the detector band is larger than the ring-up time of the cavity implies that the ADMX and SQMS sensitivity curves can only apply to small enough PBHs masses, see the discussion in Sec.~\ref{sec:MicrowaveCavities}.
The sensitivity curves plotted in Fig.~\ref{fig:PlotSingleEvents} apply to PBH inspirals with masses 
\begin{equation}
m_\text{\tiny PBH}\subset  
\begin{cases}
 (10^{-9}, 10^{-10}, 10^{-11}, 10^{-12})    M_\odot \qquad \,\, \text{ADMX}, \\
(10^{-10}, 10^{-11}, 10^{-12}, 10^{-13}) M_\odot \qquad \text{SQSM}, 
\end{cases}
\end{equation}
 respectively. The integration time is given by the merger time of the PBH inspiral at the corresponding frequency of the various detectors. In the best case scenarios, ADMX and SQSM top curves to be compared with the maximum PBH inspiral signal for $M_\text{\tiny PBH} = 10^{-9} \, M_\odot$ and $M_\text{\tiny PBH} = 10^{-10} \, M_\odot$ respectively, there is still a gap of roughly five-six orders of magnitude.

\begin{figure}[t!]
\centering
\includegraphics[width=0.8 \textwidth]{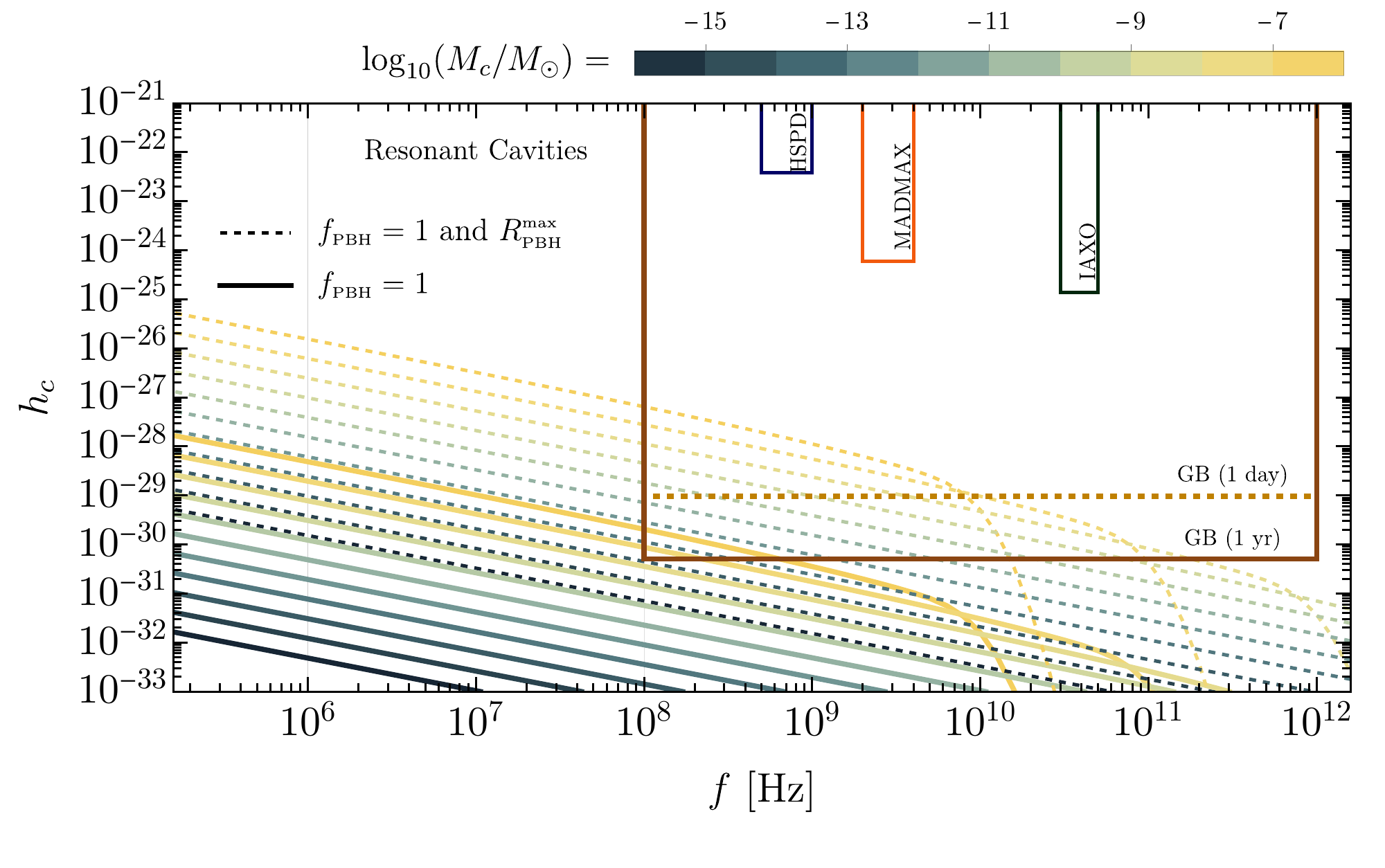}
\caption{
 The thin dashed lines correspond to the enhanced signal described in Sec.~\ref{sec:clustering}, while the thick lines indicate the SGWB obtained with $f_\text{\tiny  PBH} = 1$ and the merger rate computed in the standard Poisson scenario, as described in Sec.~\ref{sec:SGWB}. 
 See the main text for a complete description of the detector specifications.
}\label{fig:PlotStochastic}
\end{figure}

\subsubsection{Detection prospects for a SGWB from PBH mergers (Fig.~\ref{fig:PlotStochastic})}

For a SGWB, we are not limited by the signal duration. In Fig.~\ref{fig:PlotStochastic}, we assume an integration time of one year for HSPD, MADMAX and IAXO. 
We also show the sensitivity of the GB detector, as from Eq.~\eqref{eq:GBsensitivity}, assuming integration times of one day (dashed line) and one year (solid line). To plot these curves, beyond all the benchmark values for $S/N$, $\Gamma_D$, $E_0$, $B$, $L$, $\Delta A$, $\mathcal{F}$, we assume $\eta = \epsilon = 1$ and that $\Delta f/f = 10^{-6}$, so that the frequency dependence disappears. We plot the signal in the range $(10^8 \div 10^{12}) \, \text{Hz}$ because this is the range explored in the literature~\cite{Li:2000du, Li:2003tv, Li:2004df, Li:2006sx, Li:2008qr, Tong:2008rz, Stephenson:2009zz, Li:2009zzy, Li:2011zzl, Li:2013fna, Li:2014bma, Li:2015nti, Hou:05, Woods:2012upj, Ringwald:2020ist, Ringwald:2020ist}.

Assuming that the sensitivity proposed for the GB design can be achieved, such a detector may constrain PBHs in the ultra-light mass range $m_\PBH \gtrsim 10^{-13} M_\odot$, at least if a boosted merger rate close to $R_\PBH^\text{\tiny max}$ is achieved. This would already represent a novel constraint on some scenarios for the asteroidal mass PBHs as DM which is notoriously difficult to probe with other means (see e.g. Ref.~\cite{Montero-Camacho:2019jte}). 

\subsection{Discussion}
\label{sec:Discussion}

As we have seen in the previous sections, it is not possible to detect PBH-related GW signals using already operating detectors. One of the most compelling issues for many detectors is related to the short duration of the signal, which reduces the possibility of detecting transient signals in the UHF band, \textcolor{black}{as we have discussed for magnetic conversion detectors in Sec.~\ref{sec:GertEffect}.}

The currently most optimistic future prospects rely on the actual experimental implementation of the theoretical sensitivity of the GB detector, see Sec.~\ref{sec:GB}, whose feasibility has, however, been questioned~\cite{Woods:2012upj}. \textcolor{black}{Despite the benchmark LSD detector that we used for the sensitivity estimates in Fig.~\ref{fig:PlotSingleEvents} is only theoretically designed at the moment, this type of technology is currently being tested with the construction of a one-meter prototype and hence is likely to make progress in the near future. If the benchmark numbers could be achieved, this detector would be only one or two orders of magnitude away from the most optimistic PBH signal, at $10^5 \, \text{Hz}$, coming from PBH binaries with mass $(10^{-5} \div 10^{-6}) \, M_\odot$.} Furthermore, there are other interesting experimental concepts that have been proposed only very recently and might become the most promising routes forward. For instance, resonant cavities play a central role in a couple of very recent proposals~\cite{Herman:2020wao, Herman:2022fau, Berlin:2021txa, Berlin:2022hfx}. In Fig.~\ref{fig:PlotSingleEvents} and Fig.~\ref{fig:PlotStochastic}, we indicate the corresponding range of frequencies between vertical grid lines.
In Refs.~\cite{Herman:2020wao, Herman:2022fau}, the authors suggest the use of resonant cavities to resonantly amplify the electromagnetic waves generated by the passing GW through the inverse Gertsenshtein effect, as described in Sec.~\ref{sec:MicrowaveCavities}. The strain sensitivity for stochastic backgrounds reported in Refs.~\cite{Herman:2020wao, Herman:2022fau} is $h_c \sim 10^{-30}$ in the frequency range $(10^6\div10^8) \, \text{Hz}$. If such an instrument could be implemented in the lab, then we would be able to observe the stochastic signal from unresolved PBH mergers, see Fig.~\ref{fig:PlotStochastic}.\footnote{We thank Sebastien Clesse and Nicolas Herman for private discussions on these points.}
The potentially testable parameter space would include large values of the abundance of PBHs in the asteroidal mass range with masses around $m_\text{\tiny{PBH}}\simeq 10^{-11} \, M_\odot$ within the ‘standard’ merger scenario.
However, the proposal of Refs.~\cite{Herman:2020wao, Herman:2022fau} is in a quite preliminary stage, and further studies about the noise sources that could deteriorate the sensitivity are still lacking. On the other hand, other proposals suggest the use of concepts already discussed in the context of axion DM experiments~\cite{Berlin:2019ahk, Berlin:2020vrk}. Preliminary results in this direction indicate that it might be possible to reach a sensitivity for coherent signals of $h_0 \sim 10^{-23}$ in the frequency range $(10^6\div 10^8) \, \text{Hz}$. Further analyses are currently being performed to assess all the possible noise sources present in this type of detector.\footnote{We thank Diego Blas and Raffaele Tito d'Agnolo for private discussions on these points.}

\section{Conclusions}\label{sec:conclusions}

Gravitational waves in the ultra-high-frequency band represent an interesting and promising avenue for the discovery of new physics. The challenging journey that will hopefully lead to gravitational wave detection in this frequency range has just started. As we have reviewed in this paper, an intense effort is required on the experimental side to reach the required sensitivities that could probe physically interesting gravitational wave strains. From the theoretical point of view, on the other hand, it is necessary to provide precise and physically sound targets, in order to guide the work on detector concepts.

One of the best-motivated sources in the ultra-high-frequency range is light primordial black holes which may be a promising candidate to explain the dark matter in our Universe.
Building upon the recent advancements in the modeling of primordial black hole merger rates, in this paper we have used state-of-the-art techniques to provide a systematic study of the possible gravitational wave signals that can be expected from various phenomena related to primordial black holes. 
Focusing on scenarios in which primordial black holes play a role as dark matter candidates, we have shown that the specific features of transient signals in the ultra-high-frequency band render detection particularly challenging with currently operating and proposed detectors. One of the issues in this context is the short duration of the signal that does not allow for a sizeable conversion of gravitational into electromagnetic waves. This problem prevents the use of some types of detectors, described in Sec.~\ref{sec:GertEffect}, for the detection of coherent signals. Other types of concepts though, such as \textcolor{black}{levitated sensor detectors}, resonant LC circuits and microwave cavities, do not suffer from this type of problem. \textcolor{black}{If the benchmark values  reported in Sec.~\ref{sec:lsd} for levitated sensor detectors could be achieved, then the sensitivity of this type of detector would be only one order of magnitude away from the most optimistic primordial black hole inspiral signals, at $f \sim 10^5 \, \text{Hz}$. Along with levitated sensors,} resonant LC circuits represent the best current prospect in terms of detectability, with their sensitivity missing the loudest gravitational wave signal from primordial black hole inspirals at $f \sim 30 \, \text{MHz}$ for just one order of magnitude.

The issue of the duration of the signal is avoided in the case of stochastic signals or persistent signals such as those coming from superradiance effects, but in these cases the strain is such that achieving detection remains beyond the capabilities of existing proposals. The main exception is represented by the Gaussian beam detector, assuming that the sensitivity reported in the theoretical studies of this concept can be achieved in the laboratories. 
This experiment could set a novel constraint on some scenarios for the asteroidal mass primordial black hole dark matter, which is notoriously difficult to probe with other means (see e.g. Ref.~\cite{Montero-Camacho:2019jte}). 
A critical assessment of the feasibility of such detectors is a crucial theoretical and experimental step to be taken. 
Other proposals, especially those involving resonant cavities, might be promising alternative ways forward. There are various groups already involved in the effort of developing these ideas further.

We conclude with an optimistic note: in the last decades we have witnessed how the technological progress and the effort of many researchers have made it possible to overcome experimental difficulties that are similar to those now experienced in the context of ultra-high-frequency gravitational waves. 
Renowned examples are gravitational waves in the sub-$\text{kHz}$ range, where detection was finally achieved after decades of tireless efforts, and the case of axions, for which the most recent proposals would finally 
make it possible to probe almost the entire phenomenologically interesting parameter space. 
This paper provides further motivation to keep exploring ultra-high-frequency gravitational wave physics.\\

{\bf Note added:} During the completion of this work, we learned that the second version of Ref.~\cite{Domcke:2022rgu} was in preparation. 
In particular, in their App. III, the authors address some details of the computation of the merger rate of individual PBH binaries also discussed here. Where they overlap, the results are consistent with the ones presented here.

\vskip.25cm
\section*{Acknowledgements} 
The data needed to reproduce the figures in this work are available upon request to the authors.
We thank E. Thrane and P. Lasky for discussions on the GW memory and K. Kritos for discussions about dynamical binary formation channels. 
We thank Diego Blas, Sebasti\'en Clesse, Raffaele Tito d'Agnolo, Aldo Ejlli,  Nicolas Herman, Jan Sch\"utte-Engel for providing various clarifications about detector proposals. 
We thank Valerie Domcke, Sebastian Ellis, Camilo Garcia-Cely, Andreas Ringwald, Nicholas Rodd for discussions and comments on an earlier version of this work. 
We also thank Nancy Aggarwal for useful discussions about the topic of the paper.
G.F. thanks Johns Hopkins University for the kind hospitality during the completion of this project.
G. F. acknowledges financial support provided under the European Union's H2020 ERC, Starting Grant agreement no.~DarkGRA--757480, and under the MIUR PRIN and FARE programmes (GW-NEXT, CUP:~B84I20000100001), support from the Amaldi Research Center funded by the MIUR program ``Dipartimento di Eccellenza" (CUP:~B81I18001170001) and H2020-MSCA-RISE-2020 GRU. F.M. is funded
by a UKRI/EPSRC Stephen Hawking fellowship, grant
reference EP/T017279/1 and partially supported by the
STFC consolidated grant ST/P000681/1.

\appendix 

\section{GW energy emission spectrum from binary BHs}\label{app_GW_energy}

In this appendix, we report the formulas entering the computation of the energy spectrum of GW emitted during the inspiral-merger-ringdown evolution. 
Following the notation adopted in Eq.~\eqref{eq:dEdnu}, one can define
\begin{equation}
\begin{split}
    f_1 & \equiv f_1(\nu, M,\eta,\chi) = 1+\alpha_2\nu'^2 + \alpha_3 \nu'^3 \, , \\  
    f_2 & \equiv f_2(\nu, M,\eta,\chi) = 1 + \varepsilon_1 \nu' + \varepsilon_2 \nu'^2 \, ,\\
    f_3 & \equiv f(\nu, \nu_\text{\tiny ringdown}, M, \chi) = \frac{\nu}{1+\left(\frac{2(\nu-\nu_\text{\tiny ringdown})}{\sigma}\right)^2} \, ,
\end{split}
\end{equation}
with $\nu' = (\pi MG \nu / c^3)^{1/3}$. 
Additionally, the single spin parameter is defined  as 
\begin{equation}
\chi = (1+\delta)\frac{\chi_1}{2}+(1-\delta)\frac{\chi_2}{2},
\end{equation}
as a function of the relative mass difference 
$\delta \equiv (m_1-m_2)/(m_1+m_2)$, and 
\begin{equation}
    \begin{split}
\alpha_2 &= -\frac{323}{224}+\frac{451}{168}\eta \,, 
\qquad \
\alpha_3  = \left( \frac{27}{8} - \frac{11}{6}\eta \right)\chi \, , 
\\
\varepsilon_1 &= 1.4547 \chi - 1.8897 \,,
\quad
\varepsilon_2 = - 1.8153 \chi + 1.6557 \, .
\end{split}
\end{equation}
The merger, ringdown and cut frequencies, as well as $\sigma$, are approximated by
\begin{equation}
\begin{split}
    &\nu_\text{\tiny merger} = \frac{c^3}{\pi M G}  \left(1 - 4.455(1-\chi)^{0.217} + 3.521 (1-\chi)^{0.26} + \mu_\text{\tiny merger} \right) , \\
    &\nu_\text{\tiny ringdown} = \frac{c^3}{\pi M G}  \left(\left(1-0.63(1-\chi)^{0.3}\right)/2 + \mu_\text{\tiny ringdown}\right) ,\\
    &\nu_\text{\tiny cut} = \frac{c^3}{\pi M G}  \left(0.3236 + 0.04894\chi + 0.01346 \chi^2  + \mu_\text{\tiny cut}\right) , \\
    & \sigma = \frac{c^3}{\pi M G} \left( \left(1-0.63(1-\chi)^{0.3}\right)(1-\chi)^{0.45}/4 + \mu_\mathrm{\sigma} \right),
\end{split}
\label{eq:nus}
\end{equation}
where $\mu^{(ij)}_k$, with the index $k$
indicating either 
$[\text{merger},\text{ringdown},\text{cut},\sigma]$, 
is computed as
\begin{equation}
    \mu_k \equiv \mu_k(\eta,\chi) = \sum_{i=1}^{3} \sum_{j=0}^{\min(3-i,2)} y_k^{(ij)} \eta^i \chi^j \, ,
\end{equation}
with $y_k$ coefficents given in Table~1 of \cite{Ajith:2009bn}.
Finally, $\omega_1$ and $\omega_2$ are normalisation constants that guarantee continuity at the frequencies $\nu_\text{\tiny merger}$ and $\nu_\text{\tiny ringdown}$, respectively,  
\begin{equation}
 \begin{split}
    & \omega_1 = \nu_\text{\tiny merger}^{-1}f_1^2(\nu_\text{\tiny merger}, M, \eta, \chi)/f_2^2(\nu_\text{\tiny merger}, M, \eta, \chi) \, , \\
    & \omega_2 = \omega_1 \nu_\text{\tiny ringdown}^{-4/3} f_2^2(\nu_\text{\tiny ringdown}, M, \eta, \chi)  \, .
 \end{split}
\end{equation}
We will neglect the impact of spin on the GW emission as PBH are expected to be produced with very small spin in the standard scenario \cite{DeLuca:2019buf,Mirbabayi:2019uph}. 
However, we note that considering extremal spin may lead to a relative variation of the SGWB amplitude smaller than about $40\%$.

\bibliographystyle{JHEP}
\bibliography{draft}

\providecommand{\href}[2]{#2}\begingroup\raggedright\begin{thebibliography}{100}

\bibitem{Bertone:2016nfn}
G.~Bertone and D.~Hooper, {\it {History of dark matter}},  {\em Rev. Mod.
  Phys.} {\bf 90} (2018), no.~4 045002,
  [\href{http://arxiv.org/abs/1605.04909}{{\tt arXiv:1605.04909}}].

\bibitem{Hu:2000ke}
W.~Hu, R.~Barkana, and A.~Gruzinov, {\it {Cold and fuzzy dark matter}},  {\em
  Phys. Rev. Lett.} {\bf 85} (2000) 1158--1161,
  [\href{http://arxiv.org/abs/astro-ph/0003365}{{\tt astro-ph/0003365}}].

\bibitem{Amendola:2005ad}
L.~Amendola and R.~Barbieri, {\it {Dark matter from an ultra-light
  pseudo-Goldsone-boson}},  {\em Phys. Lett. B} {\bf 642} (2006) 192--196,
  [\href{http://arxiv.org/abs/hep-ph/0509257}{{\tt hep-ph/0509257}}].

\bibitem{Hui:2016ltb}
L.~Hui, J.~P. Ostriker, S.~Tremaine, and E.~Witten, {\it {Ultralight scalars as
  cosmological dark matter}},  {\em Phys. Rev. D} {\bf 95} (2017), no.~4
  043541, [\href{http://arxiv.org/abs/1610.08297}{{\tt arXiv:1610.08297}}].

\bibitem{zel1967hypothesis}
Y.~B. Zel'dovich and I.~D. Novikov, {\it The hypothesis of cores retarded
  during expansion and the hot cosmological model},  {\em Soviet Astronomy}
  {\bf 10} (1967) 602.

\bibitem{Hawking:1971ei}
S.~Hawking, {\it {Gravitationally collapsed objects of very low mass}},  {\em
  Mon. Not. Roy. Astron. Soc.} {\bf 152} (1971) 75.

\bibitem{Carr:1974nx}
B.~J. Carr and S.~W. Hawking, {\it {Black holes in the early Universe}},  {\em
  Mon. Not. Roy. Astron. Soc.} {\bf 168} (1974) 399--415.

\bibitem{Carr:1975qj}
B.~J. Carr, {\it {The Primordial black hole mass spectrum}},  {\em Astrophys.
  J.} {\bf 201} (1975) 1--19.

\bibitem{Chapline:1975ojl}
G.~F. Chapline, {\it {Cosmological effects of primordial black holes}},  {\em
  Nature} {\bf 253} (1975), no.~5489 251--252.

\bibitem{Sasaki:2018dmp}
M.~Sasaki, T.~Suyama, T.~Tanaka, and S.~Yokoyama, {\it {Primordial black
  holes---perspectives in gravitational wave astronomy}},  {\em Class. Quant.
  Grav.} {\bf 35} (2018), no.~6 063001,
  [\href{http://arxiv.org/abs/1801.05235}{{\tt arXiv:1801.05235}}].

\bibitem{Silk:1986vc}
J.~Silk and M.~S. Turner, {\it {Double Inflation}},  {\em Phys. Rev. D} {\bf
  35} (1987) 419.

\bibitem{Kawasaki:1997ju}
M.~Kawasaki, N.~Sugiyama, and T.~Yanagida, {\it {Primordial black hole
  formation in a double inflation model in supergravity}},  {\em Phys. Rev. D}
  {\bf 57} (1998) 6050--6056, [\href{http://arxiv.org/abs/hep-ph/9710259}{{\tt
  hep-ph/9710259}}].

\bibitem{Kannike:2017bxn}
K.~Kannike, L.~Marzola, M.~Raidal, and H.~Veerm\"ae, {\it {Single Field Double
  Inflation and Primordial Black Holes}},  {\em JCAP} {\bf 09} (2017) 020,
  [\href{http://arxiv.org/abs/1705.06225}{{\tt arXiv:1705.06225}}].

\bibitem{Ashoorioon:2020hln}
A.~Ashoorioon, A.~Rostami, and J.~T. Firouzjaee, {\it {Charting the Landscape
  in Our Neighborhood from the PBHs Mass Distribution and GWs}},
  \href{http://arxiv.org/abs/2012.02817}{{\tt arXiv:2012.02817}}.

\bibitem{Garcia-Bellido:1996mdl}
J.~Garcia-Bellido, A.~D. Linde, and D.~Wands, {\it {Density perturbations and
  black hole formation in hybrid inflation}},  {\em Phys. Rev. D} {\bf 54}
  (1996) 6040--6058, [\href{http://arxiv.org/abs/astro-ph/9605094}{{\tt
  astro-ph/9605094}}].

\bibitem{Alabidi:2009bk}
L.~Alabidi and K.~Kohri, {\it {Generating Primordial Black Holes Via
  Hilltop-Type Inflation Models}},  {\em Phys. Rev. D} {\bf 80} (2009) 063511,
  [\href{http://arxiv.org/abs/0906.1398}{{\tt arXiv:0906.1398}}].

\bibitem{Clesse:2015wea}
S.~Clesse and J.~Garc\'\i{}a-Bellido, {\it {Massive Primordial Black Holes from
  Hybrid Inflation as Dark Matter and the seeds of Galaxies}},  {\em Phys. Rev.
  D} {\bf 92} (2015), no.~2 023524,
  [\href{http://arxiv.org/abs/1501.07565}{{\tt arXiv:1501.07565}}].

\bibitem{Garcia-Bellido:2017mdw}
J.~Garcia-Bellido and E.~Ruiz~Morales, {\it {Primordial black holes from single
  field models of inflation}},  {\em Phys. Dark Univ.} {\bf 18} (2017) 47--54,
  [\href{http://arxiv.org/abs/1702.03901}{{\tt arXiv:1702.03901}}].

\bibitem{Ballesteros:2017fsr}
G.~Ballesteros and M.~Taoso, {\it {Primordial black hole dark matter from
  single field inflation}},  {\em Phys. Rev. D} {\bf 97} (2018), no.~2 023501,
  [\href{http://arxiv.org/abs/1709.05565}{{\tt arXiv:1709.05565}}].

\bibitem{Hertzberg:2017dkh}
M.~P. Hertzberg and M.~Yamada, {\it {Primordial Black Holes from Polynomial
  Potentials in Single Field Inflation}},  {\em Phys. Rev. D} {\bf 97} (2018),
  no.~8 083509, [\href{http://arxiv.org/abs/1712.09750}{{\tt
  arXiv:1712.09750}}].

\bibitem{Motohashi:2017kbs}
H.~Motohashi and W.~Hu, {\it {Primordial Black Holes and Slow-Roll Violation}},
   {\em Phys. Rev. D} {\bf 96} (2017), no.~6 063503,
  [\href{http://arxiv.org/abs/1706.06784}{{\tt arXiv:1706.06784}}].

\bibitem{Ballesteros:2020qam}
G.~Ballesteros, J.~Rey, M.~Taoso, and A.~Urbano, {\it {Primordial black holes
  as dark matter and gravitational waves from single-field polynomial
  inflation}},  {\em JCAP} {\bf 07} (2020) 025,
  [\href{http://arxiv.org/abs/2001.08220}{{\tt arXiv:2001.08220}}].

\bibitem{Kawasaki:2012wr}
M.~Kawasaki, N.~Kitajima, and T.~T. Yanagida, {\it {Primordial black hole
  formation from an axionlike curvaton model}},  {\em Phys. Rev. D} {\bf 87}
  (2013), no.~6 063519, [\href{http://arxiv.org/abs/1207.2550}{{\tt
  arXiv:1207.2550}}].

\bibitem{Carr:2017edp}
B.~Carr, T.~Tenkanen, and V.~Vaskonen, {\it {Primordial black holes from
  inflaton and spectator field perturbations in a matter-dominated era}},  {\em
  Phys. Rev. D} {\bf 96} (2017), no.~6 063507,
  [\href{http://arxiv.org/abs/1706.03746}{{\tt arXiv:1706.03746}}].

\bibitem{Garcia-Bellido:2016dkw}
J.~Garcia-Bellido, M.~Peloso, and C.~Unal, {\it {Gravitational waves at
  interferometer scales and primordial black holes in axion inflation}},  {\em
  JCAP} {\bf 12} (2016) 031, [\href{http://arxiv.org/abs/1610.03763}{{\tt
  arXiv:1610.03763}}].

\bibitem{Domcke:2017fix}
V.~Domcke, F.~Muia, M.~Pieroni, and L.~T. Witkowski, {\it {PBH dark matter from
  axion inflation}},  {\em JCAP} {\bf 07} (2017) 048,
  [\href{http://arxiv.org/abs/1704.03464}{{\tt arXiv:1704.03464}}].

\bibitem{Ozsoy:2018flq}
O.~\"Ozsoy, S.~Parameswaran, G.~Tasinato, and I.~Zavala, {\it {Mechanisms for
  Primordial Black Hole Production in String Theory}},  {\em JCAP} {\bf 07}
  (2018) 005, [\href{http://arxiv.org/abs/1803.07626}{{\tt arXiv:1803.07626}}].

\bibitem{Green:2000he}
A.~M. Green and K.~A. Malik, {\it {Primordial black hole production due to
  preheating}},  {\em Phys. Rev. D} {\bf 64} (2001) 021301,
  [\href{http://arxiv.org/abs/hep-ph/0008113}{{\tt hep-ph/0008113}}].

\bibitem{Bassett:2000ha}
B.~A. Bassett and S.~Tsujikawa, {\it {Inflationary preheating and primordial
  black holes}},  {\em Phys. Rev. D} {\bf 63} (2001) 123503,
  [\href{http://arxiv.org/abs/hep-ph/0008328}{{\tt hep-ph/0008328}}].

\bibitem{Martin:2019nuw}
J.~Martin, T.~Papanikolaou, and V.~Vennin, {\it {Primordial black holes from
  the preheating instability in single-field inflation}},  {\em JCAP} {\bf 01}
  (2020) 024, [\href{http://arxiv.org/abs/1907.04236}{{\tt arXiv:1907.04236}}].

\bibitem{Muia:2019coe}
F.~Muia, M.~Cicoli, K.~Clough, F.~Pedro, F.~Quevedo, and G.~P. Vacca, {\it {The
  Fate of Dense Scalar Stars}},  {\em JCAP} {\bf 07} (2019) 044,
  [\href{http://arxiv.org/abs/1906.09346}{{\tt arXiv:1906.09346}}].

\bibitem{Cotner:2019ykd}
E.~Cotner, A.~Kusenko, M.~Sasaki, and V.~Takhistov, {\it {Analytic Description
  of Primordial Black Hole Formation from Scalar Field Fragmentation}},  {\em
  JCAP} {\bf 10} (2019) 077, [\href{http://arxiv.org/abs/1907.10613}{{\tt
  arXiv:1907.10613}}].

\bibitem{Nazari:2020fmk}
Z.~Nazari, M.~Cicoli, K.~Clough, and F.~Muia, {\it {Oscillon collapse to black
  holes}},  {\em JCAP} {\bf 05} (2021) 027,
  [\href{http://arxiv.org/abs/2010.05933}{{\tt arXiv:2010.05933}}].

\bibitem{Green:1997pr}
A.~M. Green, A.~R. Liddle, and A.~Riotto, {\it {Primordial black hole
  constraints in cosmologies with early matter domination}},  {\em Phys. Rev.
  D} {\bf 56} (1997) 7559--7565,
  [\href{http://arxiv.org/abs/astro-ph/9705166}{{\tt astro-ph/9705166}}].

\bibitem{Harada:2016mhb}
T.~Harada, C.-M. Yoo, K.~Kohri, K.-i. Nakao, and S.~Jhingan, {\it {Primordial
  black hole formation in the matter-dominated phase of the Universe}},  {\em
  Astrophys. J.} {\bf 833} (2016), no.~1 61,
  [\href{http://arxiv.org/abs/1609.01588}{{\tt arXiv:1609.01588}}].

\bibitem{Krippendorf:2018tei}
S.~Krippendorf, F.~Muia, and F.~Quevedo, {\it {Moduli Stars}},  {\em JHEP} {\bf
  08} (2018) 070, [\href{http://arxiv.org/abs/1806.04690}{{\tt
  arXiv:1806.04690}}].

\bibitem{deJong:2021bbo}
E.~de~Jong, J.~C. Aurrekoetxea, and E.~A. Lim, {\it {Primordial black hole
  formation with full numerical relativity}},  {\em JCAP} {\bf 03} (2022),
  no.~03 029, [\href{http://arxiv.org/abs/2109.04896}{{\tt arXiv:2109.04896}}].

\bibitem{Padilla:2021zgm}
L.~E. Padilla, J.~C. Hidalgo, and K.~A. Malik, {\it {A new mechanism for
  primordial black hole formation during reheating}},
  \href{http://arxiv.org/abs/2110.14584}{{\tt arXiv:2110.14584}}.

\bibitem{DeLuca:2021pls}
V.~De~Luca, G.~Franciolini, A.~Kehagias, P.~Pani, and A.~Riotto, {\it
  {Primordial Black Holes in Matter-Dominated Eras: the Role of Accretion}},
  \href{http://arxiv.org/abs/2112.02534}{{\tt arXiv:2112.02534}}.

\bibitem{Das:2021wad}
S.~Das, A.~Maharana, and F.~Muia, {\it {A Faster Growth of Perturbations in an
  Early Matter Dominated Epoch: Primordial Black Holes and Gravitational
  Waves}},  \href{http://arxiv.org/abs/2112.11486}{{\tt arXiv:2112.11486}}.

\bibitem{Polnarev:1988dh}
A.~Polnarev and R.~Zembowicz, {\it {Formation of Primordial Black Holes by
  Cosmic Strings}},  {\em Phys. Rev. D} {\bf 43} (1991) 1106--1109.

\bibitem{Garriga:1993gj}
J.~Garriga and M.~Sakellariadou, {\it {Effects of friction on cosmic strings}},
   {\em Phys. Rev. D} {\bf 48} (1993) 2502--2515,
  [\href{http://arxiv.org/abs/hep-th/9303024}{{\tt hep-th/9303024}}].

\bibitem{Caldwell:1995fu}
R.~R. Caldwell and P.~Casper, {\it {Formation of black holes from collapsed
  cosmic string loops}},  {\em Phys. Rev. D} {\bf 53} (1996) 3002--3010,
  [\href{http://arxiv.org/abs/gr-qc/9509012}{{\tt gr-qc/9509012}}].

\bibitem{MacGibbon:1997pu}
J.~H. MacGibbon, R.~H. Brandenberger, and U.~F. Wichoski, {\it {Limits on black
  hole formation from cosmic string loops}},  {\em Phys. Rev. D} {\bf 57}
  (1998) 2158--2165, [\href{http://arxiv.org/abs/astro-ph/9707146}{{\tt
  astro-ph/9707146}}].

\bibitem{Helfer:2018qgv}
T.~Helfer, J.~C. Aurrekoetxea, and E.~A. Lim, {\it {Cosmic String Loop Collapse
  in Full General Relativity}},  {\em Phys. Rev. D} {\bf 99} (2019), no.~10
  104028, [\href{http://arxiv.org/abs/1808.06678}{{\tt arXiv:1808.06678}}].

\bibitem{Jenkins:2020ctp}
A.~C. Jenkins and M.~Sakellariadou, {\it {Primordial black holes from cusp
  collapse on cosmic strings}},  \href{http://arxiv.org/abs/2006.16249}{{\tt
  arXiv:2006.16249}}.

\bibitem{Chakraborty:2022mwu}
A.~Chakraborty, P.~K. Chanda, K.~L. Pandey, and S.~Das, {\it {Formation and
  Abundance of Late-forming Primordial Black Holes as Dark Matter}},  {\em
  Astrophys. J.} {\bf 932} (2022), no.~2 119,
  [\href{http://arxiv.org/abs/2204.09628}{{\tt arXiv:2204.09628}}].

\bibitem{Musco:2004ak}
I.~Musco, J.~C. Miller, and L.~Rezzolla, {\it {Computations of primordial black
  hole formation}},  {\em Class. Quant. Grav.} {\bf 22} (2005) 1405--1424,
  [\href{http://arxiv.org/abs/gr-qc/0412063}{{\tt gr-qc/0412063}}].

\bibitem{Polnarev:2006aa}
A.~G. Polnarev and I.~Musco, {\it {Curvature profiles as initial conditions for
  primordial black hole formation}},  {\em Class. Quant. Grav.} {\bf 24} (2007)
  1405--1432, [\href{http://arxiv.org/abs/gr-qc/0605122}{{\tt gr-qc/0605122}}].

\bibitem{Musco:2008hv}
I.~Musco, J.~C. Miller, and A.~G. Polnarev, {\it {Primordial black hole
  formation in the radiative era: Investigation of the critical nature of the
  collapse}},  {\em Class. Quant. Grav.} {\bf 26} (2009) 235001,
  [\href{http://arxiv.org/abs/0811.1452}{{\tt arXiv:0811.1452}}].

\bibitem{Musco:2012au}
I.~Musco and J.~C. Miller, {\it {Primordial black hole formation in the early
  universe: critical behaviour and self-similarity}},  {\em Class. Quant.
  Grav.} {\bf 30} (2013) 145009, [\href{http://arxiv.org/abs/1201.2379}{{\tt
  arXiv:1201.2379}}].

\bibitem{Musco:2018rwt}
I.~Musco, {\it {Threshold for primordial black holes: Dependence on the shape
  of the cosmological perturbations}},  {\em Phys. Rev. D} {\bf 100} (2019),
  no.~12 123524, [\href{http://arxiv.org/abs/1809.02127}{{\tt
  arXiv:1809.02127}}].

\bibitem{Kehagias:2019eil}
A.~Kehagias, I.~Musco, and A.~Riotto, {\it {Non-Gaussian Formation of
  Primordial Black Holes: Effects on the Threshold}},  {\em JCAP} {\bf 12}
  (2019) 029, [\href{http://arxiv.org/abs/1906.07135}{{\tt arXiv:1906.07135}}].

\bibitem{Musco:2020jjb}
I.~Musco, V.~De~Luca, G.~Franciolini, and A.~Riotto, {\it {The Threshold for
  Primordial Black Hole Formation: a Simple Analytic Prescription}},
  \href{http://arxiv.org/abs/2011.03014}{{\tt arXiv:2011.03014}}.

\bibitem{Musco:2021sva}
I.~Musco and T.~Papanikolaou, {\it {Primordial black hole formation for an
  anisotropic perfect fluid: initial conditions and estimation of the
  threshold}},  \href{http://arxiv.org/abs/2110.05982}{{\tt arXiv:2110.05982}}.

\bibitem{Carr:2020gox}
B.~Carr, K.~Kohri, Y.~Sendouda, and J.~Yokoyama, {\it {Constraints on
  Primordial Black Holes}},  \href{http://arxiv.org/abs/2002.12778}{{\tt
  arXiv:2002.12778}}.

\bibitem{Katz:2018zrn}
A.~Katz, J.~Kopp, S.~Sibiryakov, and W.~Xue, {\it {Femtolensing by Dark Matter
  Revisited}},  {\em JCAP} {\bf 12} (2018) 005,
  [\href{http://arxiv.org/abs/1807.11495}{{\tt arXiv:1807.11495}}].

\bibitem{Montero-Camacho:2019jte}
P.~Montero-Camacho, X.~Fang, G.~Vasquez, M.~Silva, and C.~M. Hirata, {\it
  {Revisiting constraints on asteroid-mass primordial black holes as dark
  matter candidates}},  {\em JCAP} {\bf 08} (2019) 031,
  [\href{http://arxiv.org/abs/1906.05950}{{\tt arXiv:1906.05950}}].

\bibitem{Arbey:2019vqx}
A.~Arbey, J.~Auffinger, and J.~Silk, {\it {Constraining primordial black hole
  masses with the isotropic gamma ray background}},  {\em Phys. Rev. D} {\bf
  101} (2020), no.~2 023010, [\href{http://arxiv.org/abs/1906.04750}{{\tt
  arXiv:1906.04750}}].

\bibitem{Boudaud:2018hqb}
M.~Boudaud and M.~Cirelli, {\it {Voyager 1 $e^\pm$ Further Constrain Primordial
  Black Holes as Dark Matter}},  {\em Phys. Rev. Lett.} {\bf 122} (2019), no.~4
  041104, [\href{http://arxiv.org/abs/1807.03075}{{\tt arXiv:1807.03075}}].

\bibitem{Iguaz:2021irx}
J.~Iguaz, P.~D. Serpico, and T.~Siegert, {\it {Isotropic X-ray bound on
  Primordial Black Hole Dark Matter}},  {\em Phys. Rev. D} {\bf 103} (2021),
  no.~10 103025, [\href{http://arxiv.org/abs/2104.03145}{{\tt
  arXiv:2104.03145}}].

\bibitem{Berteaud:2022tws}
J.~Berteaud, F.~Calore, J.~Iguaz, P.~D. Serpico, and T.~Siegert, {\it {Strong
  constraints on primordial black hole dark matter from 16 years of
  INTEGRAL/SPI observations}},  \href{http://arxiv.org/abs/2202.07483}{{\tt
  arXiv:2202.07483}}.

\bibitem{Carr:2009jm}
B.~J. Carr, K.~Kohri, Y.~Sendouda, and J.~Yokoyama, {\it {New cosmological
  constraints on primordial black holes}},  {\em Phys. Rev. D} {\bf 81} (2010)
  104019, [\href{http://arxiv.org/abs/0912.5297}{{\tt arXiv:0912.5297}}].

\bibitem{Ballesteros:2019exr}
G.~Ballesteros, J.~Coronado-Bl\'azquez, and D.~Gaggero, {\it {X-ray and
  gamma-ray limits on the primordial black hole abundance from Hawking
  radiation}},  {\em Phys. Lett. B} {\bf 808} (2020) 135624,
  [\href{http://arxiv.org/abs/1906.10113}{{\tt arXiv:1906.10113}}].

\bibitem{Laha:2019ssq}
R.~Laha, {\it {Primordial Black Holes as a Dark Matter Candidate Are Severely
  Constrained by the Galactic Center 511 keV $\gamma$ -Ray Line}},  {\em Phys.
  Rev. Lett.} {\bf 123} (2019), no.~25 251101,
  [\href{http://arxiv.org/abs/1906.09994}{{\tt arXiv:1906.09994}}].

\bibitem{Poulter:2019ooo}
H.~Poulter, Y.~Ali-Ha\"\i{}moud, J.~Hamann, M.~White, and A.~G. Williams, {\it
  {CMB constraints on ultra-light primordial black holes with extended mass
  distributions}},  \href{http://arxiv.org/abs/1907.06485}{{\tt
  arXiv:1907.06485}}.

\bibitem{Dasgupta:2019cae}
B.~Dasgupta, R.~Laha, and A.~Ray, {\it {Neutrino and positron constraints on
  spinning primordial black hole dark matter}},  {\em Phys. Rev. Lett.} {\bf
  125} (2020), no.~10 101101, [\href{http://arxiv.org/abs/1912.01014}{{\tt
  arXiv:1912.01014}}].

\bibitem{Laha:2020vhg}
R.~Laha, P.~Lu, and V.~Takhistov, {\it {Gas Heating from Spinning and
  Non-Spinning Evaporating Primordial Black Holes}},
  \href{http://arxiv.org/abs/2009.11837}{{\tt arXiv:2009.11837}}.

\bibitem{DeRocco:2019fjq}
W.~DeRocco and P.~W. Graham, {\it {Constraining Primordial Black Hole Abundance
  with the Galactic 511 keV Line}},  {\em Phys. Rev. Lett.} {\bf 123} (2019),
  no.~25 251102, [\href{http://arxiv.org/abs/1906.07740}{{\tt
  arXiv:1906.07740}}].

\bibitem{Laha:2020ivk}
R.~Laha, J.~B. Mu\~noz, and T.~R. Slatyer, {\it {INTEGRAL constraints on
  primordial black holes and particle dark matter}},  {\em Phys. Rev. D} {\bf
  101} (2020), no.~12 123514, [\href{http://arxiv.org/abs/2004.00627}{{\tt
  arXiv:2004.00627}}].

\bibitem{Kim:2020ngi}
H.~Kim, {\it {A constraint on light primordial black holes from the
  interstellar medium temperature}},  {\em Mon. Not. Roy. Astron. Soc.} {\bf
  504} (2021), no.~4 5475--5484, [\href{http://arxiv.org/abs/2007.07739}{{\tt
  arXiv:2007.07739}}].

\bibitem{Niikura:2017zjd}
H.~Niikura et~al., {\it {Microlensing constraints on primordial black holes
  with Subaru/HSC Andromeda observations}},  {\em Nature Astron.} {\bf 3}
  (2019), no.~6 524--534, [\href{http://arxiv.org/abs/1701.02151}{{\tt
  arXiv:1701.02151}}].

\bibitem{Smyth:2019whb}
N.~Smyth, S.~Profumo, S.~English, T.~Jeltema, K.~McKinnon, and P.~Guhathakurta,
  {\it {Updated Constraints on Asteroid-Mass Primordial Black Holes as Dark
  Matter}},  {\em Phys. Rev. D} {\bf 101} (2020), no.~6 063005,
  [\href{http://arxiv.org/abs/1910.01285}{{\tt arXiv:1910.01285}}].

\bibitem{Alcock:2000kd}
{\bf MACHO} Collaboration, C.~Alcock et~al., {\it {The MACHO project:
  microlensing detection efficiency}},  {\em Astrophys. J. Suppl.} {\bf 136}
  (2001) 439--462, [\href{http://arxiv.org/abs/astro-ph/0003392}{{\tt
  astro-ph/0003392}}].

\bibitem{Allsman:2000kg}
{\bf Macho} Collaboration, R.~A. Allsman et~al., {\it {MACHO project limits on
  black hole dark matter in the 1-30 solar mass range}},  {\em Astrophys. J.
  Lett.} {\bf 550} (2001) L169,
  [\href{http://arxiv.org/abs/astro-ph/0011506}{{\tt astro-ph/0011506}}].

\bibitem{Niikura:2019kqi}
H.~Niikura, M.~Takada, S.~Yokoyama, T.~Sumi, and S.~Masaki, {\it {Constraints
  on Earth-mass primordial black holes from OGLE 5-year microlensing events}},
  {\em Phys. Rev. D} {\bf 99} (2019), no.~8 083503,
  [\href{http://arxiv.org/abs/1901.07120}{{\tt arXiv:1901.07120}}].

\bibitem{Oguri:2017ock}
M.~Oguri, J.~M. Diego, N.~Kaiser, P.~L. Kelly, and T.~Broadhurst, {\it
  {Understanding caustic crossings in giant arcs: characteristic scales, event
  rates, and constraints on compact dark matter}},  {\em Phys. Rev.} {\bf D97}
  (2018), no.~2 023518, [\href{http://arxiv.org/abs/1710.00148}{{\tt
  arXiv:1710.00148}}].

\bibitem{Auclair:2022lcg}
P.~Auclair et~al., {\it {Cosmology with the Laser Interferometer Space
  Antenna}},  \href{http://arxiv.org/abs/2204.05434}{{\tt arXiv:2204.05434}}.

\bibitem{Petac:2022rio}
M.~Peta\v{c}, J.~Lavalle, and K.~Jedamzik, {\it {Microlensing constraints on
  clustered primordial black holes}},
  \href{http://arxiv.org/abs/2201.02521}{{\tt arXiv:2201.02521}}.

\bibitem{Gorton:2022fyb}
M.~Gorton and A.~M. Green, {\it {Effect of clustering on primordial black hole
  microlensing constraints}},  \href{http://arxiv.org/abs/2203.04209}{{\tt
  arXiv:2203.04209}}.

\bibitem{Manshanden:2018tze}
J.~Manshanden, D.~Gaggero, G.~Bertone, R.~M. Connors, and M.~Ricotti, {\it
  {Multi-wavelength astronomical searches for primordial black holes}},  {\em
  JCAP} {\bf 06} (2019) 026, [\href{http://arxiv.org/abs/1812.07967}{{\tt
  arXiv:1812.07967}}].

\bibitem{Inoue:2017csr}
Y.~Inoue and A.~Kusenko, {\it {New X-ray bound on density of primordial black
  holes}},  {\em JCAP} {\bf 1710} (2017) 034,
  [\href{http://arxiv.org/abs/1705.00791}{{\tt arXiv:1705.00791}}].

\bibitem{Ali-Haimoud:2016mbv}
Y.~Ali-Ha\"\i{}moud and M.~Kamionkowski, {\it {Cosmic microwave background
  limits on accreting primordial black holes}},  {\em Phys. Rev. D} {\bf 95}
  (2017), no.~4 043534, [\href{http://arxiv.org/abs/1612.05644}{{\tt
  arXiv:1612.05644}}].

\bibitem{Serpico:2020ehh}
P.~D. Serpico, V.~Poulin, D.~Inman, and K.~Kohri, {\it {Cosmic microwave
  background bounds on primordial black holes including dark matter halo
  accretion}},  {\em Phys. Rev. Res.} {\bf 2} (2020), no.~2 023204,
  [\href{http://arxiv.org/abs/2002.10771}{{\tt arXiv:2002.10771}}].

\bibitem{Lu:2020bmd}
P.~Lu, V.~Takhistov, G.~B. Gelmini, K.~Hayashi, Y.~Inoue, and A.~Kusenko, {\it
  {Constraining Primordial Black Holes with Dwarf Galaxy Heating}},  {\em
  Astrophys. J. Lett.} {\bf 908} (2021), no.~2 L23,
  [\href{http://arxiv.org/abs/2007.02213}{{\tt arXiv:2007.02213}}].

\bibitem{Takhistov:2021aqx}
V.~Takhistov, P.~Lu, G.~B. Gelmini, K.~Hayashi, Y.~Inoue, and A.~Kusenko, {\it
  {Interstellar Gas Heating by Primordial Black Holes}},
  \href{http://arxiv.org/abs/2105.06099}{{\tt arXiv:2105.06099}}.

\bibitem{Carr:2018rid}
B.~Carr and J.~Silk, {\it {Primordial Black Holes as Generators of Cosmic
  Structures}},  {\em Mon. Not. Roy. Astron. Soc.} {\bf 478} (2018), no.~3
  3756--3775, [\href{http://arxiv.org/abs/1801.00672}{{\tt arXiv:1801.00672}}].

\bibitem{Inomata:2016uip}
K.~Inomata, M.~Kawasaki, and Y.~Tada, {\it {Revisiting constraints on small
  scale perturbations from big-bang nucleosynthesis}},  {\em Phys. Rev. D} {\bf
  94} (2016), no.~4 043527, [\href{http://arxiv.org/abs/1605.04646}{{\tt
  arXiv:1605.04646}}].

\bibitem{Nakama:2017xvq}
T.~Nakama, B.~Carr, and J.~Silk, {\it {Limits on primordial black holes from
  $\mu$ distortions in cosmic microwave background}},  {\em Phys. Rev. D} {\bf
  97} (2018), no.~4 043525, [\href{http://arxiv.org/abs/1710.06945}{{\tt
  arXiv:1710.06945}}].

\bibitem{Ali-Haimoud:2017rtz}
Y.~Ali-Ha\"\i{}moud, E.~D. Kovetz, and M.~Kamionkowski, {\it {Merger rate of
  primordial black-hole binaries}},  {\em Phys. Rev.} {\bf D96} (2017), no.~12
  123523, [\href{http://arxiv.org/abs/1709.06576}{{\tt arXiv:1709.06576}}].

\bibitem{Raidal:2018bbj}
M.~Raidal, C.~Spethmann, V.~Vaskonen, and H.~Veerm\"ae, {\it {Formation and
  Evolution of Primordial Black Hole Binaries in the Early Universe}},  {\em
  JCAP} {\bf 02} (2019) 018, [\href{http://arxiv.org/abs/1812.01930}{{\tt
  arXiv:1812.01930}}].

\bibitem{Vaskonen:2019jpv}
V.~Vaskonen and H.~Veerm\"ae, {\it {Lower bound on the primordial black hole
  merger rate}},  {\em Phys. Rev. D} {\bf 101} (2020), no.~4 043015,
  [\href{http://arxiv.org/abs/1908.09752}{{\tt arXiv:1908.09752}}].

\bibitem{DeLuca:2020bjf}
V.~De~Luca, G.~Franciolini, P.~Pani, and A.~Riotto, {\it {The Evolution of
  Primordial Black Holes and their Final Observable Spins}},  {\em JCAP} {\bf
  04} (2020) 052, [\href{http://arxiv.org/abs/2003.02778}{{\tt
  arXiv:2003.02778}}].

\bibitem{Hall:2020daa}
A.~Hall, A.~D. Gow, and C.~T. Byrnes, {\it {Bayesian analysis of LIGO-Virgo
  mergers: Primordial vs. astrophysical black hole populations}},  {\em Phys.
  Rev. D} {\bf 102} (2020) 123524, [\href{http://arxiv.org/abs/2008.13704}{{\tt
  arXiv:2008.13704}}].

\bibitem{DeLuca:2020sae}
V.~De~Luca, V.~Desjacques, G.~Franciolini, P.~Pani, and A.~Riotto, {\it
  {GW190521 Mass Gap Event and the Primordial Black Hole Scenario}},  {\em
  Phys. Rev. Lett.} {\bf 126} (2021), no.~5 051101,
  [\href{http://arxiv.org/abs/2009.01728}{{\tt arXiv:2009.01728}}].

\bibitem{Wong:2020yig}
K.~W.~K. Wong, G.~Franciolini, V.~De~Luca, V.~Baibhav, E.~Berti, P.~Pani, and
  A.~Riotto, {\it {Constraining the primordial black hole scenario with
  Bayesian inference and machine learning: the GWTC-2 gravitational wave
  catalog}},  {\em Phys. Rev.} {\bf D103} (2021), no.~2 023026,
  [\href{http://arxiv.org/abs/2011.01865}{{\tt arXiv:2011.01865}}].

\bibitem{Hutsi:2020sol}
G.~H{\"u}tsi, M.~Raidal, V.~Vaskonen, and H.~Veerm\"ae, {\it {Two populations
  of LIGO-Virgo black holes}},  {\em JCAP} {\bf 2103} (2021) 068,
  [\href{http://arxiv.org/abs/2012.02786}{{\tt arXiv:2012.02786}}].

\bibitem{DeLuca:2021wjr}
V.~De~Luca, G.~Franciolini, P.~Pani, and A.~Riotto, {\it {Bayesian Evidence for
  Both Astrophysical and Primordial Black Holes: Mapping the GWTC-2 Catalog to
  Third-Generation Detectors}},  \href{http://arxiv.org/abs/2102.03809}{{\tt
  arXiv:2102.03809}}.

\bibitem{DeLuca:2020fpg}
V.~De~Luca, G.~Franciolini, P.~Pani, and A.~Riotto, {\it {Constraints on
  Primordial Black Holes: the Importance of Accretion}},  {\em Phys. Rev. D}
  {\bf 102} (2020), no.~4 043505, [\href{http://arxiv.org/abs/2003.12589}{{\tt
  arXiv:2003.12589}}].

\bibitem{DeLuca:2020qqa}
V.~De~Luca, G.~Franciolini, P.~Pani, and A.~Riotto, {\it {Primordial Black
  Holes Confront LIGO/Virgo data: Current situation}},  {\em JCAP} {\bf 06}
  (2020) 044, [\href{http://arxiv.org/abs/2005.05641}{{\tt arXiv:2005.05641}}].

\bibitem{TheLIGOScientific:2014jea}
{\bf LIGO Scientific} Collaboration, J.~Aasi et~al., {\it {Advanced LIGO}},
  {\em Class. Quant. Grav.} {\bf 32} (2015) 074001,
  [\href{http://arxiv.org/abs/1411.4547}{{\tt arXiv:1411.4547}}].

\bibitem{TheVirgo:2014hva}
{\bf VIRGO} Collaboration, F.~Acernese et~al., {\it {Advanced Virgo: a
  second-generation interferometric gravitational wave detector}},  {\em Class.
  Quant. Grav.} {\bf 32} (2015), no.~2 024001,
  [\href{http://arxiv.org/abs/1408.3978}{{\tt arXiv:1408.3978}}].

\bibitem{Bird:2016dcv}
S.~Bird, I.~Cholis, J.~B. Mu{\~n}oz, Y.~Ali-Ha\"\i{}moud, M.~Kamionkowski,
  E.~D. Kovetz, A.~Raccanelli, and A.~G. Riess, {\it {Did LIGO detect dark
  matter?}},  {\em Phys. Rev. Lett.} {\bf 116} (2016), no.~20 201301,
  [\href{http://arxiv.org/abs/1603.00464}{{\tt arXiv:1603.00464}}].

\bibitem{Clesse:2016vqa}
S.~Clesse and J.~Garc\'\i{}a-Bellido, {\it {The clustering of massive
  Primordial Black Holes as Dark Matter: measuring their mass distribution with
  Advanced LIGO}},  {\em Phys. Dark Univ.} {\bf 15} (2017) 142--147,
  [\href{http://arxiv.org/abs/1603.05234}{{\tt arXiv:1603.05234}}].

\bibitem{Sasaki:2016jop}
M.~Sasaki, T.~Suyama, T.~Tanaka, and S.~Yokoyama, {\it {Primordial Black Hole
  Scenario for the Gravitational-Wave Event GW150914}},  {\em Phys. Rev. Lett.}
  {\bf 117} (2016), no.~6 061101, [\href{http://arxiv.org/abs/1603.08338}{{\tt
  arXiv:1603.08338}}]. [erratum: Phys. Rev. Lett.121,no.5,059901(2018)].

\bibitem{Franciolini:2021tla}
G.~Franciolini, V.~Baibhav, V.~De~Luca, K.~K.~Y. Ng, K.~W.~K. Wong, E.~Berti,
  P.~Pani, A.~Riotto, and S.~Vitale, {\it {Quantifying the evidence for
  primordial black holes in LIGO/Virgo gravitational-wave data}},
  \href{http://arxiv.org/abs/2105.03349}{{\tt arXiv:2105.03349}}.

\bibitem{Franciolini:2021xbq}
G.~Franciolini, R.~Cotesta, N.~Loutrel, E.~Berti, P.~Pani, and A.~Riotto, {\it
  {How to assess the primordial origin of single gravitational-wave events with
  mass, spin, eccentricity, and deformability measurements}},  {\em Phys. Rev.
  D} {\bf 105} (2022), no.~6 063510,
  [\href{http://arxiv.org/abs/2112.10660}{{\tt arXiv:2112.10660}}].

\bibitem{Franciolini:2022iaa}
G.~Franciolini and P.~Pani, {\it {Searching for mass-spin correlations in the
  population of gravitational-wave events: the GWTC-3 case study}},
  \href{http://arxiv.org/abs/2201.13098}{{\tt arXiv:2201.13098}}.

\bibitem{Maggiore:1999vm}
M.~Maggiore, {\it {Gravitational wave experiments and early universe
  cosmology}},  {\em Phys. Rept.} {\bf 331} (2000) 283--367,
  [\href{http://arxiv.org/abs/gr-qc/9909001}{{\tt gr-qc/9909001}}].

\bibitem{Aggarwal:2020olq}
N.~Aggarwal et~al., {\it {Challenges and opportunities of gravitational-wave
  searches at MHz to GHz frequencies}},  {\em Living Rev. Rel.} {\bf 24}
  (2021), no.~1 4, [\href{http://arxiv.org/abs/2011.12414}{{\tt
  arXiv:2011.12414}}].

\bibitem{Green:2020jor}
A.~M. Green and B.~J. Kavanagh, {\it {Primordial Black Holes as a dark matter
  candidate}},  {\em J. Phys. G} {\bf 48} (2021), no.~4 4,
  [\href{http://arxiv.org/abs/2007.10722}{{\tt arXiv:2007.10722}}].

\bibitem{Carr:2020xqk}
B.~Carr and F.~Kuhnel, {\it {Primordial Black Holes as Dark Matter: Recent
  Developments}},  {\em Ann. Rev. Nucl. Part. Sci.} {\bf 70} (2020) 355--394,
  [\href{http://arxiv.org/abs/2006.02838}{{\tt arXiv:2006.02838}}].

\bibitem{Franciolini:2021nvv}
G.~Franciolini, {\it {Primordial Black Holes: from Theory to Gravitational Wave
  Observations}},  other thesis, 10, 2021.

\bibitem{Hawking:1974rv}
S.~W. Hawking, {\it {Black hole explosions}},  {\em Nature} {\bf 248} (1974)
  30--31.

\bibitem{Ivanov:1994pa}
P.~Ivanov, P.~Naselsky, and I.~Novikov, {\it {Inflation and primordial black
  holes as dark matter}},  {\em Phys. Rev. D} {\bf 50} (1994) 7173--7178.

\bibitem{GarciaBellido:1996qt}
J.~Garcia-Bellido, A.~D. Linde, and D.~Wands, {\it {Density perturbations and
  black hole formation in hybrid inflation}},  {\em Phys. Rev. D} {\bf 54}
  (1996) 6040--6058, [\href{http://arxiv.org/abs/astro-ph/9605094}{{\tt
  astro-ph/9605094}}].

\bibitem{Ivanov:1997ia}
P.~Ivanov, {\it {Nonlinear metric perturbations and production of primordial
  black holes}},  {\em Phys. Rev. D} {\bf 57} (1998) 7145--7154,
  [\href{http://arxiv.org/abs/astro-ph/9708224}{{\tt astro-ph/9708224}}].

\bibitem{Blinnikov:2016bxu}
S.~Blinnikov, A.~Dolgov, N.~K. Porayko, and K.~Postnov, {\it {Solving puzzles
  of GW150914 by primordial black holes}},  {\em JCAP} {\bf 1611} (2016) 036,
  [\href{http://arxiv.org/abs/1611.00541}{{\tt arXiv:1611.00541}}].

\bibitem{Escriva:2021aeh}
A.~Escriv\`a, {\it {PBH Formation from Spherically Symmetric Hydrodynamical
  Perturbations: A Review}},  {\em Universe} {\bf 8} (2022), no.~2 66,
  [\href{http://arxiv.org/abs/2111.12693}{{\tt arXiv:2111.12693}}].

\bibitem{Choptuik:1992jv}
M.~W. Choptuik, {\it {Universality and scaling in gravitational collapse of a
  massless scalar field}},  {\em Phys. Rev. Lett.} {\bf 70} (1993) 9--12.

\bibitem{Evans:1994pj}
C.~R. Evans and J.~S. Coleman, {\it {Observation of critical phenomena and
  selfsimilarity in the gravitational collapse of radiation fluid}},  {\em
  Phys. Rev. Lett.} {\bf 72} (1994) 1782--1785,
  [\href{http://arxiv.org/abs/gr-qc/9402041}{{\tt gr-qc/9402041}}].

\bibitem{Niemeyer:1997mt}
J.~C. Niemeyer and K.~Jedamzik, {\it {Near-critical gravitational collapse and
  the initial mass function of primordial black holes}},  {\em Phys. Rev.
  Lett.} {\bf 80} (1998) 5481--5484,
  [\href{http://arxiv.org/abs/astro-ph/9709072}{{\tt astro-ph/9709072}}].

\bibitem{Kalaja:2019uju}
A.~Kalaja, N.~Bellomo, N.~Bartolo, D.~Bertacca, S.~Matarrese, I.~Musco,
  A.~Raccanelli, and L.~Verde, {\it {From Primordial Black Holes Abundance to
  Primordial Curvature Power Spectrum (and back)}},  {\em JCAP} {\bf 10} (2019)
  031, [\href{http://arxiv.org/abs/1908.03596}{{\tt arXiv:1908.03596}}].

\bibitem{Escriva:2019nsa}
A.~Escriv\`a, {\it {Simulation of primordial black hole formation using
  pseudo-spectral methods}},  {\em Phys. Dark Univ.} {\bf 27} (2020) 100466,
  [\href{http://arxiv.org/abs/1907.13065}{{\tt arXiv:1907.13065}}].

\bibitem{Germani:2018jgr}
C.~Germani and I.~Musco, {\it {Abundance of Primordial Black Holes Depends on
  the Shape of the Inflationary Power Spectrum}},  {\em Phys. Rev. Lett.} {\bf
  122} (2019), no.~14 141302, [\href{http://arxiv.org/abs/1805.04087}{{\tt
  arXiv:1805.04087}}].

\bibitem{Biagetti:2021eep}
M.~Biagetti, V.~De~Luca, G.~Franciolini, A.~Kehagias, and A.~Riotto, {\it {The
  formation probability of primordial black holes}},  {\em Phys. Lett. B} {\bf
  820} (2021) 136602, [\href{http://arxiv.org/abs/2105.07810}{{\tt
  arXiv:2105.07810}}].

\bibitem{Planck:2018jri}
{\bf Planck} Collaboration, Y.~Akrami et~al., {\it {Planck 2018 results. X.
  Constraints on inflation}},  {\em Astron. Astrophys.} {\bf 641} (2020) A10,
  [\href{http://arxiv.org/abs/1807.06211}{{\tt arXiv:1807.06211}}].

\bibitem{Acquaviva:2002ud}
V.~Acquaviva, N.~Bartolo, S.~Matarrese, and A.~Riotto, {\it {Second order
  cosmological perturbations from inflation}},  {\em Nucl. Phys. B} {\bf 667}
  (2003) 119--148, [\href{http://arxiv.org/abs/astro-ph/0209156}{{\tt
  astro-ph/0209156}}].

\bibitem{Mollerach:2003nq}
S.~Mollerach, D.~Harari, and S.~Matarrese, {\it {CMB polarization from
  secondary vector and tensor modes}},  {\em Phys. Rev. D} {\bf 69} (2004)
  063002, [\href{http://arxiv.org/abs/astro-ph/0310711}{{\tt
  astro-ph/0310711}}].

\bibitem{Ananda:2006af}
K.~N. Ananda, C.~Clarkson, and D.~Wands, {\it {The Cosmological gravitational
  wave background from primordial density perturbations}},  {\em Phys. Rev. D}
  {\bf 75} (2007) 123518, [\href{http://arxiv.org/abs/gr-qc/0612013}{{\tt
  gr-qc/0612013}}].

\bibitem{Baumann:2007zm}
D.~Baumann, P.~J. Steinhardt, K.~Takahashi, and K.~Ichiki, {\it {Gravitational
  Wave Spectrum Induced by Primordial Scalar Perturbations}},  {\em Phys. Rev.
  D} {\bf 76} (2007) 084019, [\href{http://arxiv.org/abs/hep-th/0703290}{{\tt
  hep-th/0703290}}].

\bibitem{Cai:2018dig}
R.-g. Cai, S.~Pi, and M.~Sasaki, {\it {Gravitational Waves Induced by
  non-Gaussian Scalar Perturbations}},  {\em Phys. Rev. Lett.} {\bf 122}
  (2019), no.~20 201101, [\href{http://arxiv.org/abs/1810.11000}{{\tt
  arXiv:1810.11000}}].

\bibitem{Bartolo:2018rku}
N.~Bartolo, V.~De~Luca, G.~Franciolini, M.~Peloso, D.~Racco, and A.~Riotto,
  {\it {Testing primordial black holes as dark matter with LISA}},  {\em Phys.
  Rev. D} {\bf 99} (2019), no.~10 103521,
  [\href{http://arxiv.org/abs/1810.12224}{{\tt arXiv:1810.12224}}].

\bibitem{Bartolo:2018evs}
N.~Bartolo, V.~De~Luca, G.~Franciolini, A.~Lewis, M.~Peloso, and A.~Riotto,
  {\it {Primordial Black Hole Dark Matter: LISA Serendipity}},  {\em Phys. Rev.
  Lett.} {\bf 122} (2019), no.~21 211301,
  [\href{http://arxiv.org/abs/1810.12218}{{\tt arXiv:1810.12218}}].

\bibitem{Unal:2018yaa}
C.~Unal, {\it {Imprints of Primordial Non-Gaussianity on Gravitational Wave
  Spectrum}},  {\em Phys. Rev. D} {\bf 99} (2019), no.~4 041301,
  [\href{http://arxiv.org/abs/1811.09151}{{\tt arXiv:1811.09151}}].

\bibitem{Bartolo:2019zvb}
N.~Bartolo, D.~Bertacca, V.~De~Luca, G.~Franciolini, S.~Matarrese, M.~Peloso,
  A.~Ricciardone, A.~Riotto, and G.~Tasinato, {\it {Gravitational wave
  anisotropies from primordial black holes}},  {\em JCAP} {\bf 02} (2020) 028,
  [\href{http://arxiv.org/abs/1909.12619}{{\tt arXiv:1909.12619}}].

\bibitem{Wang:2019kaf}
S.~Wang, T.~Terada, and K.~Kohri, {\it {Prospective constraints on the
  primordial black hole abundance from the stochastic gravitational-wave
  backgrounds produced by coalescing events and curvature perturbations}},
  {\em Phys. Rev. D} {\bf 99} (2019), no.~10 103531,
  [\href{http://arxiv.org/abs/1903.05924}{{\tt arXiv:1903.05924}}]. [Erratum:
  Phys.Rev.D 101, 069901 (2020)].

\bibitem{Cai:2019elf}
R.-G. Cai, S.~Pi, S.-J. Wang, and X.-Y. Yang, {\it {Pulsar Timing Array
  Constraints on the Induced Gravitational Waves}},  {\em JCAP} {\bf 10} (2019)
  059, [\href{http://arxiv.org/abs/1907.06372}{{\tt arXiv:1907.06372}}].

\bibitem{DeLuca:2019ufz}
V.~De~Luca, G.~Franciolini, A.~Kehagias, and A.~Riotto, {\it {On the Gauge
  Invariance of Cosmological Gravitational Waves}},  {\em JCAP} {\bf 03} (2020)
  014, [\href{http://arxiv.org/abs/1911.09689}{{\tt arXiv:1911.09689}}].

\bibitem{Inomata:2019yww}
K.~Inomata and T.~Terada, {\it {Gauge Independence of Induced Gravitational
  Waves}},  {\em Phys. Rev. D} {\bf 101} (2020), no.~2 023523,
  [\href{http://arxiv.org/abs/1912.00785}{{\tt arXiv:1912.00785}}].

\bibitem{Yuan:2019fwv}
C.~Yuan, Z.-C. Chen, and Q.-G. Huang, {\it {Scalar induced gravitational waves
  in different gauges}},  {\em Phys. Rev. D} {\bf 101} (2020), no.~6 063018,
  [\href{http://arxiv.org/abs/1912.00885}{{\tt arXiv:1912.00885}}].

\bibitem{Pi:2020otn}
S.~Pi and M.~Sasaki, {\it {Gravitational Waves Induced by Scalar Perturbations
  with a Lognormal Peak}},  {\em JCAP} {\bf 09} (2020) 037,
  [\href{http://arxiv.org/abs/2005.12306}{{\tt arXiv:2005.12306}}].

\bibitem{Yuan:2020iwf}
C.~Yuan and Q.-G. Huang, {\it {Gravitational waves induced by the local-type
  non-Gaussian curvature perturbations}},
  \href{http://arxiv.org/abs/2007.10686}{{\tt arXiv:2007.10686}}.

\bibitem{Romero-Rodriguez:2021aws}
A.~Romero-Rodriguez, M.~Martinez, O.~Pujol\`as, M.~Sakellariadou, and
  V.~Vaskonen, {\it {Search for a Scalar Induced Stochastic Gravitational Wave
  Background in the Third LIGO-Virgo Observing Run}},  {\em Phys. Rev. Lett.}
  {\bf 128} (2022), no.~5 051301, [\href{http://arxiv.org/abs/2107.11660}{{\tt
  arXiv:2107.11660}}].

\bibitem{Balaji:2022rsy}
S.~Balaji, J.~Silk, and Y.-P. Wu, {\it {Induced gravitational waves from the
  cosmic coincidence}},  \href{http://arxiv.org/abs/2202.00700}{{\tt
  arXiv:2202.00700}}.

\bibitem{Yuan:2021qgz}
C.~Yuan and Q.-G. Huang, {\it {A topic review on probing primordial black hole
  dark matter with scalar induced gravitational waves}},
  \href{http://arxiv.org/abs/2103.04739}{{\tt arXiv:2103.04739}}.

\bibitem{Domenech:2021ztg}
G.~Dom\`enech, {\it {Scalar Induced Gravitational Waves Review}},  {\em
  Universe} {\bf 7} (2021), no.~11 398,
  [\href{http://arxiv.org/abs/2109.01398}{{\tt arXiv:2109.01398}}].

\bibitem{Saito:2009jt}
R.~Saito and J.~Yokoyama, {\it {Gravitational-Wave Constraints on the Abundance
  of Primordial Black Holes}},  {\em Prog. Theor. Phys.} {\bf 123} (2010)
  867--886, [\href{http://arxiv.org/abs/0912.5317}{{\tt arXiv:0912.5317}}].
  [Erratum: Prog.Theor.Phys. 126, 351--352 (2011)].

\bibitem{Hawking:1976de}
S.~W. Hawking, {\it {Black Holes and Thermodynamics}},  {\em Phys. Rev. D} {\bf
  13} (1976) 191--197.

\bibitem{1976PhRvD..13..198P}
D.~N. {Page}, {\it {Particle emission rates from a black hole: Massless
  particles from an uncharged, nonrotating hole}},  {\em "Phys. Rev. D"} {\bf
  13} (Jan., 1976) 198--206.

\bibitem{Dolgov:2011cq}
A.~D. Dolgov and D.~Ejlli, {\it {Relic gravitational waves from light
  primordial black holes}},  {\em Phys. Rev. D} {\bf 84} (2011) 024028,
  [\href{http://arxiv.org/abs/1105.2303}{{\tt arXiv:1105.2303}}].

\bibitem{Caprini:2018mtu}
C.~Caprini and D.~G. Figueroa, {\it {Cosmological Backgrounds of Gravitational
  Waves}},  {\em Class. Quant. Grav.} {\bf 35} (2018), no.~16 163001,
  [\href{http://arxiv.org/abs/1801.04268}{{\tt arXiv:1801.04268}}].

\bibitem{Ali-Haimoud:2018dau}
Y.~Ali-Ha\"\i{}moud, {\it {Correlation Function of High-Threshold Regions and
  Application to the Initial Small-Scale Clustering of Primordial Black
  Holes}},  {\em Phys. Rev. Lett.} {\bf 121} (2018), no.~8 081304,
  [\href{http://arxiv.org/abs/1805.05912}{{\tt arXiv:1805.05912}}].

\bibitem{Desjacques:2018wuu}
V.~Desjacques and A.~Riotto, {\it {Spatial clustering of primordial black
  holes}},  {\em Phys. Rev. D} {\bf 98} (2018), no.~12 123533,
  [\href{http://arxiv.org/abs/1806.10414}{{\tt arXiv:1806.10414}}].

\bibitem{Ballesteros:2018swv}
G.~Ballesteros, P.~D. Serpico, and M.~Taoso, {\it {On the merger rate of
  primordial black holes: effects of nearest neighbours distribution and
  clustering}},  {\em JCAP} {\bf 10} (2018) 043,
  [\href{http://arxiv.org/abs/1807.02084}{{\tt arXiv:1807.02084}}].

\bibitem{MoradinezhadDizgah:2019wjf}
A.~Moradinezhad~Dizgah, G.~Franciolini, and A.~Riotto, {\it {Primordial Black
  Holes from Broad Spectra: Abundance and Clustering}},  {\em JCAP} {\bf 11}
  (2019) 001, [\href{http://arxiv.org/abs/1906.08978}{{\tt arXiv:1906.08978}}].

\bibitem{Inman:2019wvr}
D.~Inman and Y.~Ali-Ha\"\i{}moud, {\it {Early structure formation in primordial
  black hole cosmologies}},  {\em Phys. Rev. D} {\bf 100} (2019), no.~8 083528,
  [\href{http://arxiv.org/abs/1907.08129}{{\tt arXiv:1907.08129}}].

\bibitem{Nakamura:1997sm}
T.~Nakamura, M.~Sasaki, T.~Tanaka, and K.~S. Thorne, {\it {Gravitational waves
  from coalescing black hole MACHO binaries}},  {\em Astrophys. J. Lett.} {\bf
  487} (1997) L139--L142, [\href{http://arxiv.org/abs/astro-ph/9708060}{{\tt
  astro-ph/9708060}}].

\bibitem{Ioka:1998nz}
K.~Ioka, T.~Chiba, T.~Tanaka, and T.~Nakamura, {\it {Black hole binary
  formation in the expanding universe: Three body problem approximation}},
  {\em Phys. Rev. D} {\bf 58} (1998) 063003,
  [\href{http://arxiv.org/abs/astro-ph/9807018}{{\tt astro-ph/9807018}}].

\bibitem{Kavanagh:2018ggo}
B.~J. Kavanagh, D.~Gaggero, and G.~Bertone, {\it {Merger rate of a subdominant
  population of primordial black holes}},  {\em Phys. Rev. D} {\bf 98} (2018),
  no.~2 023536, [\href{http://arxiv.org/abs/1805.09034}{{\tt
  arXiv:1805.09034}}].

\bibitem{Liu:2018ess}
L.~Liu, Z.-K. Guo, and R.-G. Cai, {\it {Effects of the surrounding primordial
  black holes on the merger rate of primordial black hole binaries}},  {\em
  Phys. Rev. D} {\bf 99} (2019), no.~6 063523,
  [\href{http://arxiv.org/abs/1812.05376}{{\tt arXiv:1812.05376}}].

\bibitem{Jedamzik:2020ypm}
K.~Jedamzik, {\it {Primordial Black Hole Dark Matter and the LIGO/Virgo
  observations}},  {\em JCAP} {\bf 09} (2020) 022,
  [\href{http://arxiv.org/abs/2006.11172}{{\tt arXiv:2006.11172}}].

\bibitem{Young:2020scc}
S.~Young and A.~S. Hamers, {\it {The impact on distant fly-bys on the rate of
  binary primordial black hole mergers}},  {\em JCAP} {\bf 10} (2020) 036,
  [\href{http://arxiv.org/abs/2006.15023}{{\tt arXiv:2006.15023}}].

\bibitem{Jedamzik:2020omx}
K.~Jedamzik, {\it {Consistency of Primordial Black Hole Dark Matter with
  LIGO/Virgo Merger Rates}},  {\em Phys. Rev. Lett.} {\bf 126} (2021), no.~5
  051302, [\href{http://arxiv.org/abs/2007.03565}{{\tt arXiv:2007.03565}}].

\bibitem{DeLuca:2020jug}
V.~De~Luca, V.~Desjacques, G.~Franciolini, and A.~Riotto, {\it {The clustering
  evolution of primordial black holes}},  {\em JCAP} {\bf 11} (2020) 028,
  [\href{http://arxiv.org/abs/2009.04731}{{\tt arXiv:2009.04731}}].

\bibitem{Trashorras:2020mwn}
M.~Trashorras, J.~Garc\'\i{}a-Bellido, and S.~Nesseris, {\it {The clustering
  dynamics of primordial black boles in $N$-body simulations}},  {\em Universe}
  {\bf 7} (2021), no.~1 18, [\href{http://arxiv.org/abs/2006.15018}{{\tt
  arXiv:2006.15018}}].

\bibitem{Tkachev:2020uin}
M.~Tkachev, S.~Pilipenko, and G.~Yepes, {\it {Dark Matter Simulations with
  Primordial Black Holes in the Early Universe}},  {\em Mon. Not. Roy. Astron.
  Soc.} {\bf 499} (2020), no.~4 4854--4862,
  [\href{http://arxiv.org/abs/2009.07813}{{\tt arXiv:2009.07813}}].

\bibitem{link}
K. Jedamzik,
  \url{https://agenda.infn.it/event/23799/contributions/125718/attachments/78986/102370/rome110221.pdf},
  2021.

\bibitem{1989ApJ...343..725Q}
G.~D. {Quinlan} and S.~L. {Shapiro}, {\it {Dynamical Evolution of Dense
  Clusters of Compact Stars}},  {\em \apj} {\bf 343} (Aug., 1989) 725.

\bibitem{Mouri:2002mc}
H.~Mouri and Y.~Taniguchi, {\it {Runaway merging of black holes: analytical
  constraint on the timescale}},  {\em Astrophys. J. Lett.} {\bf 566} (2002)
  L17--L20, [\href{http://arxiv.org/abs/astro-ph/0201102}{{\tt
  astro-ph/0201102}}].

\bibitem{Raidal:2017mfl}
M.~Raidal, V.~Vaskonen, and H.~Veerm\"ae, {\it {Gravitational Waves from
  Primordial Black Hole Mergers}},  {\em JCAP} {\bf 09} (2017) 037,
  [\href{http://arxiv.org/abs/1707.01480}{{\tt arXiv:1707.01480}}].

\bibitem{Korol:2019jud}
V.~Korol, I.~Mandel, M.~C. Miller, R.~P. Church, and M.~B. Davies, {\it {Merger
  rates in primordial black hole clusters without initial binaries}},  {\em
  Mon. Not. Roy. Astron. Soc.} {\bf 496} (2020), no.~1 994--1000,
  [\href{http://arxiv.org/abs/1911.03483}{{\tt arXiv:1911.03483}}].

\bibitem{1974ApJ...187..425P}
W.~H. {Press} and P.~{Schechter}, {\it {Formation of Galaxies and Clusters of
  Galaxies by Self-Similar Gravitational Condensation}},  {\em \apj} {\bf 187}
  (Feb., 1974) 425--438.

\bibitem{Peters:1963ux}
P.~C. Peters and J.~Mathews, {\it {Gravitational radiation from point masses in
  a Keplerian orbit}},  {\em Phys. Rev.} {\bf 131} (1963) 435--439.

\bibitem{Peters:1964zz}
P.~C. Peters, {\it {Gravitational Radiation and the Motion of Two Point
  Masses}},  {\em Phys. Rev.} {\bf 136} (1964) B1224--B1232.

\bibitem{binn}
J.~Binney and S.~Tremaine, {\em {Galactic dynamics}}.
\newblock Princeton University Press, 1987.

\bibitem{landau1976mechanics}
L.~Landau, E.~Lifshitz, J.~Sykes, and J.~Bell, {\em Mechanics: Volume 1}.
\newblock Course of theoretical physics. Elsevier Science, 1976.

\bibitem{Ricotti:2007au}
M.~Ricotti, J.~P. Ostriker, and K.~J. Mack, {\it {Effect of Primordial Black
  Holes on the Cosmic Microwave Background and Cosmological Parameter
  Estimates}},  {\em Astrophys. J.} {\bf 680} (2008) 829,
  [\href{http://arxiv.org/abs/0709.0524}{{\tt arXiv:0709.0524}}].

\bibitem{Ricotti:2007jk}
M.~Ricotti, {\it {Bondi accretion in the early universe}},  {\em Astrophys. J.}
  {\bf 662} (2007) 53--61, [\href{http://arxiv.org/abs/0706.0864}{{\tt
  arXiv:0706.0864}}].

\bibitem{Adamek:2019gns}
J.~Adamek, C.~T. Byrnes, M.~Gosenca, and S.~Hotchkiss, {\it {WIMPs and
  stellar-mass primordial black holes are incompatible}},  {\em Phys. Rev.}
  {\bf D100} (2019), no.~2 023506, [\href{http://arxiv.org/abs/1901.08528}{{\tt
  arXiv:1901.08528}}].

\bibitem{Tada:2015noa}
Y.~Tada and S.~Yokoyama, {\it {Primordial black holes as biased tracers}},
  {\em Phys. Rev. D} {\bf 91} (2015), no.~12 123534,
  [\href{http://arxiv.org/abs/1502.01124}{{\tt arXiv:1502.01124}}].

\bibitem{Young:2015kda}
S.~Young and C.~T. Byrnes, {\it {Signatures of non-gaussianity in the
  isocurvature modes of primordial black hole dark matter}},  {\em JCAP} {\bf
  04} (2015) 034, [\href{http://arxiv.org/abs/1503.01505}{{\tt
  arXiv:1503.01505}}].

\bibitem{Young:2019gfc}
S.~Young and C.~T. Byrnes, {\it {Initial clustering and the primordial black
  hole merger rate}},  {\em JCAP} {\bf 03} (2020) 004,
  [\href{http://arxiv.org/abs/1910.06077}{{\tt arXiv:1910.06077}}].

\bibitem{Atal:2020igj}
V.~Atal, A.~Sanglas, and N.~Triantafyllou, {\it {LIGO/Virgo black holes and
  dark matter: The effect of spatial clustering}},  {\em JCAP} {\bf 11} (2020)
  036, [\href{http://arxiv.org/abs/2007.07212}{{\tt arXiv:2007.07212}}].

\bibitem{DeLuca:2021hcf}
V.~De~Luca, G.~Franciolini, and A.~Riotto, {\it {Constraining the Initial
  Primordial Black Hole Clustering with CMB-distortion}},
  \href{http://arxiv.org/abs/2103.16369}{{\tt arXiv:2103.16369}}.

\bibitem{Ivanova:2005mi}
N.~Ivanova, K.~Belczynski, J.~M. Fregeau, and F.~A. Rasio, {\it {The Evolution
  of binary fractions in globular clusters}},  {\em Mon. Not. Roy. Astron.
  Soc.} {\bf 358} (2005) 572--584,
  [\href{http://arxiv.org/abs/astro-ph/0501131}{{\tt astro-ph/0501131}}].

\bibitem{2010ApJ...717..948I}
N.~Ivanova, S.~Chaichenets, J.~Fregeau, C.~O. Heinke, J.~C. Lombardi, Jr., and
  T.~Woods, {\it {Formation of black-hole X-ray binaries in globular
  clusters}},  {\em Astrophys. J.} {\bf 717} (2010) 948--957,
  [\href{http://arxiv.org/abs/1001.1767}{{\tt arXiv:1001.1767}}].

\bibitem{Rodriguez:2021qhl}
C.~L. Rodriguez et~al., {\it {Modeling Dense Star Clusters in the Milky Way and
  beyond with the Cluster Monte Carlo Code}},  {\em Astrophys. J. Supp.} {\bf
  258} (2022), no.~2 22, [\href{http://arxiv.org/abs/2106.02643}{{\tt
  arXiv:2106.02643}}].

\bibitem{Kritos:2020wcl}
K.~Kritos, V.~De~Luca, G.~Franciolini, A.~Kehagias, and A.~Riotto, {\it {The
  Astro-Primordial Black Hole Merger Rates: a Reappraisal}},  {\em JCAP} {\bf
  05} (2021) 039, [\href{http://arxiv.org/abs/2012.03585}{{\tt
  arXiv:2012.03585}}].

\bibitem{Garcia-Bellido:2017knh}
J.~Garc\'\i{}a-Bellido and S.~Nesseris, {\it {Gravitational wave energy
  emission and detection rates of Primordial Black Hole hyperbolic
  encounters}},  {\em Phys. Dark Univ.} {\bf 21} (2018) 61--69,
  [\href{http://arxiv.org/abs/1711.09702}{{\tt arXiv:1711.09702}}].

\bibitem{Garcia-Bellido:2021jlq}
J.~Garc\'\i{}a-Bellido, S.~Jaraba, and S.~Kuroyanagi, {\it {The stochastic
  gravitational wave background from close hyperbolic encounters of primordial
  black holes in dense clusters}},  \href{http://arxiv.org/abs/2109.11376}{{\tt
  arXiv:2109.11376}}.

\bibitem{Morras:2021atg}
G.~Morr\'as, J.~Garc\'\i{}a-Bellido, and S.~Nesseris, {\it {Search for black
  hole hyperbolic encounters with gravitational wave detectors}},  {\em Phys.
  Dark Univ.} {\bf 35} (2022) 100932,
  [\href{http://arxiv.org/abs/2110.08000}{{\tt arXiv:2110.08000}}].

\bibitem{Suyama:2019cst}
T.~Suyama and S.~Yokoyama, {\it {Clustering of primordial black holes with
  non-Gaussian initial fluctuations}},  {\em PTEP} {\bf 2019} (2019), no.~10
  103E02, [\href{http://arxiv.org/abs/1906.04958}{{\tt arXiv:1906.04958}}].

\bibitem{Atal:2020yic}
V.~Atal, A.~Sanglas, and N.~Triantafyllou, {\it {NANOGrav signal as mergers of
  Stupendously Large Primordial Black Holes}},
  \href{http://arxiv.org/abs/2012.14721}{{\tt arXiv:2012.14721}}.

\bibitem{DeLuca:2021hde}
V.~De~Luca, G.~Franciolini, P.~Pani, and A.~Riotto, {\it {The Minimum Testable
  Abundance of Primordial Black Holes at Future Gravitational-Wave Detectors}},
   \href{http://arxiv.org/abs/2106.13769}{{\tt arXiv:2106.13769}}.

\bibitem{DES:2020mpv}
{\bf DES} Collaboration, A.~Chen et~al., {\it {Constraints on dark matter to
  dark radiation conversion in the late universe with DES-Y1 and external
  data}},  {\em Phys. Rev. D} {\bf 103} (2021), no.~12 123528,
  [\href{http://arxiv.org/abs/2011.04606}{{\tt arXiv:2011.04606}}].

\bibitem{Pujolas:2021yaw}
O.~Pujolas, V.~Vaskonen, and H.~Veerm\"ae, {\it {Prospects for probing
  gravitational waves from primordial black hole binaries}},  {\em Phys. Rev.
  D} {\bf 104} (2021), no.~8 083521,
  [\href{http://arxiv.org/abs/2107.03379}{{\tt arXiv:2107.03379}}].

\bibitem{Navarro:1995iw}
J.~F. Navarro, C.~S. Frenk, and S.~D.~M. White, {\it {The Structure of cold
  dark matter halos}},  {\em Astrophys. J.} {\bf 462} (1996) 563--575,
  [\href{http://arxiv.org/abs/astro-ph/9508025}{{\tt astro-ph/9508025}}].

\bibitem{Navarro:1996gj}
J.~F. Navarro, C.~S. Frenk, and S.~D.~M. White, {\it {A Universal density
  profile from hierarchical clustering}},  {\em Astrophys. J.} {\bf 490} (1997)
  493--508, [\href{http://arxiv.org/abs/astro-ph/9611107}{{\tt
  astro-ph/9611107}}].

\bibitem{Cautun:2019eaf}
M.~Cautun, A.~Benitez-Llambay, A.~J. Deason, C.~S. Frenk, A.~Fattahi, F.~A.
  G\'omez, R.~J.~J. Grand, K.~A. Oman, J.~F. Navarro, and C.~M. Simpson, {\it
  {The Milky Way total mass profile as inferred from Gaia DR2}},  {\em Mon.
  Not. Roy. Astron. Soc.} {\bf 494} (2020), no.~3 4291--4313,
  [\href{http://arxiv.org/abs/1911.04557}{{\tt arXiv:1911.04557}}].

\bibitem{Maggiore:1900zz}
M.~Maggiore, {\em {Gravitational Waves. Vol. 1: Theory and Experiments}}.
\newblock Oxford Master Series in Physics. Oxford University Press, 2007.

\bibitem{Sathyaprakash:2009xs}
B.~S. Sathyaprakash and B.~F. Schutz, {\it {Physics, Astrophysics and Cosmology
  with Gravitational Waves}},  {\em Living Rev. Rel.} {\bf 12} (2009) 2,
  [\href{http://arxiv.org/abs/0903.0338}{{\tt arXiv:0903.0338}}].

\bibitem{Moore:2014lga}
C.~J. Moore, R.~H. Cole, and C.~P.~L. Berry, {\it {Gravitational-wave
  sensitivity curves}},  {\em Class. Quant. Grav.} {\bf 32} (2015), no.~1
  015014, [\href{http://arxiv.org/abs/1408.0740}{{\tt arXiv:1408.0740}}].

\bibitem{Harry_2010}
G.~M.~H. and, {\it Advanced {LIGO}: the next generation of gravitational wave
  detectors},  {\em Classical and Quantum Gravity} {\bf 27} (apr, 2010) 084006.

\bibitem{Hild:2010id}
S.~Hild et~al., {\it {Sensitivity Studies for Third-Generation Gravitational
  Wave Observatories}},  {\em Class. Quant. Grav.} {\bf 28} (2011) 094013,
  [\href{http://arxiv.org/abs/1012.0908}{{\tt arXiv:1012.0908}}].

\bibitem{Robson:2018ifk}
T.~Robson, N.~J. Cornish, and C.~Liu, {\it {The construction and use of LISA
  sensitivity curves}},  {\em Class. Quant. Grav.} {\bf 36} (2019), no.~10
  105011, [\href{http://arxiv.org/abs/1803.01944}{{\tt arXiv:1803.01944}}].

\bibitem{Berlin:2021txa}
A.~Berlin, D.~Blas, R.~Tito~D'Agnolo, S.~A.~R. Ellis, R.~Harnik, Y.~Kahn, and
  J.~Sch\"utte-Engel, {\it {Detecting High-Frequency Gravitational Waves with
  Microwave Cavities}},  \href{http://arxiv.org/abs/2112.11465}{{\tt
  arXiv:2112.11465}}.

\bibitem{Domcke:2022rgu}
V.~Domcke, C.~Garcia-Cely, and N.~L. Rodd, {\it {A novel search for
  high-frequency gravitational waves with low-mass axion haloscopes}},
  \href{http://arxiv.org/abs/2202.00695}{{\tt arXiv:2202.00695}}.

\bibitem{Berlin:2022hfx}
A.~Berlin et~al., {\it {Searches for New Particles, Dark Matter, and
  Gravitational Waves with SRF Cavities}},  in {\em {2022 Snowmass Summer
  Study}}, 3, 2022.
\newblock \href{http://arxiv.org/abs/2203.12714}{{\tt arXiv:2203.12714}}.

\bibitem{Dolgov:1992pu}
A.~Dolgov and J.~Silk, {\it {Baryon isocurvature fluctuations at small scales
  and baryonic dark matter}},  {\em Phys. Rev. D} {\bf 47} (1993) 4244--4255.

\bibitem{Carr:2017jsz}
B.~Carr, M.~Raidal, T.~Tenkanen, V.~Vaskonen, and H.~Veerm\"ae, {\it
  {Primordial black hole constraints for extended mass functions}},  {\em Phys.
  Rev.} {\bf D96} (2017), no.~2 023514,
  [\href{http://arxiv.org/abs/1705.05567}{{\tt arXiv:1705.05567}}].

\bibitem{Garcia-Bellido:2017fdg}
J.~Garc\'ia-Bellido, {\it {Massive Primordial Black Holes as Dark Matter and
  their detection with Gravitational Waves}},  {\em J. Phys. Conf. Ser.} {\bf
  840} [\href{http://arxiv.org/abs/1702.08275}{{\tt arXiv:1702.08275}}].

\bibitem{Gow:2019pok}
A.~D. Gow, C.~T. Byrnes, A.~Hall, and J.~A. Peacock, {\it {Primordial black
  hole merger rates: distributions for multiple LIGO observables}},  {\em JCAP}
  {\bf 01} (2020) 031, [\href{http://arxiv.org/abs/1911.12685}{{\tt
  arXiv:1911.12685}}].

\bibitem{Ajith:2009bn}
P.~Ajith et~al., {\it {Inspiral-merger-ringdown waveforms for black-hole
  binaries with non-precessing spins}},  {\em Phys. Rev. Lett.} {\bf 106}
  (2011) 241101, [\href{http://arxiv.org/abs/0909.2867}{{\tt
  arXiv:0909.2867}}].

\bibitem{Zhu:2011bd}
X.-J. Zhu, E.~Howell, T.~Regimbau, D.~Blair, and Z.-H. Zhu, {\it {Stochastic
  Gravitational Wave Background from Coalescing Binary Black Holes}},  {\em
  Astrophys. J.} {\bf 739} (2011) 86,
  [\href{http://arxiv.org/abs/1104.3565}{{\tt arXiv:1104.3565}}].

\bibitem{Ringwald:2020ist}
A.~Ringwald, J.~Sch\"utte-Engel, and C.~Tamarit, {\it {Gravitational Waves as a
  Big Bang Thermometer}},  {\em JCAP} {\bf 03} (2021) 054,
  [\href{http://arxiv.org/abs/2011.04731}{{\tt arXiv:2011.04731}}].

\bibitem{Bavera:2021wmw}
S.~S. Bavera, G.~Franciolini, G.~Cusin, A.~Riotto, M.~Zevin, and T.~Fragos,
  {\it {Stochastic gravitational-wave background as a tool to investigate
  multi-channel astrophysical and primordial black-hole mergers}},
  \href{http://arxiv.org/abs/2109.05836}{{\tt arXiv:2109.05836}}.

\bibitem{Christodoulou:1991cr}
D.~Christodoulou, {\it {Nonlinear nature of gravitation and gravitational wave
  experiments}},  {\em Phys. Rev. Lett.} {\bf 67} (1991) 1486--1489.

\bibitem{PhysRevD.44.R2945}
A.~G. Wiseman and C.~M. Will, {\it Christodoulou's nonlinear gravitational-wave
  memory: Evaluation in the quadrupole approximation},  {\em Phys. Rev. D} {\bf
  44} (Nov, 1991) R2945--R2949.

\bibitem{Blanchet:1992br}
L.~Blanchet and T.~Damour, {\it {Hereditary effects in gravitational
  radiation}},  {\em Phys. Rev. D} {\bf 46} (1992) 4304--4319.

\bibitem{Favata:2008ti}
M.~Favata, {\it {Gravitational-wave memory revisited: memory from the merger
  and recoil of binary black holes}},  {\em J. Phys. Conf. Ser.} {\bf 154}
  (2009) 012043, [\href{http://arxiv.org/abs/0811.3451}{{\tt
  arXiv:0811.3451}}].

\bibitem{Favata:2009ii}
M.~Favata, {\it {Nonlinear gravitational-wave memory from binary black hole
  mergers}},  {\em Astrophys. J. Lett.} {\bf 696} (2009) L159--L162,
  [\href{http://arxiv.org/abs/0902.3660}{{\tt arXiv:0902.3660}}].

\bibitem{Pollney:2010hs}
D.~Pollney and C.~Reisswig, {\it {Gravitational memory in binary black hole
  mergers}},  {\em Astrophys. J. Lett.} {\bf 732} (2011) L13,
  [\href{http://arxiv.org/abs/1004.4209}{{\tt arXiv:1004.4209}}].

\bibitem{Lasky:2016knh}
P.~D. Lasky, E.~Thrane, Y.~Levin, J.~Blackman, and Y.~Chen, {\it {Detecting
  gravitational-wave memory with LIGO: implications of GW150914}},  {\em Phys.
  Rev. Lett.} {\bf 117} (2016), no.~6 061102,
  [\href{http://arxiv.org/abs/1605.01415}{{\tt arXiv:1605.01415}}].

\bibitem{Hubner:2019sly}
M.~H\"ubner, C.~Talbot, P.~D. Lasky, and E.~Thrane, {\it {Measuring
  gravitational-wave memory in the first LIGO/Virgo gravitational-wave
  transient catalog}},  {\em Phys. Rev. D} {\bf 101} (2020), no.~2 023011,
  [\href{http://arxiv.org/abs/1911.12496}{{\tt arXiv:1911.12496}}].

\bibitem{Ebersold:2020zah}
M.~Ebersold and S.~Tiwari, {\it {Search for nonlinear memory from subsolar mass
  compact binary mergers}},  {\em Phys. Rev. D} {\bf 101} (2020), no.~10
  104041, [\href{http://arxiv.org/abs/2005.03306}{{\tt arXiv:2005.03306}}].

\bibitem{Zhao:2021hmx}
Z.-C. Zhao, X.~Liu, Z.~Cao, and X.~He, {\it {Gravitational wave memory of the
  binary black hole events in GWTC-2}},  {\em Phys. Rev. D} {\bf 104} (2021),
  no.~6 064056, [\href{http://arxiv.org/abs/2111.13882}{{\tt
  arXiv:2111.13882}}].

\bibitem{PhysRevD.45.520}
K.~S. Thorne, {\it Gravitational-wave bursts with memory: The christodoulou
  effect},  {\em Phys. Rev. D} {\bf 45} (Jan, 1992) 520--524.

\bibitem{Johnson:2018xly}
A.~D. Johnson, S.~J. Kapadia, A.~Osborne, A.~Hixon, and D.~Kennefick, {\it
  {Prospects of detecting the nonlinear gravitational wave memory}},  {\em
  Phys. Rev. D} {\bf 99} (2019), no.~4 044045,
  [\href{http://arxiv.org/abs/1810.09563}{{\tt arXiv:1810.09563}}].

\bibitem{PhysRevLett.118.181103}
L.~O. McNeill, E.~Thrane, and P.~D. Lasky, {\it Detecting gravitational wave
  memory without parent signals},  {\em Phys. Rev. Lett.} {\bf 118} (May, 2017)
  181103.

\bibitem{Domenech:2021odz}
G.~Dom\`enech, {\it {Were recently reported MHz events planet mass primordial
  black hole mergers?}},  {\em Eur. Phys. J. C} {\bf 81} (2021), no.~11 1042,
  [\href{http://arxiv.org/abs/2110.00550}{{\tt arXiv:2110.00550}}].

\bibitem{Lasky:2021naa}
P.~D. Lasky and E.~Thrane, {\it {Did Goryachev et al. detect megahertz
  gravitational waves?}},  {\em Phys. Rev. D} {\bf 104} (2021), no.~10 103017,
  [\href{http://arxiv.org/abs/2110.13319}{{\tt arXiv:2110.13319}}].

\bibitem{Peccei:1977hh}
R.~D. Peccei and H.~R. Quinn, {\it {CP Conservation in the Presence of
  Instantons}},  {\em Phys. Rev. Lett.} {\bf 38} (1977) 1440--1443.

\bibitem{Wilczek:1977pj}
F.~Wilczek, {\it {Problem of Strong $P$ and $T$ Invariance in the Presence of
  Instantons}},  {\em Phys. Rev. Lett.} {\bf 40} (1978) 279--282.

\bibitem{Weinberg:1977ma}
S.~Weinberg, {\it {A New Light Boson?}},  {\em Phys. Rev. Lett.} {\bf 40}
  (1978) 223--226.

\bibitem{Arvanitaki:2009fg}
A.~Arvanitaki, S.~Dimopoulos, S.~Dubovsky, N.~Kaloper, and J.~March-Russell,
  {\it {String Axiverse}},  {\em Phys. Rev. D} {\bf 81} (2010) 123530,
  [\href{http://arxiv.org/abs/0905.4720}{{\tt arXiv:0905.4720}}].

\bibitem{Ternov:1978gq}
I.~M. Ternov, V.~R. Khalilov, G.~A. Chizhov, and A.~B. Gaina, {\it {Finite
  motion of massive particles in the Kerr and Schwarzschild fields}},  {\em
  Sov. Phys. J.} {\bf 21} (1978) 1200--1204.

\bibitem{Zouros:1979iw}
T.~J.~M. Zouros and D.~M. Eardley, {\it {INSTABILITIES OF MASSIVE SCALAR
  PERTURBATIONS OF A ROTATING BLACK HOLE}},  {\em Annals Phys.} {\bf 118}
  (1979) 139--155.

\bibitem{Arvanitaki:2010sy}
A.~Arvanitaki and S.~Dubovsky, {\it {Exploring the String Axiverse with
  Precision Black Hole Physics}},  {\em Phys. Rev. D} {\bf 83} (2011) 044026,
  [\href{http://arxiv.org/abs/1004.3558}{{\tt arXiv:1004.3558}}].

\bibitem{Arvanitaki:2012cn}
A.~Arvanitaki and A.~A. Geraci, {\it {Detecting high-frequency gravitational
  waves with optically-levitated sensors}},  {\em Phys. Rev. Lett.} {\bf 110}
  (2013), no.~7 071105, [\href{http://arxiv.org/abs/1207.5320}{{\tt
  arXiv:1207.5320}}].

\bibitem{arXiv:2010.13157}
N.~Aggarwal, G.~P. Winstone, M.~Teo, M.~Baryakhtar, S.~L. Larson, V.~Kalogera,
  and A.~A. Geraci, {\it {Searching for new physics with a
  levitated-sensor-based gravitational-wave detector}},
  \href{http://arxiv.org/abs/2010.13157}{{\tt arXiv:2010.13157}}.

\bibitem{Detweiler:1980uk}
S.~L. Detweiler, {\it {KLEIN-GORDON EQUATION AND ROTATING BLACK HOLES}},  {\em
  Phys. Rev. D} {\bf 22} (1980) 2323--2326.

\bibitem{Yoshino:2013ofa}
H.~Yoshino and H.~Kodama, {\it {Gravitational radiation from an axion cloud
  around a black hole: Superradiant phase}},  {\em PTEP} {\bf 2014} (2014)
  043E02, [\href{http://arxiv.org/abs/1312.2326}{{\tt arXiv:1312.2326}}].

\bibitem{Arvanitaki:2014wva}
A.~Arvanitaki, M.~Baryakhtar, and X.~Huang, {\it {Discovering the QCD Axion
  with Black Holes and Gravitational Waves}},  {\em Phys. Rev. D} {\bf 91}
  (2015), no.~8 084011, [\href{http://arxiv.org/abs/1411.2263}{{\tt
  arXiv:1411.2263}}].

\bibitem{Brito:2014wla}
R.~Brito, V.~Cardoso, and P.~Pani, {\it {Black holes as particle detectors:
  evolution of superradiant instabilities}},  {\em Class. Quant. Grav.} {\bf
  32} (2015), no.~13 134001, [\href{http://arxiv.org/abs/1411.0686}{{\tt
  arXiv:1411.0686}}].

\bibitem{Brito:2015oca}
R.~Brito, V.~Cardoso, and P.~Pani, {\it {Superradiance}},  {\em Lect. Notes
  Phys.} {\bf 906} (2015) pp.1--237,
  [\href{http://arxiv.org/abs/1501.06570}{{\tt arXiv:1501.06570}}].

\bibitem{DeLuca:2019buf}
V.~De~Luca, V.~Desjacques, G.~Franciolini, A.~Malhotra, and A.~Riotto, {\it
  {The initial spin probability distribution of primordial black holes}},  {\em
  JCAP} {\bf 05} (2019) 018, [\href{http://arxiv.org/abs/1903.01179}{{\tt
  arXiv:1903.01179}}].

\bibitem{Mirbabayi:2019uph}
M.~Mirbabayi, A.~Gruzinov, and J.~Nore{\~n}a, {\it {Spin of Primordial Black
  Holes}},  {\em JCAP} {\bf 2003} (2020) 017,
  [\href{http://arxiv.org/abs/1901.05963}{{\tt arXiv:1901.05963}}].

\bibitem{Harada:2017fjm}
T.~Harada, C.-M. Yoo, K.~Kohri, and K.-I. Nakao, {\it {Spins of primordial
  black holes formed in the matter-dominated phase of the Universe}},  {\em
  Phys. Rev. D} {\bf 96} (2017), no.~8 083517,
  [\href{http://arxiv.org/abs/1707.03595}{{\tt arXiv:1707.03595}}]. [Erratum:
  Phys.Rev.D 99, 069904 (2019)].

\bibitem{Flores:2021tmc}
M.~M. Flores and A.~Kusenko, {\it {Spins of primordial black holes formed in
  different cosmological scenarios}},  {\em Phys. Rev. D} {\bf 104} (2021),
  no.~6 063008, [\href{http://arxiv.org/abs/2106.03237}{{\tt
  arXiv:2106.03237}}].

\bibitem{Dvali:2021byy}
G.~Dvali, F.~K\"uhnel, and M.~Zantedeschi, {\it {Primordial Black Holes from
  Confinement}},  \href{http://arxiv.org/abs/2108.09471}{{\tt
  arXiv:2108.09471}}.

\bibitem{Eroshenko:2021sez}
Y.~N. Eroshenko, {\it {Spin of primordial black holes in the model with
  collapsing domain walls}},  \href{http://arxiv.org/abs/2111.03403}{{\tt
  arXiv:2111.03403}}.

\bibitem{Chongchitnan:2021ehn}
S.~Chongchitnan and J.~Silk, {\it {Extreme-value statistics of the spin of
  primordial black holes}},  {\em Phys. Rev. D} {\bf 104} (2021), no.~8 083018,
  [\href{http://arxiv.org/abs/2109.12268}{{\tt arXiv:2109.12268}}].

\bibitem{deFreitasPacheco:2020wdg}
J.~A. de~Freitas~Pacheco and J.~Silk, {\it {Primordial Rotating Black Holes}},
  {\em Phys. Rev. D} {\bf 101} (2020), no.~8 083022,
  [\href{http://arxiv.org/abs/2003.12072}{{\tt arXiv:2003.12072}}].

\bibitem{Barausse:2009uz}
E.~Barausse and L.~Rezzolla, {\it {Predicting the direction of the final spin
  from the coalescence of two black holes}},  {\em Astrophys. J. Lett.} {\bf
  704} (2009) L40--L44, [\href{http://arxiv.org/abs/0904.2577}{{\tt
  arXiv:0904.2577}}].

\bibitem{Aggarwal:2020umq}
N.~Aggarwal, G.~P. Winstone, M.~Teo, M.~Baryakhtar, S.~L. Larson, V.~Kalogera,
  and A.~A. Geraci, {\it {Searching for new physics with a
  levitated-sensor-based gravitational-wave detector}},
  \href{http://arxiv.org/abs/2010.13157}{{\tt arXiv:2010.13157}}.

\bibitem{Aguiar:2010kn}
O.~D. Aguiar, {\it {The Past, Present and Future of the Resonant-Mass
  Gravitational Wave Detectors}},  {\em Res. Astron. Astrophys.} {\bf 11}
  (2011) 1--42, [\href{http://arxiv.org/abs/1009.1138}{{\tt arXiv:1009.1138}}].

\bibitem{Harry:1996gh}
G.~M. Harry, T.~R. Stevenson, and H.~J. Paik, {\it {Detectability of
  gravitational wave events by spherical resonant mass antennas}},  {\em Phys.
  Rev. D} {\bf 54} (1996) 2409--2420,
  [\href{http://arxiv.org/abs/gr-qc/9602018}{{\tt gr-qc/9602018}}].

\bibitem{Goryachev:2014yra}
M.~Goryachev and M.~E. Tobar, {\it {Gravitational Wave Detection with High
  Frequency Phonon Trapping Acoustic Cavities}},  {\em Phys. Rev. D} {\bf 90}
  (2014), no.~10 102005, [\href{http://arxiv.org/abs/1410.2334}{{\tt
  arXiv:1410.2334}}].

\bibitem{Page:2020zbr}
M.~A. Page et~al., {\it {Gravitational wave detectors with broadband high
  frequency sensitivity}},  {\em Commun. Phys.} {\bf 4} (2021) 27,
  [\href{http://arxiv.org/abs/2007.08766}{{\tt arXiv:2007.08766}}].

\bibitem{Goryachev:2021zzn}
M.~Goryachev, W.~M. Campbell, I.~S. Heng, S.~Galliou, E.~N. Ivanov, and M.~E.
  Tobar, {\it {Rare Events Detected with a Bulk Acoustic Wave High Frequency
  Gravitational Wave Antenna}},  {\em Phys. Rev. Lett.} {\bf 127} (2021), no.~7
  071102, [\href{http://arxiv.org/abs/2102.05859}{{\tt arXiv:2102.05859}}].

\bibitem{Herman:2020wao}
N.~Herman, A.~F\"uzfa, S.~Clesse, and L.~Lehoucq, {\it {Detecting
  Planetary-mass Primordial Black Holes with Resonant Electromagnetic
  Gravitational Wave Detectors}},  \href{http://arxiv.org/abs/2012.12189}{{\tt
  arXiv:2012.12189}}.

\bibitem{Herman:2022fau}
N.~Herman, L.~Lehoucq, and A.~F\'{u}zfa, {\it {Electromagnetic Antennas for the
  Resonant Detection of the Stochastic Gravitational Wave Background}},
  \href{http://arxiv.org/abs/2203.15668}{{\tt arXiv:2203.15668}}.

\bibitem{Holometer:2016qoh}
{\bf Holometer} Collaboration, A.~S. Chou et~al., {\it {MHz Gravitational Wave
  Constraints with Decameter Michelson Interferometers}},  {\em Phys. Rev. D}
  {\bf 95} (2017), no.~6 063002, [\href{http://arxiv.org/abs/1611.05560}{{\tt
  arXiv:1611.05560}}].

\bibitem{Akutsu:2008qv}
T.~Akutsu et~al., {\it {Search for a stochastic background of 100-MHz
  gravitational waves with laser interferometers}},  {\em Phys. Rev. Lett.}
  {\bf 101} (2008) 101101, [\href{http://arxiv.org/abs/0803.4094}{{\tt
  arXiv:0803.4094}}].

\bibitem{Ito:2019wcb}
A.~Ito, T.~Ikeda, K.~Miuchi, and J.~Soda, {\it {Probing GHz gravitational waves
  with graviton\textendash{}magnon resonance}},  {\em Eur. Phys. J. C} {\bf 80}
  (2020), no.~3 179, [\href{http://arxiv.org/abs/1903.04843}{{\tt
  arXiv:1903.04843}}].

\bibitem{Fixsen:2009xn}
D.~J. Fixsen et~al., {\it {ARCADE 2 Measurement of the Extra-Galactic Sky
  Temperature at 3-90 GHz}},  {\em Astrophys. J.} {\bf 734} (2011) 5,
  [\href{http://arxiv.org/abs/0901.0555}{{\tt arXiv:0901.0555}}].

\bibitem{Bowman:2018yin}
J.~D. Bowman, A.~E.~E. Rogers, R.~A. Monsalve, T.~J. Mozdzen, and N.~Mahesh,
  {\it {An absorption profile centred at 78 megahertz in the sky-averaged
  spectrum}},  {\em Nature} {\bf 555} (2018), no.~7694 67--70,
  [\href{http://arxiv.org/abs/1810.05912}{{\tt arXiv:1810.05912}}].

\bibitem{Domcke:2020yzq}
V.~Domcke and C.~Garcia-Cely, {\it {Potential of radio telescopes as
  high-frequency gravitational wave detectors}},  {\em Phys. Rev. Lett.} {\bf
  126} (2021), no.~2 021104, [\href{http://arxiv.org/abs/2006.01161}{{\tt
  arXiv:2006.01161}}].

\bibitem{gertsenshtein1962wave}
M.~Gertsenshtein, {\it Wave resonance of light and gravitional waves},  {\em
  Sov Phys JETP, 1962, 14: 84} {\bf 85} (1962).

\bibitem{Boccaletti1970ConversionOP}
D.~Boccaletti, V.~de~Sabbata, P.~Fortini, and C.~B. Gualdi, {\it Conversion of
  photons into gravitons and vice versa in a static electromagnetic field},
  {\em Il Nuovo Cimento B (1965-1970)} {\bf 70} (1970) 129--146.

\bibitem{Zioutas:1998cc}
K.~Zioutas et~al., {\it {A Decommissioned LHC model magnet as an axion
  telescope}},  {\em Nucl. Instrum. Meth. A} {\bf 425} (1999) 480--489,
  [\href{http://arxiv.org/abs/astro-ph/9801176}{{\tt astro-ph/9801176}}].

\bibitem{GraciaGarza:2015sos}
J.~Gracia~Garza, {\it {Micromegas for the search of solar axions in CAST and
  low-mass WIMPs in TREX-DM}},  other thesis, U. Zaragoza, LFNAE, 11, 2015.

\bibitem{Ruz:2018omp}
J.~Ruz et~al., {\it {Next Generation Search for Axion and ALP Dark Matter with
  the International Axion Observatory}},  in {\em {2018 IEEE Nuclear Science
  Symposium and Medical Imaging Conference}}, p.~8824640, 2018.

\bibitem{OSQAR:2015qdv}
{\bf OSQAR} Collaboration, R.~Ballou et~al., {\it {New exclusion limits on
  scalar and pseudoscalar axionlike particles from light shining through a
  wall}},  {\em Phys. Rev. D} {\bf 92} (2015), no.~9 092002,
  [\href{http://arxiv.org/abs/1506.08082}{{\tt arXiv:1506.08082}}].

\bibitem{OSQAR:2013jqp}
{\bf OSQAR} Collaboration, P.~Pugnat et~al., {\it {Search for weakly
  interacting sub-eV particles with the OSQAR laser-based experiment: results
  and perspectives}},  {\em Eur. Phys. J. C} {\bf 74} (2014), no.~8 3027,
  [\href{http://arxiv.org/abs/1306.0443}{{\tt arXiv:1306.0443}}].

\bibitem{ALPS:2009des}
{\bf ALPS} Collaboration, K.~Ehret et~al., {\it {Resonant laser power build-up
  in ALPS: A 'Light-shining-through-walls' experiment}},  {\em Nucl. Instrum.
  Meth. A} {\bf 612} (2009) 83--96, [\href{http://arxiv.org/abs/0905.4159}{{\tt
  arXiv:0905.4159}}].

\bibitem{Ehret:2010mh}
K.~Ehret et~al., {\it {New ALPS Results on Hidden-Sector Lightweights}},  {\em
  Phys. Lett. B} {\bf 689} (2010) 149--155,
  [\href{http://arxiv.org/abs/1004.1313}{{\tt arXiv:1004.1313}}].

\bibitem{Bahre:2013ywa}
R.~B\"ahre et~al., {\it {Any light particle search II \textemdash{}Technical
  Design Report}},  {\em JINST} {\bf 8} (2013) T09001,
  [\href{http://arxiv.org/abs/1302.5647}{{\tt arXiv:1302.5647}}].

\bibitem{Albrecht:2020ntd}
C.~Albrecht, S.~Barbanotti, H.~Hintz, K.~Jensch, R.~Klos, W.~Maschmann,
  O.~Sawlanski, M.~Stolper, and D.~Trines, {\it {Straightening of
  Superconducting HERA Dipoles for the Any-Light-Particle-Search Experiment
  ALPS II}},  {\em EPJ Tech. Instrum.} {\bf 8} (2021), no.~1 5,
  [\href{http://arxiv.org/abs/2004.13441}{{\tt arXiv:2004.13441}}].

\bibitem{Beacham:2019nyx}
J.~Beacham et~al., {\it {Physics Beyond Colliders at CERN: Beyond the Standard
  Model Working Group Report}},  {\em J. Phys. G} {\bf 47} (2020), no.~1
  010501, [\href{http://arxiv.org/abs/1901.09966}{{\tt arXiv:1901.09966}}].

\bibitem{Ejlli:2019bqj}
A.~Ejlli, D.~Ejlli, A.~M. Cruise, G.~Pisano, and H.~Grote, {\it {Upper limits
  on the amplitude of ultra-high-frequency gravitational waves from graviton to
  photon conversion}},  {\em Eur. Phys. J. C} {\bf 79} (2019), no.~12 1032,
  [\href{http://arxiv.org/abs/1908.00232}{{\tt arXiv:1908.00232}}].

\bibitem{Li:2000du}
F.-Y. Li, M.-X. Tang, J.~Luo, and Y.-C. Li, {\it {Electrodynamical response of
  a high-energy photon flux to a gravitational wave}},  {\em Phys. Rev. D} {\bf
  62} (2000) 044018.

\bibitem{Li:2003tv}
F.-Y. Li, M.-X. Tang, and D.-P. Shi, {\it {Electromagnetic response of a
  Gaussian beam to high frequency relic gravitational waves in quintessential
  inflationary models}},  {\em Phys. Rev. D} {\bf 67} (2003) 104008,
  [\href{http://arxiv.org/abs/gr-qc/0306092}{{\tt gr-qc/0306092}}].

\bibitem{Li:2004df}
F.-Y. Li and N.~Yang, {\it {Resonant interaction between a weak gravitational
  wave and a microwave beam in the double polarized states through a static
  magnetic field}},  {\em Chin. Phys. Lett.} {\bf 21} (2004) 2113--2116,
  [\href{http://arxiv.org/abs/gr-qc/0410060}{{\tt gr-qc/0410060}}].

\bibitem{Li:2006sx}
F.~Li, R.~M.~L. Baker, Jr., and Z.~Chen, {\it {Perturbative photon flux
  generated by high-frequency relic gravitational waves and utilization of them
  for their detection}},  \href{http://arxiv.org/abs/gr-qc/0604109}{{\tt
  gr-qc/0604109}}.

\bibitem{Li:2008qr}
F.~Li, R.~M.~L. Baker, Jr., Z.~Fang, G.~V. Stephenson, and Z.~Chen, {\it
  {Perturbative Photon Fluxes Generated by High-Frequency Gravitational Waves
  and Their Physical Effects}},  {\em Eur. Phys. J. C} {\bf 56} (2008)
  407--423, [\href{http://arxiv.org/abs/0806.1989}{{\tt arXiv:0806.1989}}].

\bibitem{Tong:2008rz}
M.-l. Tong, Y.~Zhang, and F.-Y. Li, {\it {Using polarized maser to detect
  high-frequency relic gravitational waves}},  {\em Phys. Rev. D} {\bf 78}
  (2008) 024041, [\href{http://arxiv.org/abs/0807.0885}{{\tt
  arXiv:0807.0885}}].

\bibitem{Stephenson:2009zz}
G.~V. Stephenson, {\it {The standard quantum limit for the Li-Baker HFGW
  detector}},  {\em AIP Conf. Proc.} {\bf 1103} (2009), no.~1 542--547.

\bibitem{Li:2009zzy}
F.~Li, N.~Yang, Z.~Fang, R.~M.~L. Baker, Jr., G.~V. Stephenson, and H.~Wen,
  {\it {Signal Photon Flux and Background Noise in a Coupling Electromagnetic
  Detecting System for High Frequency Gravitational Waves}},  {\em Phys. Rev.
  D} {\bf 80} (2009) 064013, [\href{http://arxiv.org/abs/0909.4118}{{\tt
  arXiv:0909.4118}}].

\bibitem{Li:2011zzl}
J.~Li, K.~Lin, F.~Li, and Y.~Zhong, {\it {The signal photon flux, background
  photons and shot noise in electromagnetic response of high-frequency relic
  gravitational waves}},  {\em Gen. Rel. Grav.} {\bf 43} (2011) 2209--2222.

\bibitem{Li:2013fna}
F.-Y. Li, H.~Wen, and Z.-Y. Fang, {\it {High-frequency gravitational waves
  having large spectral densities and their electromagnetic response}},  {\em
  Chin. Phys. B} {\bf 22} (2013), no.~12 120402.

\bibitem{Li:2014bma}
J.~Li, L.~Zhang, K.~Lin, and H.~Wen, {\it {Resonance of Gaussian
  electromagnetic field to the high frequency gravitational waves}},  {\em Int.
  J. Theor. Phys.} {\bf 55} (2016), no.~8 3506--3514,
  [\href{http://arxiv.org/abs/1411.1811}{{\tt arXiv:1411.1811}}].

\bibitem{Li:2015nti}
J.~Li, L.~Zhang, and H.~Wen, {\it {Optimization of the Electromagnetic (EM)
  Perturbative Effects Produced by High-Frequency Gravitational Waves}},  {\em
  Int. J. Theor. Phys.} {\bf 55} (2016), no.~3 1871--1881.

\bibitem{PhysRevLett.89.223901}
W.~Wen, L.~Zhou, J.~Li, W.~Ge, C.~T. Chan, and P.~Sheng, {\it Subwavelength
  photonic band gaps from planar fractals},  {\em Phys. Rev. Lett.} {\bf 89}
  (Nov, 2002) 223901.

\bibitem{doi:10.1063/1.1553993}
L.~Zhou, W.~Wen, C.~T. Chan, and P.~Sheng, {\it Reflectivity of planar metallic
  fractal patterns},  {\em Applied Physics Letters} {\bf 82} (2003), no.~7
  1012--1014,
  [\href{http://arxiv.org/abs/https://doi.org/10.1063/1.1553993}{{\tt
  https://doi.org/10.1063/1.1553993}}].

\bibitem{Hou:05}
B.~Hou, G.~Xu, H.~K. Wong, and W.~Wen, {\it Tuning of photonic bandgaps by a
  field-induced structural change of fractal metamaterials},  {\em Opt.
  Express} {\bf 13} (Nov, 2005) 9149--9154.

\bibitem{Woods:2012upj}
R.~C. Woods, {\it {Diffraction from Embedded Reflectors in Li-Baker HFGW
  Detector}},  {\em Phys. Procedia} {\bf 38} (2012) 66--76.

\bibitem{Kahn:2016aff}
Y.~Kahn, B.~R. Safdi, and J.~Thaler, {\it {Broadband and Resonant Approaches to
  Axion Dark Matter Detection}},  {\em Phys. Rev. Lett.} {\bf 117} (2016),
  no.~14 141801, [\href{http://arxiv.org/abs/1602.01086}{{\tt
  arXiv:1602.01086}}].

\bibitem{Ouellet:2018beu}
J.~L. Ouellet et~al., {\it {First Results from ABRACADABRA-10 cm: A Search for
  Sub-$\mu$eV Axion Dark Matter}},  {\em Phys. Rev. Lett.} {\bf 122} (2019),
  no.~12 121802, [\href{http://arxiv.org/abs/1810.12257}{{\tt
  arXiv:1810.12257}}].

\bibitem{Ouellet:2019tlz}
J.~L. Ouellet et~al., {\it {Design and implementation of the ABRACADABRA-10 cm
  axion dark matter search}},  {\em Phys. Rev. D} {\bf 99} (2019), no.~5
  052012, [\href{http://arxiv.org/abs/1901.10652}{{\tt arXiv:1901.10652}}].

\bibitem{Salemi:2021gck}
C.~P. Salemi et~al., {\it {Search for Low-Mass Axion Dark Matter with
  ABRACADABRA-10~cm}},  {\em Phys. Rev. Lett.} {\bf 127} (2021), no.~8 081801,
  [\href{http://arxiv.org/abs/2102.06722}{{\tt arXiv:2102.06722}}].

\bibitem{Gramolin:2020ict}
A.~V. Gramolin, D.~Aybas, D.~Johnson, J.~Adam, and A.~O. Sushkov, {\it {Search
  for axion-like dark matter with ferromagnets}},  {\em Nature Phys.} {\bf 17}
  (2021), no.~1 79--84, [\href{http://arxiv.org/abs/2003.03348}{{\tt
  arXiv:2003.03348}}].

\bibitem{Chaudhuri:2014dla}
S.~Chaudhuri, P.~W. Graham, K.~Irwin, J.~Mardon, S.~Rajendran, and Y.~Zhao,
  {\it {Radio for hidden-photon dark matter detection}},  {\em Phys. Rev. D}
  {\bf 92} (2015), no.~7 075012, [\href{http://arxiv.org/abs/1411.7382}{{\tt
  arXiv:1411.7382}}].

\bibitem{Silva-Feaver:2016qhh}
M.~Silva-Feaver et~al., {\it {Design Overview of DM Radio Pathfinder
  Experiment}},  {\em IEEE Trans. Appl. Supercond.} {\bf 27} (2017), no.~4
  1400204, [\href{http://arxiv.org/abs/1610.09344}{{\tt arXiv:1610.09344}}].

\bibitem{ADMX:2021nhd}
{\bf ADMX} Collaboration, C.~Bartram et~al., {\it {Search for Invisible Axion
  Dark Matter in the 3.3\textendash{}4.2\,\,\ensuremath{\mu}eV Mass Range}},
  {\em Phys. Rev. Lett.} {\bf 127} (2021), no.~26 261803,
  [\href{http://arxiv.org/abs/2110.06096}{{\tt arXiv:2110.06096}}].

\bibitem{ADMX:2019uok}
{\bf ADMX} Collaboration, T.~Braine et~al., {\it {Extended Search for the
  Invisible Axion with the Axion Dark Matter Experiment}},  {\em Phys. Rev.
  Lett.} {\bf 124} (2020), no.~10 101303,
  [\href{http://arxiv.org/abs/1910.08638}{{\tt arXiv:1910.08638}}].

\bibitem{ADMX:2018ogs}
{\bf ADMX} Collaboration, C.~Boutan et~al., {\it {Piezoelectrically Tuned
  Multimode Cavity Search for Axion Dark Matter}},  {\em Phys. Rev. Lett.} {\bf
  121} (2018), no.~26 261302, [\href{http://arxiv.org/abs/1901.00920}{{\tt
  arXiv:1901.00920}}].

\bibitem{HAYSTAC:2018rwy}
{\bf HAYSTAC} Collaboration, L.~Zhong et~al., {\it {Results from phase 1 of the
  HAYSTAC microwave cavity axion experiment}},  {\em Phys. Rev. D} {\bf 97}
  (2018), no.~9 092001, [\href{http://arxiv.org/abs/1803.03690}{{\tt
  arXiv:1803.03690}}].

\bibitem{Lee:2020cfj}
S.~Lee, S.~Ahn, J.~Choi, B.~R. Ko, and Y.~K. Semertzidis, {\it {Axion Dark
  Matter Search around 6.7 $\mu$eV}},  {\em Phys. Rev. Lett.} {\bf 124} (2020),
  no.~10 101802, [\href{http://arxiv.org/abs/2001.05102}{{\tt
  arXiv:2001.05102}}].

\bibitem{McAllister:2017lkb}
B.~T. McAllister, G.~Flower, E.~N. Ivanov, M.~Goryachev, J.~Bourhill, and M.~E.
  Tobar, {\it {The ORGAN Experiment: An axion haloscope above 15 GHz}},  {\em
  Phys. Dark Univ.} {\bf 18} (2017) 67--72,
  [\href{http://arxiv.org/abs/1706.00209}{{\tt arXiv:1706.00209}}].

\bibitem{Berlin:2019ahk}
A.~Berlin, R.~T. D'Agnolo, S.~A.~R. Ellis, C.~Nantista, J.~Neilson,
  P.~Schuster, S.~Tantawi, N.~Toro, and K.~Zhou, {\it {Axion Dark Matter
  Detection by Superconducting Resonant Frequency Conversion}},  {\em JHEP}
  {\bf 07} (2020), no.~07 088, [\href{http://arxiv.org/abs/1912.11048}{{\tt
  arXiv:1912.11048}}].

\bibitem{Berlin:2020vrk}
A.~Berlin, R.~T. D'Agnolo, S.~A.~R. Ellis, and K.~Zhou, {\it {Heterodyne
  broadband detection of axion dark matter}},  {\em Phys. Rev. D} {\bf 104}
  (2021), no.~11 L111701, [\href{http://arxiv.org/abs/2007.15656}{{\tt
  arXiv:2007.15656}}].

\end{thebibliography}\endgroup

\end{document}